\documentclass[review]{elsarticle}
\usepackage[]{natbib}
\usepackage{lineno,hyperref}
\biboptions{numbers,sort&compress}
\usepackage{amssymb}
\usepackage{subcaption}
\usepackage{multirow}
\usepackage{amsmath}
\usepackage{graphicx}
\usepackage{xcolor}
\usepackage{multirow}
\usepackage{commath}
\usepackage[top=0.75in, bottom=0.75in, left=0.875in, right=0.875in]{geometry}
\def\be{\begin{equation}}\def\ee{\end{equation}}
\def\ba{\begin{array}}\def\ea{\end{array}}
\def\bfg{\begin{figure}}\def\efg{\end{figure}}
\begin{document}

 \begin{frontmatter}
\title{Crack Dynamics in Rotating, Initially Stressed Material Strips: A Mathematical Approach}

\author[label1]{Soniya Chaudhary*}
\author[label1]{Diksha}
\author[label1]{Pawan Kumar Sharma}

\cortext[cor1]{Corresponding author: soniyachaudhary18@gmail.com}
\address[label1]{Department of Mathematics and Scientific Computing, National Institute of Technology Hamirpur, Himachal Pradesh, 177005, India}
\end{frontmatter}

%%%%%%%%*******************    ABSTRACT           ************************************************************************
\noindent {\bf Abstract:}
The current study explores the analysis of crack in initially stressed, rotating material strips, drawing insights from singular integral equations. In this work, a self-reinforced material strip with finite thickness and infinite extent, subjected to initial stress and rotational motion, has been considered to examine the Griffith fracture. The edges of the strip are pushed by constant loads from punches moving alongside it. This study makes waves in the material that affect the fracture's movement. A distinct mathematical technique is utilized to streamline the resolution of a pair of singular integral equations featuring First-order singularities. These obtained equations help us understand how the fracture behaves. The force acting at the fracture's edge is modeled using the Dirac delta function. Then, the Hilbert transformation method calculates the stress intensity factor (SIF) at the fracture's edge. Additionally, the study explores various scenarios, including constant intensity force without punch pressure, rotation parameter, initial stress, and isotropy in the strip, deduced from the SIF expression. Numerical computations and graphical analyses are conducted to assess the influence of various factors on SIF in the study. Finally, a comparison is made between the behavior of fractures in the initially stressed and rotating reinforced material strip and those in a standard material strip to identify any differences.

    \vspace*{0.05in}
    
 \section{Introduction}
 In recent years, significant focus has been directed toward investigating the mechanical stress and strain configurations within elastic solids encompassing fractures of confined size. Determining fracture and material failure requires understanding the stress field near a crack in the medium or material under consideration. 
A comprehensive comprehension of the material/medium's strength and stiffness, alongside the structural considerations regarding the load-bearing capacity of elastic solid bodies, whether with or without initial cracks, is facilitated through mathematical modeling of fracture problems in elastic media. In the field of fracture mechanics, the modeling of a moving fracture in elastic solids is undoubtedly fascinating. Many theoretical and analytical techniques in fracture or solid mechanics can be used to explain the forces that cause a fracture/crack and to show how the material resists the crack/fracture. The monograph provides a thorough explanation of dynamic fracture mechanics within the framework of mathematical physics \cite{michel1992freund}.

 Numerous engineering disciplines, including geology, rock mechanics, earthquake engineering, mechanical engineering, oil production, and civil engineering, have been significantly affected by the issues of moving cracks caused by elastic wave propagation. Achenbach et al. \cite{achenbach1976elastodynamic} investigated the stress field analysis around a fracture edge between two different isotropic elastic solids.
Williams \cite{williams1957stress}, \cite{williams1959stresses} addressed the issue of stress and displacement patterns in proximity to crack edges by employing a complex variable method.

There is a growing demand for obtaining analytical insights into the behavior of cracked or deformed bodies, prompting the consideration of realistic constitutive relations for anisotropic elastic mediums. In these materials, which display varying mechanical properties in different directions, the movement of cracks introduces complexity, particularly in its influence on wave propagation within the material. This interaction complicates the mathematical modeling necessary for analysis. Researchers employ sophisticated computational techniques and theoretical models to deal with this challenge, often combining finite element analysis or boundary element methods with appropriate constitutive relations for anisotropic materials. This approach aims to enhance our understanding and predictive capabilities regarding the mechanical responses of cracked anisotropic elastic bodies, which is crucial for various engineering applications such as structural integrity assessments, material design, and failure analysis. 

% The theory of linear elasticity for transversely isotropic materials explains the mechanical behavior of fiber-reinforced composite materials, particularly when the preferred direction aligns with the fiber direction, and the fibers are arranged in parallel straight lines.

Reinforced materials outperform traditional structural materials in civil and mechanical engineering. Reinforced materials are composites made by fortifying polymer fibers. Examples include alumina, graphite, and cement. Fiber-reinforced composite materials can become self-reinforced by reinforcing a matrix with identical fibers at specified pressures and temperatures. Due to the diverse composition of soft and hard rocks constituting the Earth's crust, which may possess inherent self-reinforcing characteristics, we opted to examine this self-reinforced material in our study. Eminent researchers focus on self-reinforced material medium issues due to the emergence of novel features linked to internal instability. Spencer \cite{spencer1972deformations} proposed a constitutive relation for a linearly anisotropic material with a preferred orientation, delineating it through mathematical formulations. 
Belfield et al. \cite{belfield1983stress}  explored integrating reinforcement into an elastic body, arranging fibers in concentric circles to bolster material strength. Verma and Rana \cite{verma1983rotation} explored the rotational dynamics of a circular cylindrical tube within the context of a similar fiber-reinforced model, wherein the reinforcement is dispersed along helical patterns. Several researchers have carried out a number of fascinating analyses to determine the characteristics of surface waves in self-reinforced materials \cite{sengupta2001surface,modi2016reflection,chattopadhyay2012torsional,ala2020shear}.
The stress intensity factor (SIF) is an essential factor in fracture mechanics that calculates the amount of stress near the crack edge due to mechanical loading or residual stresses. The Stress Intensity Factor (SIF) holds fundamental importance in fracture mechanics, serving as a crucial parameter governing crack growth rates and providing a basis for failure criteria. It plays a significant role in material safety analysis, structural stability assessments, and design analyses. Carpinteri \cite{carpinteri1992crack}, Rubio-Gonzalez and Mason \cite{rubio2000dynamic}, Ma and Hou \cite{ma1991transient}, and Viola et al. \cite{viola1989crack} have independently investigated and established expressions for SIF in various scenarios, including moving crack edges in anisotropic materials, crack propagation under dynamic loading, and fractures in orthotropic materials.
Several studies have explored moving cracks in diverse elastic solids exhibiting specific anisotropic characteristics, yielding formulations for stress intensity factor (SIF) expressions that consider many dynamic loading scenarios. The stress and displacement components of a Griffith crack were also studied by Awasthi et al. \cite{awasthi2022griffith}.
%%%===================================%%%%%%%%%%%%%%%%%%%
% Numerous distinguished researchers have devoted significant efforts to addressing the challenges associated with studying the impact of initial stress \cite{biot1940influence,  sharma2005effect,wang2007effect,gupta2010effect,zhang2013effects,kumar2014effect,shams2016effect,kundu2019effect,othmani2020effects, kumar2023effect} and rotation parameter \cite{ahmad2001effect,auriault2004body,roychoudhuri2005thermoelastic,zembaty2009tutorial,abd2011effect,gupta2014effect, sreelakshmi2015effect,choudhury2020wave,singh2020effect,hafed2023effect} on wave propagation phenomena.

Many researchers have analyzed wave propagation problems in initially stressed layered structures \cite{biot1940influence,  sharma2005effect,wang2007effect,gupta2010effect,zhang2013effects,kumar2014effect,shams2016effect,kundu2019effect,othmani2020effects, kumar2023effect} and rotating layered structures \cite{ahmad2001effect,auriault2004body,roychoudhuri2005thermoelastic,zembaty2009tutorial,abd2011effect,gupta2014effect, sreelakshmi2015effect,choudhury2020wave,singh2020effect,hafed2023effect} and observed that how initial stress and rotation parameters affect wave propagation phenomena.
To the extent of the authors' awareness, there has been no prior research examination investigating the effects of magnetoelastic plane waves on Griffith fracture propagation within a self-reinforced strip subjected to initial stress and rotations.\\
The present research investigates the fracture dynamics in material strips under initial stress and rotational forces, employing insights from singular integral equations. The study focuses on a self-reinforced material strip of finite thickness and infinite length to analyze Griffith fractures, subject to initial stress and rotational motion. Constant loads applied by moving punches along the strip's edges induce waves in the material, affecting fracture propagation. A specific mathematical approach is employed to solve a pair of singular integral equations characterized by First-order singularities, facilitating comprehension of fracture behavior. The force exerted at the fracture's edge is represented using the Dirac delta function, while the Hilbert transformation method computes the stress intensity factor (SIF) at the fracture edge. Furthermore, the research explores various scenarios, including constant force intensity without punch pressure, rotational parameters, initial stress levels, and strip isotropy inferred from the SIF expression. Numerical computations and graphical analyses are conducted to evaluate the influence of these factors on SIF. Lastly, a comparative analysis is performed between fractures in initially stressed and rotating reinforced material strips and those in standard material strips, aiming to discern any disparities in behavior. 

\section{Basic equation}

The equations that govern the behavior of a self-reinforced linearly elastic model under initial stress and rotation can be expressed as follows (Belfield et al., 1983):

\be
\left.
\ba{lll}
\displaystyle \tau_{lm}=&-P(\delta_{lm}+w_{lm})+\lambda e_{nn}\delta_{lm}+2\mu_{2} e_{lm}+a_1(b_{n}b_{o}e_{no}\delta_{lm}+e_{nn}b_{l}b_{m})\\
&+2(\mu_{1}-\mu_{2})(b_{l}b_{n}e_{nm}+b_{m}b_{n}e_{nl})+a_2 b_{n}b_{o}e_{no}b_{l}b_{m},\hspace{0.2cm} \:\:l,m,n,o=1,2,3
\ea
\right.
\label{eq_1}
\ee
here $\tau_{lm}$ represents the stress components, $e_{lm}$ denotes the infinitesimal strain components, P stands for the initial compressive stress, $ \Omega=(\Omega_1,\Omega_2,\Omega_3)$ represent the uniform angular velocity, $\delta_{lm}$ is Kronecker delta and $\vec{b}$=$(b_1,b_2,b_3)$ is a unit vector indicating the direction of reinforcement with ${\left|\vec{b}\right|}=1$. Additionally, the coefficients $a_1$, $a_2$, and $\lambda$ are elastic constants with stress dimensions, while  $\mu_{1}$ and $\mu_{2}$ represent the longitudinal and transverse shear moduli perpendicular and parallel to the preferred direction, respectively, and the expression of $w_{lm}$ and $e_{lm}$ are as
\be
\left.
\ba{lll}
\displaystyle
w_{lm} =\frac{1}{2}(d_{l,m}-d_{m,l}),\hspace{0.2cm} e_{lm}=\frac{1}{2}(d_{l,m}+d_{m,l}).
\ea
\right.
\label{eq_2}
\ee

\begin{figure}
    \centering
    \includegraphics[width=0.8\linewidth]{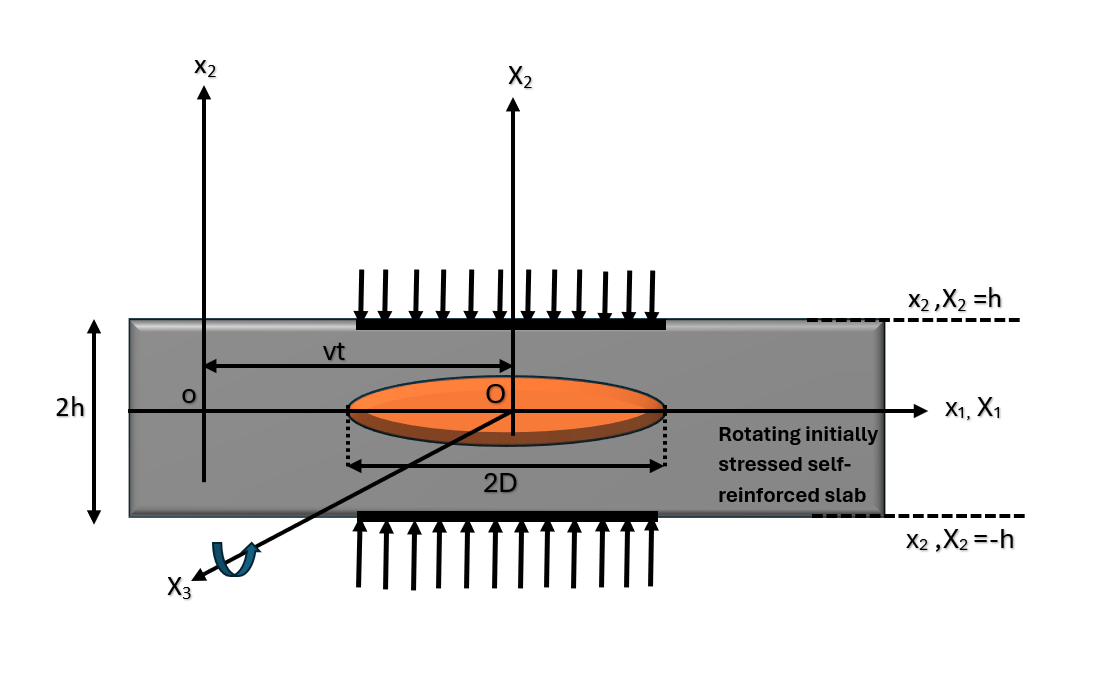}
    \caption{Geometry of the problem}
    \label{Figure 0}
\end{figure}
The equations of motion for the considered problem is
\be
\left.
\ba{lll}
\displaystyle
    \tau_{lm}+\vec{F_{l}}=\rho(\ddot{d_{l}}+\Omega_{m}d_{m}\Omega_{l}-\Omega^{2} d_{l}-2\epsilon_{lmn}\Omega_{m} \dot{d_{n}})
\ea
\right.
\label{eq_3}
\ee
where
\begin{equation*}
\vec{F_l}=(\vec{J}\times\vec{M})_l.   
\end{equation*}

The equation governing the electromagnetic field, according to Maxwell's foundational equation \cite{dunkin1963propagation,maugin1981wave} can be expressed as:

\be
\left.
\ba{lll}
\displaystyle
    \vec{\nabla} \cdot \vec{M} =0,\hspace{0.2cm}
    \vec{\nabla}\times\vec{E}=-\frac{\partial\vec{M}}{\partial t},\hspace{0.2cm} 
   \vec{\nabla}\times\vec{H} ={\vec{J}},\hspace{0.2cm} 
    \vec{M}=\mu_{e}\vec{H}.
\ea
\right.
\label{eq_4}
\ee

Integrating the generalized Ohm's law into the framework of a deformable continuum,
\be
\left.
\ba{lll}
\displaystyle
 \vec{J}=\sigma\left(\frac{\partial{\vec{d}}}{\partial{t}}\times\vec{M} + \vec{E} \right).
\ea
\right.
\label{eq_5}
\ee
In this equation, $\vec{J}$ denotes the electric current density, $\vec{M}$ represents the magnetic flux density, $\vec{E}$ signifies 
the electric intensity, $\mu_e$ symbolizes the magnetic permeability, $\sigma$ indicates the electric conductivity, $\vec{d}=(d_1,d_2,d_3)$ represents displacement components, and $t$ stands for time, and $\vec{H}$ denotes the magnetic field intensity. Additionally, the symbol $\vec{\nabla}$ is defined as $\frac{\partial }{\partial x_1}\hat{\imath} + \frac{\partial }{\partial x_2}\hat{\jmath} +\frac{\partial }{\partial x_3}\hat{k}$.
The equations that describe the motion of elastic vibrations within a highly conductive rotating self-reinforced initially stressed material medium, having electromagnetic effect represented by $\vec{J}\times\vec{M}$ (the Lorentz force), as the sole body force, are:
\be
\left.
\ba{lll}
\displaystyle
\vspace{0.2cm}
\frac{\partial{\tau_{11}}}{\partial{x_{1}}}+
    \frac{\partial{\tau_{12}}}{\partial{x_{2}}}+
    \frac{\partial{\tau_{13}}}{\partial{x_{3}}}+(\vec{J}\times\vec{M})_{x_{1}}=\rho \frac{\partial^2{d_1}}{\partial{t^{2}}}+ \rho[\Omega^{2}_1\ d_1+\Omega_1 \Omega_2 d_2+\Omega_1 \Omega_3 d_3-\Omega^{2}d_1-2\Omega_2\dot{d_{3}}+2\Omega_3\dot{d_2}],\\
    \vspace{0.2cm}
 \displaystyle \frac{\partial{\tau_{21}}}{\partial{x_{1}}}+
    \frac{\partial{\tau_{22}}}{\partial{x_{2}}}+
    \frac{\partial{\tau_{23}}}{\partial{x_{3}}}+(\vec{J}\times\vec{M})_{x_{2}}=\rho \frac{\partial^2{d_2}}{\partial{t^{2}}}+\rho[\Omega_2 \Omega_1 d_1+\Omega^2_{2}d_2+\Omega_2 \Omega_3 d_3-\Omega^{2}d_2-2\Omega_3\dot{d_{1}}+2\Omega_1\dot{d_3}],\\
    \vspace{0.2cm}
\displaystyle \frac{\partial{\tau_{31}}}{\partial{x_{1}}}+
    \frac{\partial{\tau_{32}}}{\partial{x_{2}}}+
    \frac{\partial{\tau_{33}}}{\partial{x_{3}}}+(\vec{J}\times\vec{M})_{x_{3}}=\rho \frac{\partial^2{d_3}}{\partial{t^{2}}}+\rho[\Omega_3 \Omega_1 d_1+\Omega_3 \Omega_2 d_2+\Omega^{2}_3{d_3}-\Omega^{2}d_3-2\Omega_1\dot{d_{2}}+2\Omega_2\dot{d_1}].
\ea
\right.
\label{eq_6}
\ee
In this context, {$((\vec{J}\times\vec{M})_{x_{1}},(\vec{J}\times\vec{M})_{x_{2}},(\vec{J}\times\vec{M})_{x_{3}})$} refers to the constituents of the body force along the $(x_1,x_2,x_3)$ axis directions, respectively. Additionally, $\rho$ denotes the mass density of the medium.
%%%%%%%%%%%%%%%%%%%%%%%%%%%%%%%%%%%%%%%%%%%%%%%%

 \section{Problem formulation} 

 This problem examines a fracture scenario in a finite-length, two-dimensional setting. It involves propagating a Magnetic wave in an elastic medium at the midline of an infinite rotating initial stressed self-reinforced slab with boundless and bounded thickness (2h). The slab is punched simultaneously on both faces in a parallel manner, precisely at the position corresponding to the fraction.
A system of Cartesian coordinates \( ox_1x_2x_3 \) has been established such that the slab can be characterized by \( -\infty < x_1 < \infty \) and \( -h \leq x_2 \leq h \). The crack is delimited by \( |\vec{x}_1| \leq D \) on \( x_2 = 0 \), while the smooth parallel punches of load apply within \( |\vec{x}_1| \leq D \) on \( x_2 = \pm h \). 
The assumption is made that the crack moves alongside the parallel punches, which exert force on the surfaces of the slab, at a uniform speed $v$ and without altering its length in the direction of positive $x_1$-axis. This problem configuration is depicted in Figure \ref{Figure 0}.
Plane-strain deformation condition concerning the fracture's motion induced by the propagation of a magneto-elastic plane wave within the self-reinforced strip, aligned with the $x_1x_2$-plane, is articulated as:

\be
\left.
\ba{lll}
\displaystyle
d_1=d_1(x_1,x_2,t),\hspace{0.2cm} d_2=d_2(x_1,x_2,t),\hspace{0.2cm} 
d_3=0, \hspace{0.2cm}
\frac{\partial}{\partial{x_3}}=0, \hspace{0.2cm} \Omega=\Omega(0,0,1).
\ea
\right.
\label{eq_7}
\ee

Since the $x_1x_2$-plane is chosen as the plane of symmetry for the self-reinforced medium, the self-reinforced components are oriented such that $b = (b_1, b_2, 0)$. Consequently, the pertinent components of the stress tensors, considering Eq. \ref{eq_1} to Eq. \ref{eq_7}, are expressed as:
 
\be
\left.
\ba{lll}
\displaystyle
\vspace{0.2cm}
\tau_{11}=-P+
 c_1\frac{\partial{d_1}}{\partial{x_1}}+
 c_2\frac{\partial{d_2}}{\partial{x_2}}+
 c_3\left(\frac{\partial{d_1}}{\partial{x_2}}+\frac{\partial{d_2}}{\partial{x_1}}\right),\\
 \vspace{0.2cm}
\displaystyle \tau_{12}=
     c_3\frac{\partial{d_1}}{\partial{x_1}}+
     c_4\frac{\partial{d_2}}{\partial{x_2}}+
     \left(c_5+\frac{P}{2}\right)\frac{\partial{d_1}}{\partial{x_2}}+
     \left(c_5-\frac{P}{2}\right)\frac{\partial{d_2}}{\partial{x_1}},\\
     \vspace{0.2cm}
\displaystyle     \tau_{22 }=-P+
    c_2\frac{\partial{d_1}}{\partial{x_1}}+
    c_6\frac{\partial{d_2}}{\partial{x_2}}+
    c_4\left(\frac{\partial{d_1}}{\partial{x_2}}+\frac{\partial{d_2}}{\partial{x_1}}\right)
\ea
\right.
\label{eq_8}
\ee
where
\begin{subequations}
    \begin{align}
     c_1&=\lambda+2\mu_{2}+2b^2_{1}a_1+4(\mu_{1}-\mu_{2})b^2_{1}+a_2 b^4_{1},\\
    c_2&=\lambda+a_1(b^{2}_{1}+b^{2}_{2})+a_2 b^2_{1}b^2_{2},\\
    c_3&= a_1 b_1 b_2+2(\mu_{1}-\mu_{2})b_1 b_2+a_2 b^3_1 b_2,\\
    c_4&=a_1 b_1b_2+2(\mu_{1}-\mu_{2})b_1 b_2+a_2 b_1b^2_3,\\
    c_5&= \mu_2+(\mu_{1}-\mu_{2})(b^2_1+b^2_2)+a_2 b^2_1 b^2_2,\\
    c_6&=\lambda+2\mu_2+2a_1 b^2_2+4(\mu_{1}-\mu_{2})b^2_2+a_2 b^4_2.
    \end{align}
\end{subequations}

It is posited that the induced magnetic field is represented by $\vec{h}=(h_{x_1},h_{x_2},h_{x_3})$, originating from the presence of the initial magnetic field $\vec{H_0}=(0,0,H_0)$, in addition to the perturbation produced by the transmission of a consider wave in a rotating magnetoelastic initial stressed self-reinforced slab. The combined magnetic flux density, denoted as $\vec{H}=(H_{x_1},H_{x_2},H_{x_3})$, encompasses both the primary and induced magnetic components. Regarding Eq. \ref{eq_5}, Eq. \ref{eq_4} yields:

\be
\left.
\ba{lll}
\displaystyle
    \nabla{^2{\vec{H}}}=\sigma \mu_{e}\left(\frac{\partial{\vec{H}}}{\partial{t}}-\vec{\nabla}\times\left(\frac{\partial \vec{d}}{\partial{t}}\times\vec{H}\right)\right)
\ea
\right.
\label{eq_10}
\ee

where
\be
\left.
\ba{lll}
\displaystyle
\vec{\nabla}\times\left(\frac{\partial \vec{d}}{\partial{t}}\times\vec{H}\right)=\left(0,0,-H_0\left\{\frac{\partial}{\partial{x_1}} \left(\frac{\partial{d_1}}{\partial{t}}\right)+
\frac{\partial}{\partial{}x_2} \left(\frac{\partial{d_2}}{\partial{t}}\right) \right\}\right).

\ea
\right.
\label{eq_11}
\ee
It's assumed that the joule heating effect is sufficiently negligible, such that the energy equation remains inactive. When dealing with a perfectly conducting medium (where conductivity, represented by $\sigma$, tends towards infinity), Equation \ref{eq_10} simplifies into the following component equation:

\be
\left.
\ba{lll}
\displaystyle
\left.
\frac{\partial{H_{x_1}}}{\partial{t}}=0,\hspace{0.2cm} \frac{\partial{H_{x_2}}}{\partial{t}}=0,\hspace{0.2cm} \frac{\partial{H_{x_3}}}{\partial{t}}=-H_0\left\{\frac{\partial}{\partial{x_1}} \left( \frac{\partial{d_1}}{\partial{t}}\right)+
\frac{\partial}{\partial{}x_2}\left(\frac{\partial{d_2}}{\partial{t}}\right)\right\}.
\right .
\ea
\right.
\label{eq_12}
\ee

The initial assumption is that the primary magnetic field remains uniform across the medium. Consequently, according to Eq. \ref{eq_12}, there are no disturbances in $H_{x_1}$ and $H_{x_2}$. However, there might be a perturbation in $H_{x_3}$, denoted by the small perturbation $h_{x_3}$. Assuming the initial value of $h$ to be zero, Eq. \ref{eq_12} can be expressed as:

\be
\ba{lll}
\displaystyle  
\frac{\partial{h_{x_1}}}{\partial{t}}=0,\hspace{0.3cm} \frac{\partial{h_{x_2}}}{\partial{t}}=0,\hspace{0.3cm} \frac{\partial{h_{x_3}}}{\partial{t}}=-H_0\left\{\frac{\partial}{\partial{x_1}} \left( \frac{\partial{d_1}}{\partial{t}}\right)+
\frac{\partial}{\partial{}x_2}\left(\frac{\partial{d_2}}{\partial{t}}\right)\right\}.
\ea
\label{eq_13}
\ee
Upon integration of Eq. \ref{eq_13}, we obtain the components of $\vec{h}$ as:

\be
\ba{lll}
\displaystyle
    h_{x_1}=0,\hspace{0.1cm} h_{x_2}=0,\hspace{0.1cm} h_{x_3}=-H_{0}\left\{\frac{\partial{d_1}}{\partial{x_1}}+\frac{\partial{d_2}}{\partial{x_2}}\right\}.
\ea
\label{eq_14}
\ee
Reducing the electromagnetic body force $\vec{J}\times\vec{M}=\mu_e ((\nabla\times\vec{h})\times\vec{H})$
in the component along $x_1$, $x_2 $ and $x_3$ directions leads to the following equations
\be
\ba{lll}
\displaystyle
  (\vec{J}\times\vec{M})_{x_1}=\mu_e H^2_{0}\left(\frac{\partial^{2}d_1}{\partial{x^2_1}}+\frac{\partial^2{d_2}}{\partial{x_1}\partial{x_2}}\right),\hspace{0.1cm} (\vec{J}\times\vec{M})_{x_2}=\mu_e H^2_{0}\left(\frac{\partial^{2}d_1}{\partial{x_1}\partial{x_2}}+\frac{\partial^2{d_2}}{\partial{x^2_2}}\right),\hspace{0.1cm} 
     (\vec{J}\times\vec{M})_{x_3}=0.
\ea
\label{eq_15}
\ee
%%%=========================================================
Utilizing equations \ref{eq_7}, \ref{eq_8}, and \ref{eq_15}, the dynamic equation governing a crack in motion, specifically equations \ref{eq_6}, leads to the derivation of two field equations

\begin{equation}
 \begin{aligned}
 \displaystyle
(c_1+\mu_e H^2_{0})\frac{\partial^2{d_1}}{\partial{x^2_1}}+(c_5-\frac{P}{2})\frac{\partial^2{d_1}}{\partial{x^2_{2}}}+2c_3\frac{\partial^2{d_1}}{\partial{x_1}\partial{x_2}}+c_3\frac{\partial^2{d_2}}{\partial{x^2_1}}+c_4\frac{\partial^2{d_2}}{\partial{x^2_2}}+(c_2+c_5+\frac{P}{2}+\mu_e H^2_{0})\frac{\partial^2{d_2}}{\partial{x_1}\partial{x_2}} \\
    =\rho\frac{\partial^2{d_1}}{\partial{t^2}}+\rho\left(-\Omega^2 d_1+2\Omega\frac{\partial{d_2}}{\partial{t}}\right),
\end{aligned}
\label{eq_16}
\end{equation}

%%===========================================
    
\begin{equation}
\begin{aligned}
\displaystyle
    c_3\frac{\partial^2{d_1}}{\partial{x^2_1}}+c_4\frac{\partial^2{d_1}}{\partial{x^2_2}}+\left(c_2+c_5+\frac{P}{2}+\mu_e H^2_0\right)\frac{\partial^2{d_1}}{\partial{x_1}\partial{x_2}}+\left(c_5-\frac{P}{2}\right)\frac{\partial^2{d_2}}{\partial{x^2_1}}+(c_6+\mu_e H^2_0)\frac{\partial^2{d_2}}{\partial{x^2_2}}+c_4\frac{\partial^2{d_2}}{\partial{x_1}\partial{x_2}}\\=\rho\frac{\partial^2{d_2}}{\partial{t^2}}+\rho\left(-\Omega^2 d_2-2\Omega\frac{\partial{d_1}}{\partial{t}}\right).
\end{aligned}
\label{eq_17}
\end{equation}
%%===========================================
The assumption is that the crack moves steadily at a uniform speed $v$ with constant length, resulting in a steady stress pattern. We consider replacing the coordinate system $(x_1, x_2, x_3)$ with $(X_1, X_2, X_3)$, which is fixed to the propagating crack, as illustrated in the Figure \ref{Figure 0}. Thus, according 
 
\be
X_1=x_1-vt,\hspace{0.2cm} X_2=x_2, \hspace{0.2cm}X_3=x_3 .
\label{eq_18}
\ee
Utilizing the equation $X_1 = x_1 - vt$, we find that $\frac{\partial}{\partial{x_1}} = \frac{\partial}{\partial{X_1}}$ and $\frac{\partial}{\partial{t}} = v\frac{\partial}{\partial{X_1}}$, while the displacement becomes a function of $X_1$, $X_2$, and $t$. Consequently, equations \ref{eq_16} and \ref{eq_17} can be expressed in terms of the moving coordinate as:

\begin{equation}
\begin{aligned}
\displaystyle
    (c_1+\mu_e H^2_{0}-v^2\rho)\frac{\partial^2{d_1}}{\partial{X^2_1}}+\left(c_5-\frac{P}{2}\right)\frac{\partial^2{d_1}}{\partial{X^2_{2}}}+2c_3\frac{\partial^2{d_1}}{\partial{X_1}\partial{X_2}}+c_3\frac{\partial^2{d_2}}{\partial{X^2_1}}
    +\left(c_2+c_5+\frac{P}{2}+\mu_e H^2_{0}\right)\frac{\partial^2{d_2}}{\partial{X_1}\partial{X_2}}\\
   +c_4\frac{\partial^2{d_2}}{\partial{X^2_2}} +\rho\Omega^2 d_1-2\Omega\rho v\frac{\partial{d_2}}{\partial{X_1}}=0,
\end{aligned}
\label{eq_19}
\end{equation}

%%===========================================
    
\begin{equation}
\begin{aligned}
\displaystyle
    c_3\frac{\partial^2{d_1}}{\partial{X^2_1}}+c_4\frac{\partial^2{d_1}}{\partial{X^2_2}}+\left(c_2+c_5+\frac{P}{2}+\mu_e H^2_0\right)\frac{\partial^2{d_1}}{\partial{X_1}\partial{X_2}}+\left(c_5-\frac{P}{2}-v^2\rho \right)\frac{\partial^2{d_2}}{\partial{X^2_1}}+(c_6+\mu_e H^2_0)\frac{\partial^2{d_2}}{\partial{X^2_2}}\\+c_4\frac{\partial^2{d_2}}{\partial{X_1}\partial{X_2}}
    +\rho\Omega^2 d_2+2\Omega\rho v\frac{\partial{d_1}}{\partial{X_1}}=0.
\end{aligned}
\label{eq_20}
\end{equation}
%%===========================================

\section{Boundary conditions}
 Concerning the problem, the Griffith crack moves within the middle plane of the rotating self-reinforced initial stressed strip. The inertial frame of the fracture's movement is established within $|X_1| \leq D$ at $X_2 = 0$. This region experiences internal normal stress denoted by $p(X_1)$ and a normal pressure load $q(X_1)$ applied over $|X_1| \leq D$ at $X_2 = \pm h$. Considering symmetry about the $X_1$-axis, analysis of the upper half of the slab, i.e. $0 \leq X_2 \leq h$, suffices. The prescribed boundary conditions for the considered model within the dynamic coordinate system are delineated as follows:\\
(I) The boundary conditions at $X_2=h$ can be expressed as follows:
\begin{subequations}
\begin{align}
 &(1) \begin{aligned}[t]
     d_1(X_1,h) &= 0,\hspace{0.2cm} |X_1| \le D,
 \end{aligned} \label{21a} \\
 &(2) \begin{aligned}[t]
     d_2(X_1,h) &= 0, \hspace{0.2cm}|X_1| \le D,
 \end{aligned} \label{21b} \\
 &(3) \begin{aligned}[t]
     \tau_{12}(X_1,h) &= 0, \hspace{0.2cm} |X_1| < \infty,
 \end{aligned} \label{21c} \\
 &(4) \begin{aligned}[t]
     \tau_{22}(X_1,h) &= -q(X_1), \hspace{0.2cm} D < |X_1| < \infty.
 \end{aligned} \label{21d}
\end{align}
\end{subequations}
(II) The boundary conditions at $X_2$=0 are given by
 \begin{subequations}
\begin{align}
 &(1) \begin{aligned}[t]
     d_1(X_1,0) &= 0,\hspace{0.2cm} |X_1| \le D,
 \end{aligned} \label{22a} \\
 &(2) \begin{aligned}[t]
     d_2(X_1,0) &= 0, \hspace{0.2cm}|X_1| \le D,
 \end{aligned} \label{22b} \\
 &(3) \begin{aligned}[t]
     \tau_{12}(X_1,0) &= 0, \hspace{0.2cm} |X_1| < \infty,
 \end{aligned} \label{22c} \\
 &(4) \begin{aligned}[t]
     \tau_{22}(X_1,0) &= -p(X_1), \hspace{0.2cm} D < |X_1| < \infty.
 \end{aligned} \label{22d}
\end{align}
\end{subequations}
%%%========================================================
\section{Solution of the problem}
The appropriate solutions in integral form for equations \ref{eq_19} and \ref{eq_20} are expected to assume the following formats:
%%%========================================================
\be
\ba{lll}
\displaystyle
  d_1(X_1,X_2)=\int_0^\infty{A_1(\phi,X_2)\sin(\phi \: X_1) d\phi},
\ea
\label{eq_23}
\ee
%%%========================================================
\be
\ba{lll}
\displaystyle
 d_2(X_1,X_2)=\int_0^\infty{A_2(\phi,X_2)\cos(\phi \: X_1 )d\phi}
 \ea
\label{eq_24}
\ee
%%%=========================================================
where $A_1(\phi)$ and $A_2(\phi)$ represent the undetermined functions to be established based on the specified requirements (boundary conditions). 

Considering equations \ref{eq_23} and \ref{eq_24}, equations \ref{eq_19} and \ref{eq_20} yield the following relationships governing the functions $A_1(\phi)$ and $A_2(\phi)$:
\be
% \ba{lll}
\displaystyle
\left(c_5-\frac{P}{2}\right)\frac{\partial^2{A_1}}{\partial{X^2_2}}-(c_2+c_5+\frac{P}{2}+\mu_e H^2_{0})\phi \frac{\partial{A_2}}{\partial{X_2}}-\left(c_1+\mu_e H^2_{0}-v^2\rho\right)\phi^2A_1+\rho\Omega^2 A_1+2\Omega\rho v \phi A_2=0
% \ea
\label{eq_25}
\ee
%%%=========================================================
and
\be
\left.
\ba{lll}
\displaystyle
(c_6+\mu_e H^2_0)\frac{\partial^2{A_2}}{\partial{X^2_2}}+\left(c_2+c_5+\frac{P}{2}+\mu_e H^2_0\right)\phi\frac{\partial{A_1}}{\partial{X_2}}-\left(c_5-\frac{P}{2}-v^2\rho \right)\phi^2 A_2+\rho\Omega^2A_2+2\Omega\rho v \phi A_1=0
\ea
\right.
\label{eq_26}
\ee
%%%=========================================================
Equation \ref{eq_25} and \ref{eq_26} admits solution in the form as
\be
\left.
\ba{lll}
\displaystyle
\left.
A_1(\phi,X_2)=\gamma_1(\phi)\cosh(\alpha_1\phi X_2)+
\gamma_2(\phi)\cosh(\alpha_2\phi X_2)+
\gamma_3(\phi)\sinh(\alpha_1\phi X_2)+
\gamma_4(\phi)\sinh(\alpha_2\phi X_2)
\right \}
\ea
\right.
\label{eq_27}
\ee
%%%=========================================================
and
\be
\left.
\ba{lll}
\displaystyle
\left.
A_2(\phi,X_2)=\xi_1(\phi)\sinh(\alpha_1\phi X_2)+
\xi_2(\phi)\sinh(\alpha_2\phi X_2)+\xi_3(\phi)\cosh(\alpha_1\phi X_2)+\xi_4(\phi)\cosh(\alpha_2\phi X_2)
\right \}
\ea
\right.
\label{eq_28}
\ee
where $\alpha_1$ and $\alpha_2$ are positive roots of the following biquadratic equation

%%========================
\be
\left.
\begin{aligned}
\displaystyle
   \alpha^4_j R_1 +
\alpha^2_j \left[ \left( R_2 + R_5\right)^2- R_1 \left(R_4-\frac{v^2}{\beta^2}- R_3 \right) 
+R_3 -\left( 1-\frac{v^2}{\beta^2}\right) \right]+ 
\left[- 4 \: R_3 \: \frac{v^2}{\beta^2} \right. \\
\left.+\left(R_4-\frac{v^2}{\beta^2}-R_3\right) \left( 1-\frac{v^2}{\beta^2}-R_3\right)\right]=0
\end{aligned}
\right.
\label{eq_29}
\ee
%%%=========================================================
where
\begin{equation*}
  R_1 =  \frac{c_6+\mu_e H^2_0}{c_5-\frac{P}{2}}, \: R_2 = \frac{c_2+\mu_e H^2_0}{c_5-\frac{P}{2}}, \: R_3 = \frac{\rho \Omega^2}{(c_5-\frac{P}{2})\phi^2}, \: R_4 = \frac{c_1+\mu_e H^2_0}{c_5-\frac{P}{2}}, \: R_5 = \frac{c_5+\frac{P}{2}}{c_5-\frac{P}{2}}, \: \beta= \sqrt{\frac{c_5-\frac{P}{2}}{\rho}};
\end{equation*}
 $\gamma_m$ and $\xi_m$ (m=1,2,3,4) are arbitrary functions in which $\xi_m$ is related to $\gamma_m$ by the relation

\be
\left.
\ba{lll}
\displaystyle
\xi_m(\phi)=\frac{\varphi_1}{\alpha_1}\gamma_m(\phi),\hspace{0.2cm} m=1,3 \ \ and \ \

\displaystyle \xi_m(\phi)=\frac{\varphi_2}{\alpha_2}\gamma_m(\phi),\hspace{0.2cm}m=2,4
\ea
\right.
\label{eq_30}
\ee
%%%=========================================================
with
\be
\left.
\ba{lll}
\displaystyle
\varphi_l=\frac{\varphi_{11} \: \alpha_l}{\varphi_{12} },\hspace{0.2cm} l=1,2
\ea
\right.
\label{eq_31}
\ee
where 
\begin{equation*}
   \varphi_{11} = \alpha^2_l-\left(\frac{c_1+\mu_e H^2_0+\rho\Omega^2}{c_5-\frac{P}{2}}-\frac{v^2}{\beta^2}\right), \:  \varphi_{12} = \left( \frac{c_2+\mu_e H^2_0}{c_5-\frac{P}{2}}+\frac{c_5+\frac{P}{2}}{c_5-\frac{P}{2}}\right)\alpha_l-\frac{2\Omega\rho v}{(c_5-\frac{P}{2})\phi}.
\end{equation*}
Using Eq. \ref{eq_30}, the stress components defined in Eq. \ref{eq_8} take form in the connective coordinate as
\begin{equation}
% \left.
\begin{aligned}
% \displaystyle
    \tau_{12}=&\int_0^\infty\left[(c_3+c_4\varphi_1)\gamma_1 \cosh(\alpha_1\phi X_2)+(c_3+c_4\varphi_2)\gamma_2 \cosh(\alpha_2\phi X_2)+(c_3+c_4\varphi_1)\gamma_3 \sinh(\alpha_1\phi X_2)\right.\\
    &+\left.(c_3+c_4\varphi_2)\gamma_4 \sinh(\alpha_2\phi X_2)\right]\phi \cos(\phi X_1)d\phi+\int_0^\infty\left[\chi_1\gamma_1 \sinh(\alpha_1\phi X_2)+\chi_2\gamma_2 \sinh(\alpha_2\phi X_2)\right.\\
   &+\left.\chi_1\gamma_3 \cosh(\alpha_1\phi X_2)
    +\chi_2\gamma_4 \cosh(\alpha_2\phi X_2)\right]\phi \sin(\phi X_1)d\phi
\end{aligned}
% \right.
\label{eq_32}
\end{equation}
%%======================================
and
\begin{equation}
% \left.
\begin{aligned}
\displaystyle
   \tau_{22}=-&P+\int_0^\infty\left[
    (c_2+c_6\varphi_1)\gamma_1 \cosh(\alpha_1\phi X_2)+(c_2+c_6\varphi_2)\gamma_2 \cosh(\alpha_2\phi X_2)+(c_2+c_6\varphi_1)\gamma_3 \sinh(\alpha_1\phi X_2)\right.\\
  +& \left. (c_2+c_6\varphi_2)\gamma_4 \sinh(\alpha_2\phi X_2)\right]
    \phi \cos(\phi X_1)d\phi+
    \frac{c_4}{c_5}\int_0^\infty\left[\left(\chi_1+\frac{P}{2}\left(\alpha_1+\frac{\varphi_1}{\alpha_1}\right)\right)\gamma_1 \sinh(\alpha_1\phi X_2)\right.\\
+& \left. \left(\chi_2+\frac{P}{2}\left(\alpha_2+\frac{\varphi_2}{\alpha_2}\right)\right)\gamma_2 \sinh(\alpha_2\phi X_2)+
    \left(\chi_1+\frac{P}{2}\left(\alpha_1+\frac{\varphi_1}{\alpha_1}\right)\right)\gamma_3 \cosh(\alpha_1\phi X_2)\right.\\
    +&\left.\left(\chi_2+\frac{P}{2}\left(\alpha_2+\frac{\varphi_2}{\alpha_2}\right)\right)\gamma_4 \cosh(\alpha_2\phi X_2)\right]\phi \sin(\phi X_1)d\phi
\end{aligned}
% \right.
\label{eq_33}
\end{equation}

where $\chi_m=\left[(c_5-\frac{P}{2})\alpha_m-\frac{\varphi_m}{\alpha_m}(c_5+\frac{P}{2})\right]$; m=1,2.\\
Considering the boundary condition in Eq. \ref{22c}, Eq. \ref{eq_32} yields
\be
\gamma_4=-\frac{\chi_1}{\chi_2}\gamma_3.
\label{eq_34}
\ee

Utilizing the boundary conditions in Eqs. \ref{22b}, \ref{eq_24}, and \ref{eq_28} results in
\be
\int_0^\infty\gamma_3 \cos (\phi \: X_1) d\phi=0,\hspace{0.4cm} |X_1|\le D.
\label{eq_35}
\ee
Taking into account Eqs. \ref{eq_24} along with \ref{eq_28} within the boundary condition Eq. \ref{21b} yields

\be
\left.
% \ba{lll}
\begin{aligned}
% \displaystyle
    \int_0^\infty\left[\frac{\varphi_1}{\alpha_1}\gamma_1(\phi) \sinh(\alpha_1\phi h)+\frac{\varphi_2}{\alpha_2}\gamma_2(\phi) \sinh(\alpha_2\phi h)+
\frac{\varphi_1}{\alpha_1}\gamma_3(\phi) \cosh(\alpha_1\phi h) \right.\\+
\left.\frac{\varphi_2}{\alpha_2}\gamma_4(\phi) \cosh(\alpha_2\phi h)\right] \cos(\phi X_1)d\phi=0,\hspace{0.2cm} |X_1|\le D.
% \ea
\end{aligned}
\right.
\label{eq_36}
\ee

The values of the undetermined functions, as expressed in integral form, have been derived from Eqs. \ref{eq_35} and \ref{eq_36} as
% To ascertain the values of the undetermined functions found in Eqs. (35) and (36), we establish the following equations:
\be
\gamma_3(\phi)=\frac{1}{\phi}\int_0^D q_1(t)sin(\phi t)dt
\label{eq_37}
\ee
and
\be
\left.
\begin{aligned}
% \ba{lll}
\displaystyle
  \frac{\varphi_1}{\alpha_1}\gamma_1(\phi) \sinh(\alpha_1\phi h)+\frac{\varphi_2}{\alpha_2}\gamma_2(\phi) \sinh(\alpha_2\phi h)
  +\frac{\varphi_1}{\alpha_1}\gamma_3(\phi) \cosh(\alpha_1\phi h) 
  +\frac{\varphi_2}{\alpha_2}\gamma_4(\phi) \cosh(\alpha_2\phi h)\\
  =\frac{1}{\phi}\int_0^D q_2(t) \sin(\phi t)dt
% \ea
\end{aligned}
\right.
\label{eq_38}
\ee
where $ q_1(t)$ and $q_2(t)$ represent functions that are not yet known. Equations \ref{eq_36} and \ref{eq_37} fully adhere to the integral property, meaning
\begin{equation*}
 \int_0^\infty\frac{ \sin\phi t. \: \cos\phi t}{\phi}d\phi=
\begin{cases}
\frac{\pi}{2}, & t > X_1, \\
0, & t < X_1.
\end{cases}
\end{equation*}

Solving Eq. \ref{eq_36} alongside the equation derived from the boundary condition in Eq. \ref{21c} yields the subsequent expressions:
\be
\gamma_1(\phi)=-\frac{[1+\psi_1(\phi)]}{\phi}\int_0^D q_1(t) \sin(\phi t)dt+\frac{\psi_2(\phi)}{\phi}\int_0^D q_2(t) \sin(\phi t)dt,
\label{eq_39}
\ee
%%========================
\be
\gamma_2(\phi)=\frac{\chi_1}{\chi_2}\frac{[1+\psi_3(\phi)]}{\phi}\int_0^D q_1(t) \sin(\phi t)dt-\frac{\psi_4(\phi)}{\phi}\int_0^D q_2(t) \sin(\phi t)dt
\label{eq_40}
\ee
where
\be
\psi_1(\phi)=\frac{e^{-\alpha_1 \phi h}}{\sinh(\alpha_1\phi h)},\hspace{0.2cm}
\psi_2(\phi)=\frac{E_1}{E_2 \sinh(\alpha_1\phi h)},\hspace{0.2cm}
\psi_3(\phi)=\frac{e^{-\alpha_2 \phi h}}{\sinh(\alpha_2\phi h)},\hspace{0.2cm}
\psi_4(\phi)=\frac{\chi_1}{\chi_2}\frac{E_1}{E_2 \sinh(\alpha_2\phi h)}
\ee
with
\be
E_1=\frac{\chi_2}{E^*},\hspace{0.2cm}E_2=\frac{1}{E^*}\left(\frac{\varphi_1\chi_2}{\alpha_1}-\frac{\varphi_2 \chi_1}{\alpha_2}\right),\hspace{0.2cm} E^*=\chi_1\epsilon_2-\chi_2\epsilon_1,\hspace{0.2cm} \epsilon_m=c_2+c_6\varphi_m;\hspace{0.2cm}m=1,2.
\label{eq_42}
\ee
Utilizing the boundary conditions in Eqs. \ref{21d} and \ref{22d} alongside Eqs. \ref{eq_32} and \ref{eq_33}, respectively, lead to the following form:
%%====================================
\begin{equation}
% \left.
\begin{aligned}
% \displaystyle
-P+\int_0^\infty(\epsilon_1\gamma_1+\epsilon_2\gamma_2)\phi \cos\phi X_1d\phi+\frac{c_4}{c_5}\int_0^\infty\left[\left(\chi_1+\frac{P}{2}\left(\alpha_1+\frac{\varphi_1}{\alpha_1}\right)\right)c_1+\left(\chi_2+\frac{P}{2}\left(\alpha_2+\frac{\varphi_2}{\alpha_2}\right)\right)c_2\right]\\
\phi \sin\phi X_1d\phi
=-p(X_1), \hspace{0.2cm}
D<|X_1|<\infty
\end{aligned}
% \right.
\label{eq_43}
\end{equation}
%%====================================
and
\begin{equation}
% \left.
\begin{aligned}
\displaystyle
    -P+\int_0^\infty\left[\epsilon_1\gamma_1 \cosh(\alpha_1\phi h)+\epsilon_2\gamma_2 \cosh(\alpha_2\phi h)+\epsilon_1\gamma_3 \sinh(\alpha_1\phi h)+\epsilon_2\gamma_4 \sinh(\alpha_2\phi h))\phi \cos(\phi X_1)\right]d\phi\\
    +\frac{c_4}{c_5}\int_0^\infty\left[\left(\chi_1+\frac{P}{2}\left(\alpha_1+\frac{\varphi_1}{\alpha_1}\right)\right)\gamma_1 \sinh(\alpha_1\phi h)
    +\left(\chi_2+\frac{P}{2}\left(\alpha_2+\frac{\varphi_2}{\alpha_2}\right)\right)\gamma_2 \sinh(\alpha_2\phi h)\right.\\
    \left.+\left(\chi_1+\frac{P}{2}\left(\alpha_1+\frac{\varphi_1}{\alpha_1}\right)\right)\gamma_3 \cosh(\alpha_1\phi h)
    +\left(\chi_2+\frac{P}{2}\left(\alpha_2+\frac{\varphi_2}{\alpha_2}\right)\right)\gamma_4 \cosh(\alpha_2\phi h)\right]\phi \sin(\phi X_1)d\phi\\
    =-q(X_1),\hspace{0.2cm}
D<|X_1|<\infty.
\end{aligned}
% \right.
\label{eq_44}
\end{equation}
%%====================================
Substituting the values of $\gamma_1$, $\gamma_2$, $\gamma_3$, and $\gamma_4$ from Eqs. \ref{eq_34}, \ref{eq_37}, \ref{eq_39} and \ref{eq_40} into Eqs. \ref{eq_43} and \ref{eq_44} gives rise to the subsequent integral equations:
\begin{equation}
\begin{aligned}
% \displaystyle
    \int_0^D\frac{t}{t^2-X^2_1} q_1(t)dt+\frac{1}{2}\int_0^DF_{11}(X_1,t)q_1(t)dt+\frac{1}{2}\int_0^\infty F_{12}(X_1,t)q_2(t)dt\\
     +\frac{c_4 E_1}{c_5\chi_2}\int_0^\infty \int_0^D E_4 \sin(\phi t) \sin(\phi X_1)q_1(t)d\phi dt=-E_1 p(X_1),\hspace{0.2cm}D<|X_1|<\infty,
\end{aligned}
\label{eq_45}
\end{equation}

%%========================================
\be
\left.
\begin{aligned}
\displaystyle
    P E_2+\int_0^D\frac{t}{t^2-X^2_1}q_2(t)dt+\frac{1}{2}\int_0^DF_{21}(X_1,t)q_1(t)dt+\frac{1}{2}\int_0^DF_{22}(X_1,t)q_2(t)dt\\-\frac{c_4}{c_5 E^*}\int_0^\infty\int_0^D E_4 sin(\phi t) \sin(\phi X_1) d\phi dt=E_2q(X_1),\hspace{0.2cm}D<|X_1|<\infty
\end{aligned}
\right.
\label{eq_46}
\ee
%%======================================
where
\be
\left.
\ba{lll}
\displaystyle
F_{lm}(X_1,t)=\int_0^\infty k_{ij}(\phi)[\sin\phi(t+X_1)+\sin\phi(t-X_1)]d\phi; \hspace{0.2cm} (l,m=1,2),
\ea
\right.
\label{eq_47}
\ee
%%==================================
\begin{subequations}
\begin{align}
    k_{11}(\phi)&=k_{22}(\phi)=\frac{1}{E^*}\left[\chi_1 \epsilon_2 \psi_3(\phi)-\chi_2 \epsilon_1 \psi_1(\phi)\right],\\
    k_{12}(\phi)&=\frac{\chi_2}{E^*}\left[\epsilon_1\psi_2(\phi)-\epsilon_2 \psi_4(\phi)\right], \hspace{0.2cm} k_{21}(\phi)=E_2 \left[\psi_5(\phi)-\frac{\chi_1}{\chi_2}\psi_6(\phi)\right],\\
\psi_5(\phi)&=\epsilon_1\left[e^{-\alpha_1\phi h}+\psi_1(\phi)\cosh(\alpha_1\phi h)\right],
    \hspace{0.2cm}
    \psi_6(\phi)=\epsilon_2\left[e^{-\alpha_2\phi h}+\psi_3(\phi)\cosh(\alpha_2\phi h)\right],\\
    E_4&=\frac{P}{2}\left[\left(\alpha_1+\frac{\varphi_1}{\alpha_1}\right)\chi_2-\left(\alpha_2+\frac{\varphi_2}{\alpha_2}\right)\chi_1 \right].
\end{align}
\end{subequations}
To estimate $F_{lm}(X_1,t)$, we employed the first-order approximation of $\psi_l(\phi)$ for large $h$, leading to the following outcome:
%%=======================
\be
\left.
\begin{aligned}
\displaystyle
F_{11}(X_1,t)&=-\frac{2\chi_2\epsilon_1}{E^*}\zeta_1(X_1,t)
+\frac{2\chi_1\epsilon_2}{E^*}\zeta_2(X_1,t)=F_{22}(X_1,t),\\
F_{12}(X_1,t)&=\frac{2\chi^2_2\epsilon_1}{E_2 E^{*2}}\eta_1(X_1,t)-\frac{2\chi_2\chi_1\epsilon_2}{E_2 E^{*2}}\eta_2(X_1,t),\\
F_{21}(X_1,t)&=2\epsilon_1 E_2\eta_1(X_1,t)-\frac{2E_2\chi_1\epsilon_2}{\chi_2}\eta_2(X_1,t)
\end{aligned}
\right.
\label{eq_49}
\ee
%%===================================================
where
\be
\left.
\begin{aligned}
\displaystyle
   \zeta_m(X_1,t)&=\frac{J_1}{4\alpha^2_m h^2+(J_1)^2}+\frac{J_2}{4\alpha^2_m h^2+(J_2)^2}+
    \frac{J_1}{16\alpha^2_m h^2+(J_1)^2}+\frac{J_2}{16\alpha^2_m h^2+(J_2)^2}+...\\ \text{and}\\
    \eta_m(X_1,t)&=\frac{J_1}{\alpha^2_m h^2+(J_1)^2}+\frac{J_2}{\alpha^2_m h^2+(J_2)^2}+
    \frac{J_1}{9\alpha^2_m h^2+(J_1)^2}+\frac{J_2}{9\alpha^2_m h^2+(J_2)^2}+...
\end{aligned}
\right.
\label{eq_50}
\ee
where $J_1 = t+X_1$,\: $J_2 = t-X_1$, $m=1,2$  and extending the expressions for $\zeta_m(X_1,t)$ and $\eta_m(X_1,t)$ by employing the series expansion of

$\frac{\pi^2}{8}=\sum_{x=1}^\infty\frac{1}{(2x-1)^2}$, $\frac{\pi^2}{12}=\sum_{x=1}^\infty\frac{(-1)^{x+1}}{x^2}$ and $\frac{\pi^2}{6}=\sum_{x=1}^\infty\frac{1}{x^2}$ in terms of $\frac{1}{h}$, Eq. \ref{eq_49} take the form
\be
\left.
\ba{lll}
\displaystyle
   F_{11}(X_1,t)= F_{22}(X_1,t)=-\frac{E_3\pi^2}{6h^2}t,
\ea
\right.
\label{eq_51}
\ee
\be
\left.
\ba{lll}
\displaystyle
   F_{12}(X_1,t)=\frac{\chi_2 E_3 \pi^2}{2E^*E_2 h^2}t,
\ea
\right.
\label{eq_52}
\ee
\be
\left.
\ba{lll}
\displaystyle
  F_{21}(X_1,t)=\frac{E_2 E_3 E^*\pi^2}{2\chi_2 h^2}t
\ea
\right.
\label{eq_53}
\ee
   where $E_3=\frac{1}{E^*}\left( \frac{\chi_2\epsilon_1}{\alpha^2_1}-\frac{\chi_1\epsilon_2}{\alpha^2_2}\right)$.
        
Let's explore the asymptotic expansion of $q_l(t)$ using the following format:
\be
\left.
\ba{lll}
\displaystyle
   q_l(t)=q^{(0)}_l(t)+\frac{1}{h^2}q^{(1)}_l(t)+O\left(\frac{1}{h^4}\right).
\ea
\right.
\label{eq_54}
\ee
% Using Eqs. (54)-(57) and comparing the coefficients of constant terms and terms containing $\frac{1}{h^2}$ from Eqs. (45) and (56) yields the following integral equations:
Utilizing Eqs. \ref{eq_51}-\ref{eq_54} and analyzing the coefficients of constant and $\frac{1}{h^2}$ terms from Eqs. \ref{eq_45} and \ref{eq_46} lead to the derivation of the subsequent integral equations:

   \begin{align}
   \int_0^D\frac{2tq^{(0)}_1(t)}{t^2-X^2_1}dt&=-2E_1p(X_1), \label{eq_55}\\
   \int_0^D\frac{2tq^{(0)}_2(t)}{t^2-X^2_1}dt&=2E_2q(X_1),\label{eq_56}\\
    \int_0^D\frac{2tq^{(1)}_1(t)}{t^2-X^2_1}dt&=\frac{\pi^2 E_3}{6}\int_0^D tq^{(0)}_1(t)dt-\frac{\pi^2\chi_2E_3}{2E_2E^*}\int_0^D tq^{(0)}_2(t)dt,\label{eq_57}\\
    \int_0^D\frac{2tq^{(1)}_2(t)}{t^2-X^2_1}dt&=\frac{\pi^2 E_3}{6}\int_0^D tq^{(0)}_2(t)dt-\frac{\pi^2 E_2E^*E_3}{2\chi_2}\int_0^D tq^{(0)}_1(t)dt.\label{eq_58}
   \end{align}
    Let's assume that the edge \( X_1=0 \) of the crack is subjected to a constant force p at \( X_1=X^{'}_{1} \) where \(X^{'}_{1}\) represents any point within the slab. Thus, we proposed the scenario for point-loading
\be
   p(X_1)=p\delta(X_1-X^{'}_{1})
   \label{eq_59}
   \ee
   where $\delta(X_1-X^{'}_{1})$ is the unit impulse function.\\
   
   By utilizing Eq. \ref{eq_54} alongside the findings presented in \cite{tricomi1951finite} for solving Eqs. \ref{eq_55}-\ref{eq_58}, we deduce the following relationships.
   \begin{align}
   q_1(t)&=\frac{4E_1p}{\pi^2}\frac{t X^{'}_{1}\sqrt{D^2-X^{'2}_{1}}}{\sqrt{D^2-t^2}(X^{'2}_{1}-t^2)}+\frac{2E_1 p}{\pi}\frac{t}{\sqrt{D^2-t^2}}+\frac{D^2}{h^2}\frac{U_1 t}{\pi\sqrt{D^2-t^2}}, \label{eq_60}\\
   q_2(t)&=\frac{2E_2q }{\pi}\frac{t}{\sqrt{D^2-t^2}}+\frac{D^2}{h^2}\frac{U_2 t}{\pi \sqrt{D^2-t^2}} \label{eq_61}
   \end{align} 
   with\\
   
   $U_1=\frac{\pi^2E_1E_3p}{12}-\frac{\pi^2\chi_2 E_3q}{4E^*}$ and $U_2=\frac{\pi^2E_3E_2q}{6}-\frac{\pi^2 E_1E_2E_3E^*p}{2\chi_2}$ where the applied load pressure is presumed to remain constant throughout i.e $q(X_1)=q$.
   
   % The stress intensity factor (SIF) at the crack edge \(X_1 = D\) for the point loading is determined using the following relation:
The stress intensity factor (SIF) at the crack edge \(X_1 = D\) is determined using the following relation:

\be
\left.
\ba{lll}
\displaystyle
   K_1=\lim_{x\to D^+} \sqrt{2(X_1-D)}\tau_{22}(X_1,0)
\ea
\right.
\label{eq_62}
\ee 
   and is calculated as

\be
\left.
\ba{lll}
\displaystyle
    \frac{K_1}{p\sqrt{D}}=\frac{2X^{'}_1}{\pi\sqrt{D^2-X^{'2}_1}}-\frac{U_1D^2}{2E_1h^2p}-1+\frac{c_4 E_4}{c_5\chi_2}\left(\frac{2X^{'}_1E_1}{\pi\sqrt{D^2-X^{'2}_1}}-E_1-\frac{D^2 U_1}{2E_1 h^2p}\right)
\ea
\right.
\label{eq_63}
\ee
Eq. \ref{eq_63} represents the stress intensity factor (SIF) expression at the edge of a crack in motion, subject to continuous point load, in a rotating self-reinforced initially stressed slab.

\section{Graphical representation}

%%%%%%%%%%%%%%%%%%%%%%table  1%%%%%%%%%%%%%%%%%%%%%%

% %%%%%%%%%%%%%%%%%%%%%%table 2 %%%%%%%%%%%%%%%%%%%%%%%%%%%
% \begin{table}
% \centering
% \caption{Material constants for isotropic materials \cite{gubbins1990seismology}}
% \begin{tabular}{c c c  c} 
% 			\hline
%    Material constant &  Units&  Isotropic material\\
%    \hline
%    $\mu$ & $N/m^2$ & $19.87*10^9$\\
%    $\lambda$ & $N/m^2$ & $25.1*10^9$\\
%    $\rho$ & $kg/m^3$ & $4705$\\
%    \hline
%    \end{tabular}
% \label{material constant}
% \end{table}

Numerical computations explore how various factors affect the dimensionless stress intensity factor ($K/p\sqrt{D}$) which is obtained in Eq. \ref{eq_63}. Parameters include magnetoelastic coupling (${\mu_e H^2_0}/{c_5}$), punch pressure(q/p), crack length($D/h$), point load position($X^{'}_1/D$), initial stress (P), angular velocity ($\Omega$), and crack speed($v/\beta$). Results are shown in Figures \ref{Figure 1}-\ref{Figure 6}.\\ 
Observations from Fig. \ref{Figure 1} indicate that the Stress Intensity Factor (SIF) magnitude rises as crack velocity ($v/\beta \to 1$) increases in case of steel. Additionally, the figure reveals that the SIF magnitude escalates with various parameters such as dimensionless crack length, punch pressure, and point load positions (see fig \ref{Fig_2}, fig \ref{Fig_6} and fig \ref{Fig_7} respectively). Conversely, the SIF magnitude demonstrates an inverse relationship with the initial compressive stress magnitude, uniform angular velocity, and magnetoelastic coupling parameter (see fig \ref{Fig_3}-fig \ref{Fig_5}).
Figure \ref{Figure 2} illustrates the impact of the crack velocity on the Stress Intensity Factor (SIF) for carbon fiber. It is evident that the SIF magnitude escalates as the velocity of the crack increases. Notably, it is observed that the SIF exhibits a rising trend with respect to the all parameters which is considered in the present study.
From looking at Figure \ref{Figure 3}, it is clear that when the crack moves faster (i.e., when $v/\beta \to 1$), the Stress Intensity Factor (SIF) gets bigger. Additionally, in Figure \ref{Figure 3}, we see that as the crack length, punch pressure, and point load positions increase, the SIF also increases. Conversely, the SIF decreases when the initial compressive stress, uniform angular velocity, and magnetoelastic coupling parameter increase.
According to Fig. \ref{Figure 4}, it's noted that as the uniform angular velocity ($\Omega$) increases, the stress intensity factor (SIF) decreases. Furthermore, the same figure shows that the SIF magnitude rises with increasing dimensionless crack length, magnetoelastic coupling parameter, punch pressure, and different point load positions. Conversely, as the crack's initial compressive stress and dimensionless velocity increase, the SIF magnitude decreases.
The graphical representation shows how the stress intensity factor affects a crack that moves under a continuous point load in a rotating strip with initial stresses.
It includes numerical values for material parameters carbon fiber resin, steel, and isotropic material in $(N/m^2)$ and $(kg/m^3)$ see references (\cite{markham1970measurement,hool1924reinforced, gubbins1990seismology} in table \ref{material constant}.

\begin{table}
\centering
\caption{Considered material constants for the present study
\cite{markham1970measurement,hool1924reinforced, gubbins1990seismology}}
\begin{tabular}{c c c  c } 
			\hline
   \textbf{Material constant} &    \textbf{Carbon fibre resin} & \textbf{Steel} & \textbf{Isotropic material}\\
   \hline
   $\mu_1$  &  $5.66*10^9$  & $2.45*10^9$  & $-$\\
    $\mu_2$  &  $2.46*10^9$ & $1.89*10^9$ & $-$\\
    $\mu$  &  $-$ & $-$ & $19.87*10^9$  \\
     $\lambda$ &  $5.65*10^9$  & $7.59*10^9$ & $25.1*10^9$ \\
    $a_1$ &  $-1.28*10^9$  & $-1.28*10^9$ & $-$\\
     $a_2$ &  $220.9*10^9$ & $0.32*10^9$ & $-$\\
      $\rho$ &  $1600$ & $7800$ & $4705$ \\
   \hline
   \end{tabular}
\label{material constant}
\end{table}
Based on the findings in Fig. \ref{Figure 5}, it's noted that as the uniform angular velocity ($\Omega$) increases, the stress intensity factor (SIF) decreases. Additionally, the same figure indicates that the SIF magnitude rises with increasing dimensionless crack length, punch pressure, and various point load positions. Conversely, as the initial compressive stress, dimensionless velocity of crack, and magnetoelastic coupling parameter increase, the SIF magnitude decreases. From the observations in Fig. \ref{Figure 6}, it's evident that the stress intensity factor (SIF) usually goes up as the uniform angular velocity ($\Omega$) increases, except in one case shown in Fig. \ref{Fig_35} where the magnetoelastic parameter changes.  Additionally, Fig. \ref{Figure 6}, it's shown that the SIF gets bigger with longer crack lengths, higher punch pressures, and different positions of point loads, but when the initial compressive stress, crack velocity, or magnetoelastic coupling parameter increase, the SIF tends to decrease.

%%%=======Enter steel Graphs=================================
\bfg[htbp]
\centering
\begin{subfigure}[b] {0.38\textwidth}
\includegraphics[width=\textwidth ]{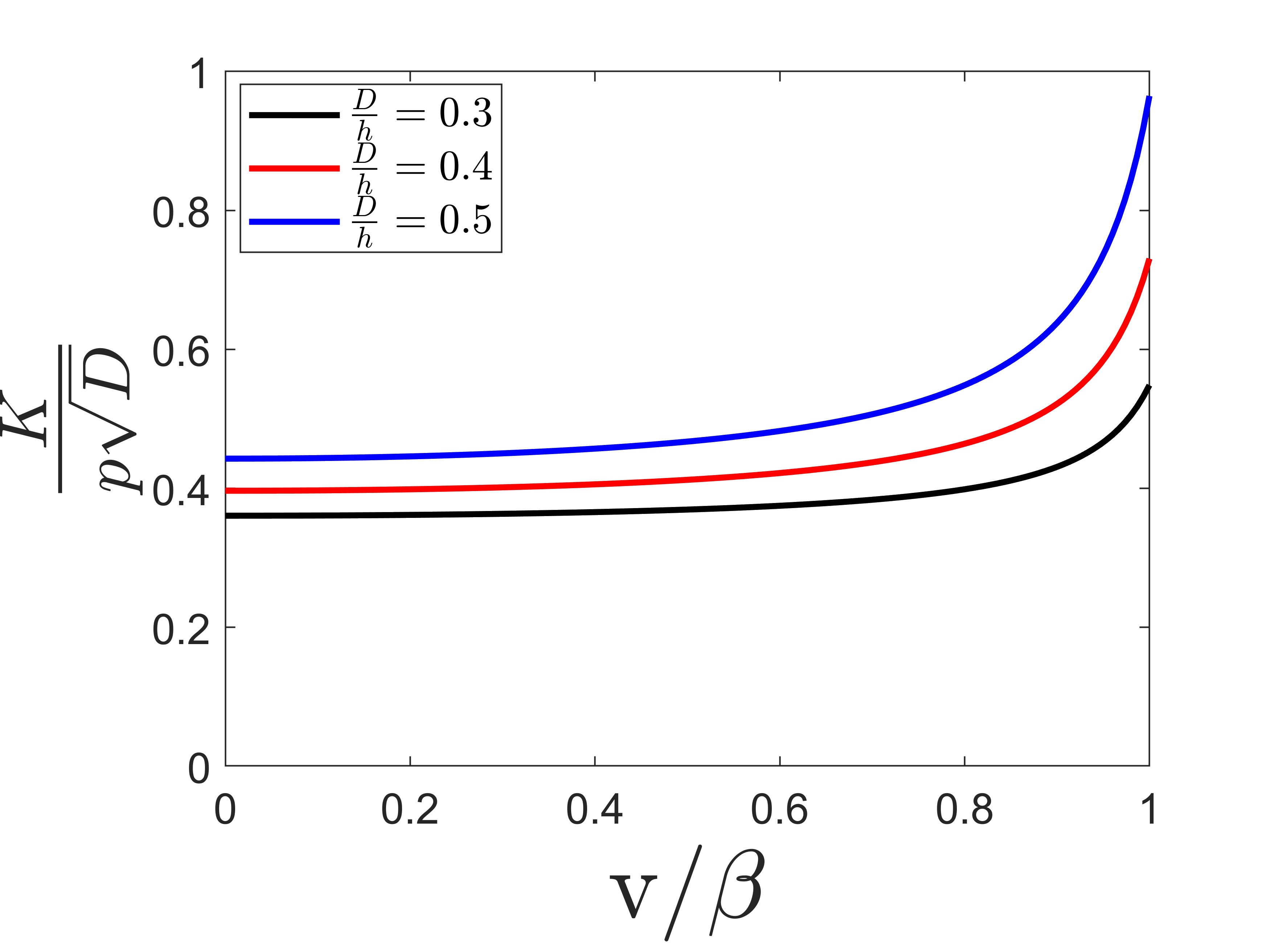}
\caption{}
\label{Fig_2}
\end{subfigure}
~
\begin{subfigure}[b] {0.38\textwidth}
\includegraphics[width=\textwidth ]{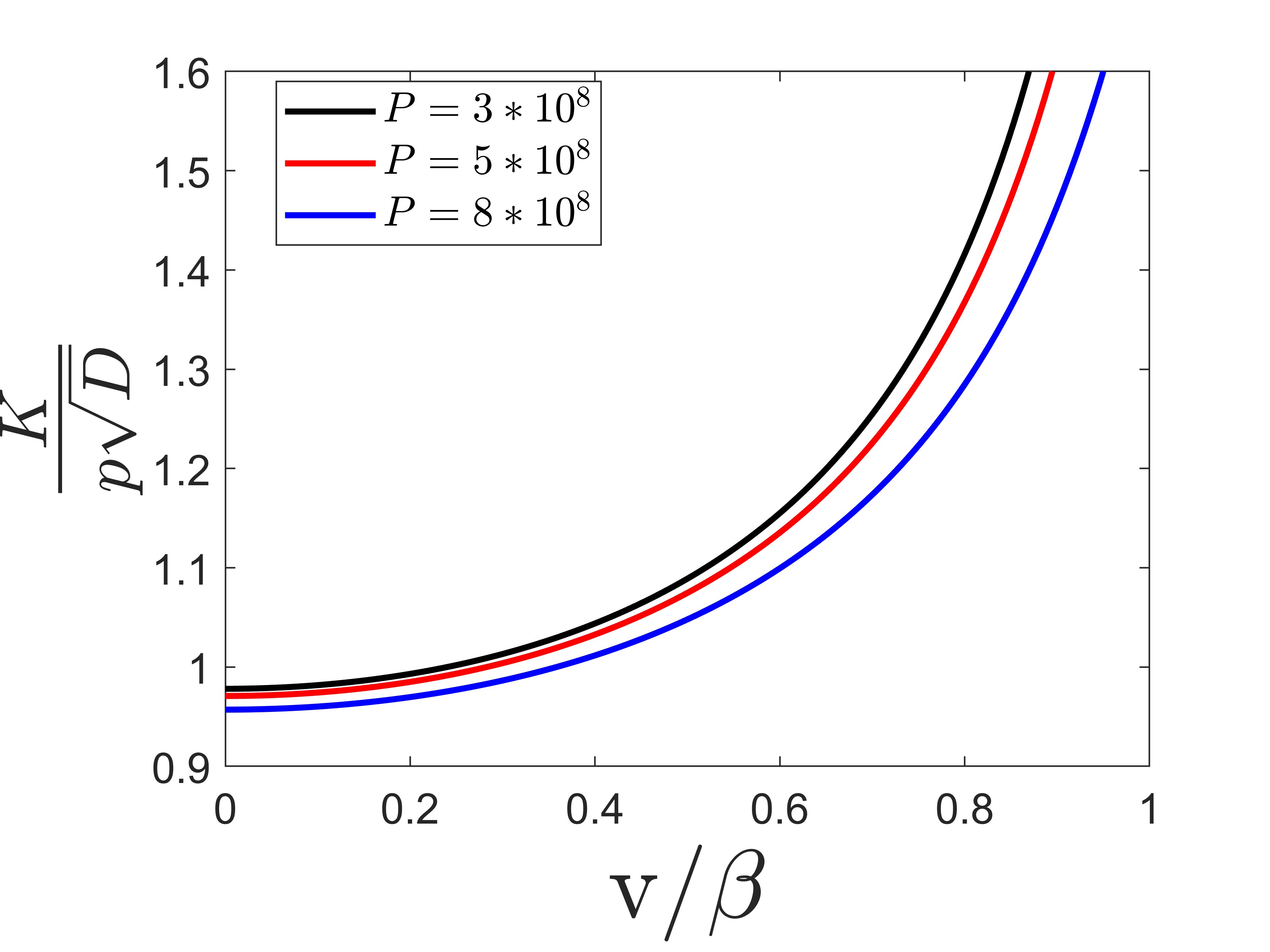}
\caption{}
\label{Fig_3}
\end{subfigure}
~
\begin{subfigure}[b] {0.38\textwidth}
\includegraphics[width=\textwidth ]{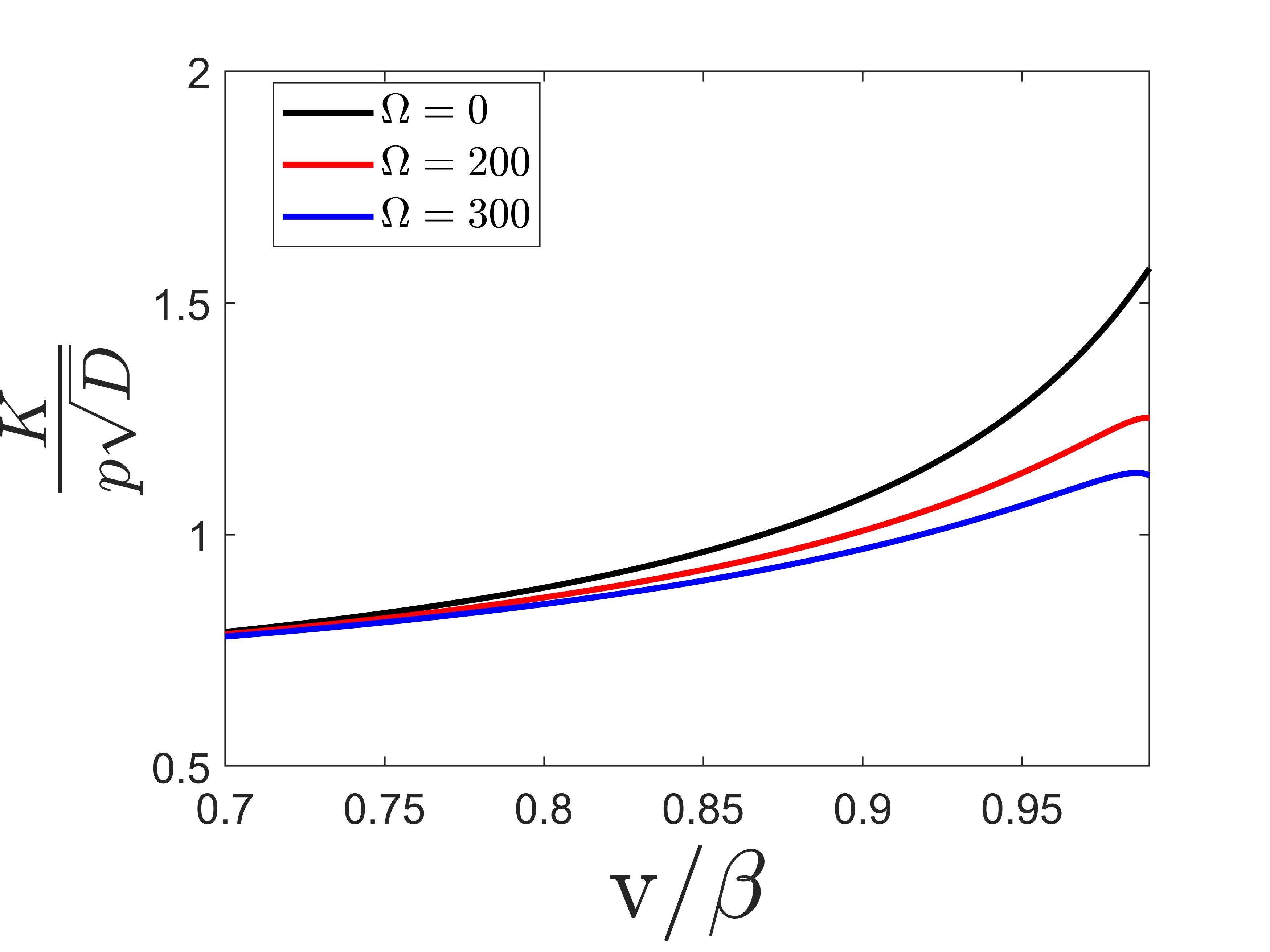}
\caption{}
\label{Fig_4}
\end{subfigure}
~
\begin{subfigure}[b] {0.38\textwidth}
\includegraphics[width=\textwidth ]{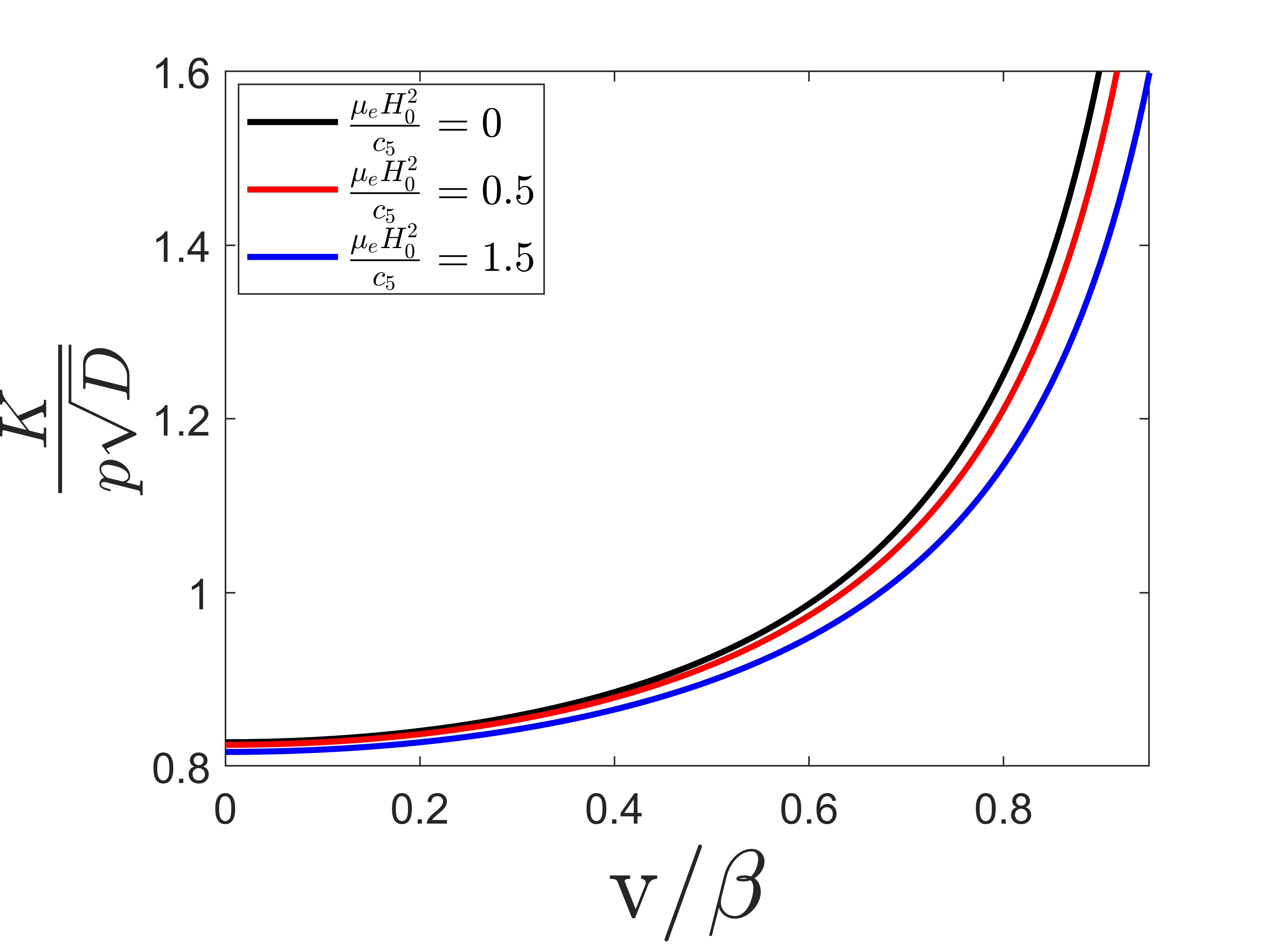}
\caption{}
\label{Fig_5}
\end{subfigure}
~
\begin{subfigure}[b] {0.38\textwidth}
\includegraphics[width=\textwidth ]{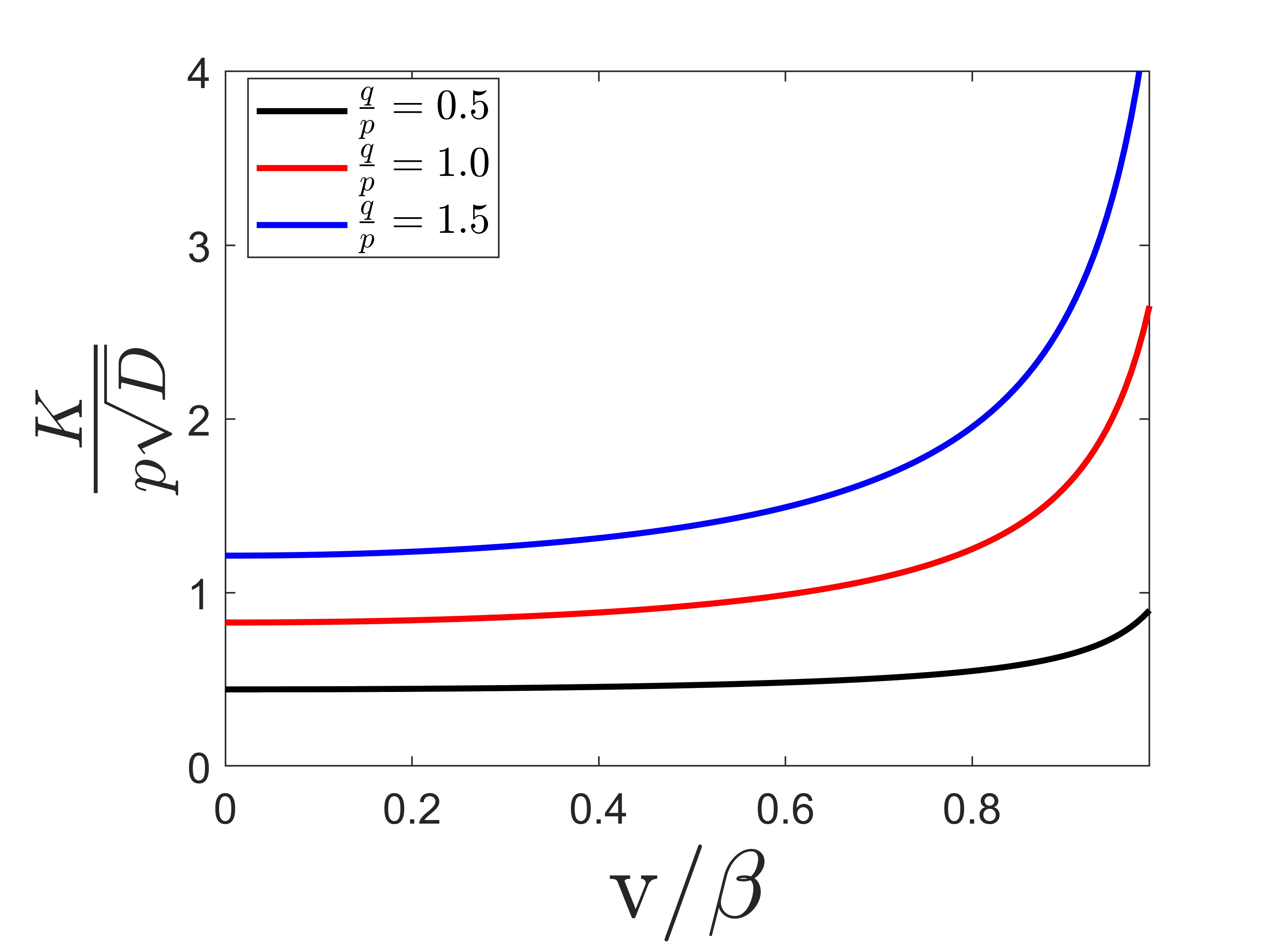}
\caption{}
\label{Fig_6}
\end{subfigure}
~
\begin{subfigure}[b] {0.38\textwidth}
\includegraphics[width=\textwidth ]{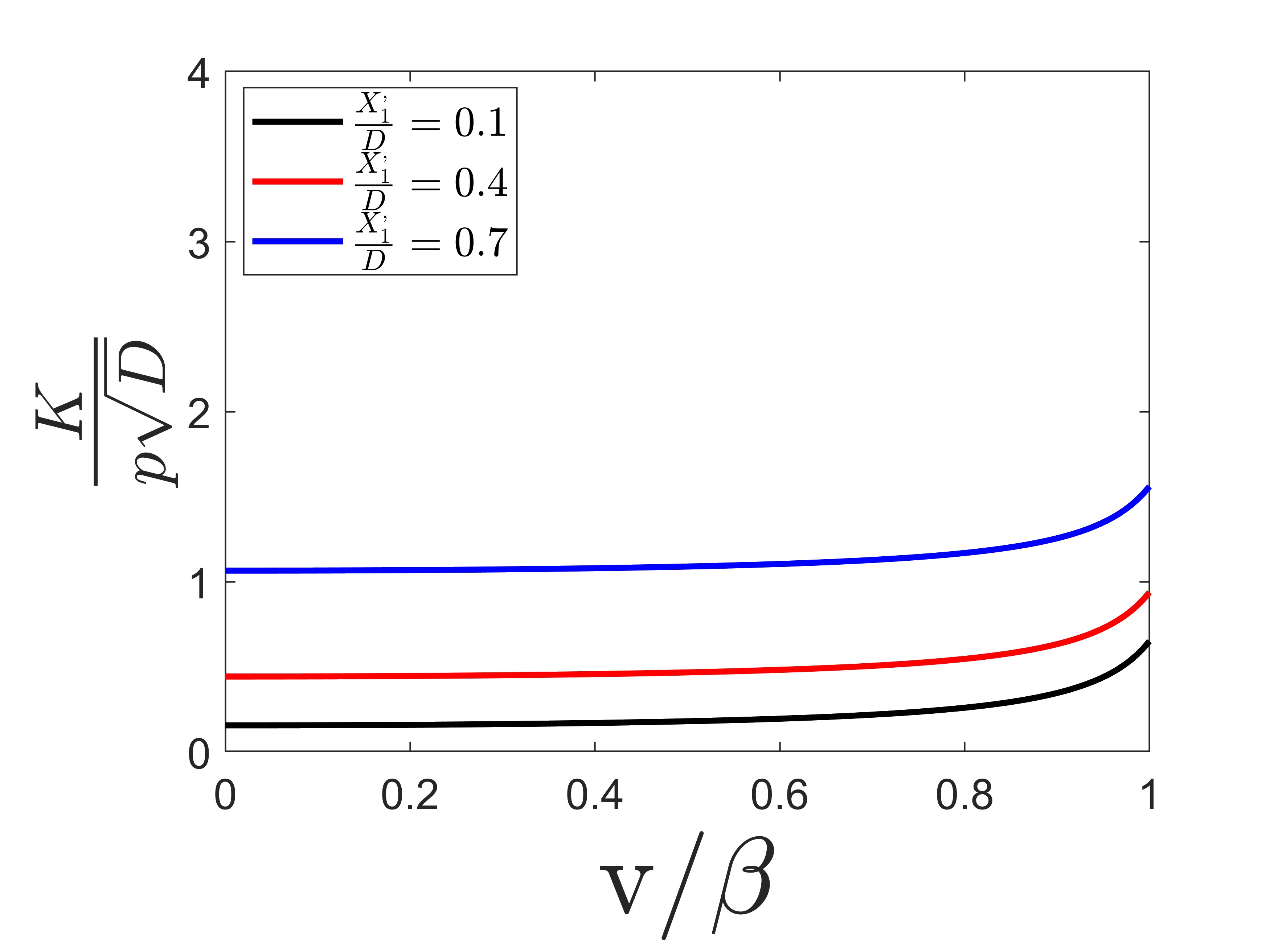}
\caption{}
\label{Fig_7}
\end{subfigure}
\caption{Impact of dimensionless stress intensity factor ($K/p\sqrt{D}$) with dimensionless crack velocity ($v/\beta$) in case of steel, illustrating the influence of dimensionless parameters such as (i) crack length $(D/h)$, (ii) initial compressive stress $(P)$, (iii) uniform angular velocity $(\Omega)$, (iv) magnetoelastic coupling parameter (${\mu_e H^2_0}/{c_5}$), (v) punch pressure $(q/p)$, and (vi) point load position ($X^{'}_1/D$). } 

\label{Figure 1}
\efg
%%%==========Carbon graphs=====================================
\bfg[htbp]
\centering
\begin{subfigure}[b] {0.38\textwidth}
\includegraphics[width=\textwidth ]{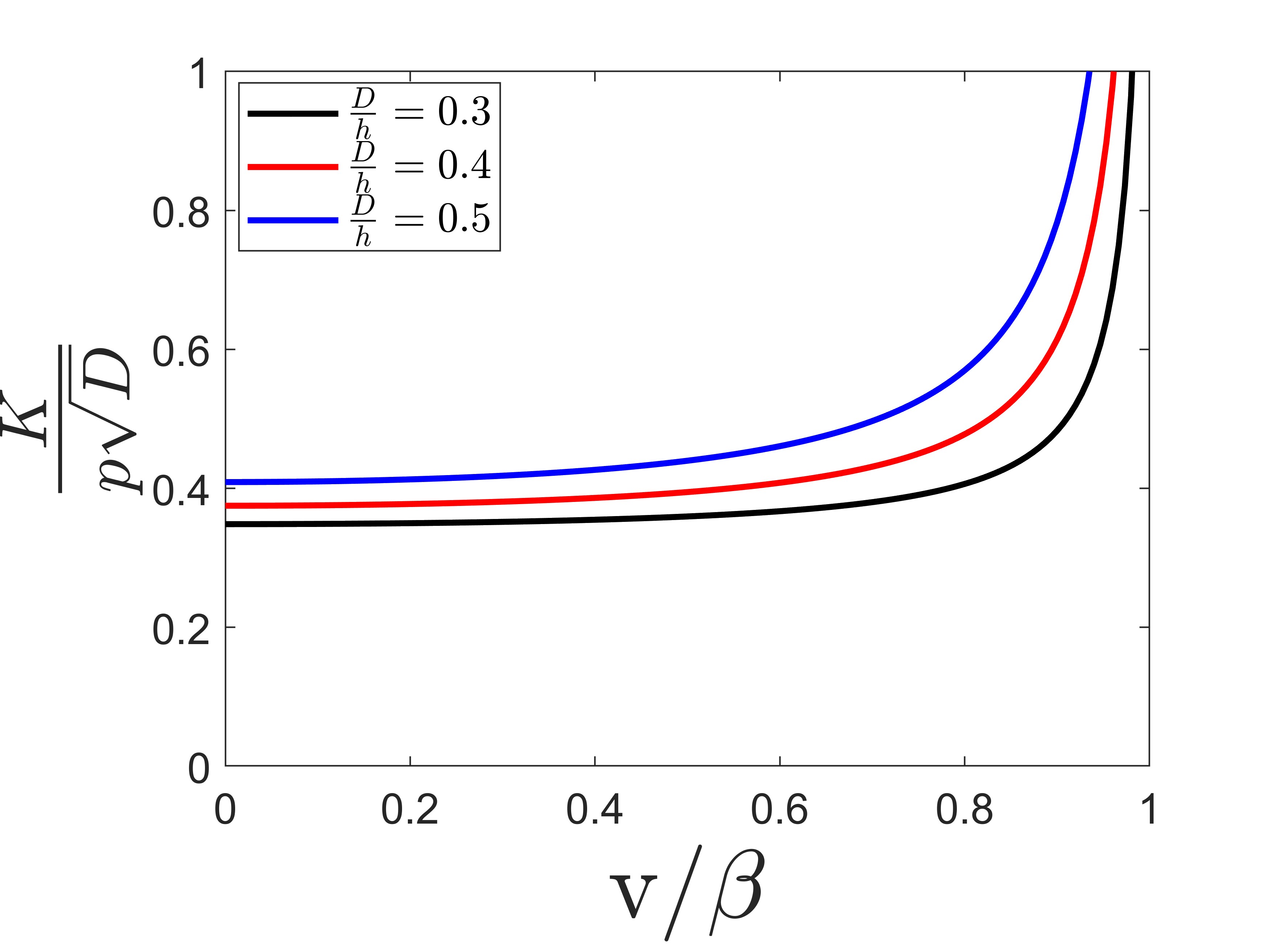}
\caption{}
\label{Fig_8}
\end{subfigure}
~
\begin{subfigure}[b] {0.38\textwidth}
\includegraphics[width=\textwidth ]{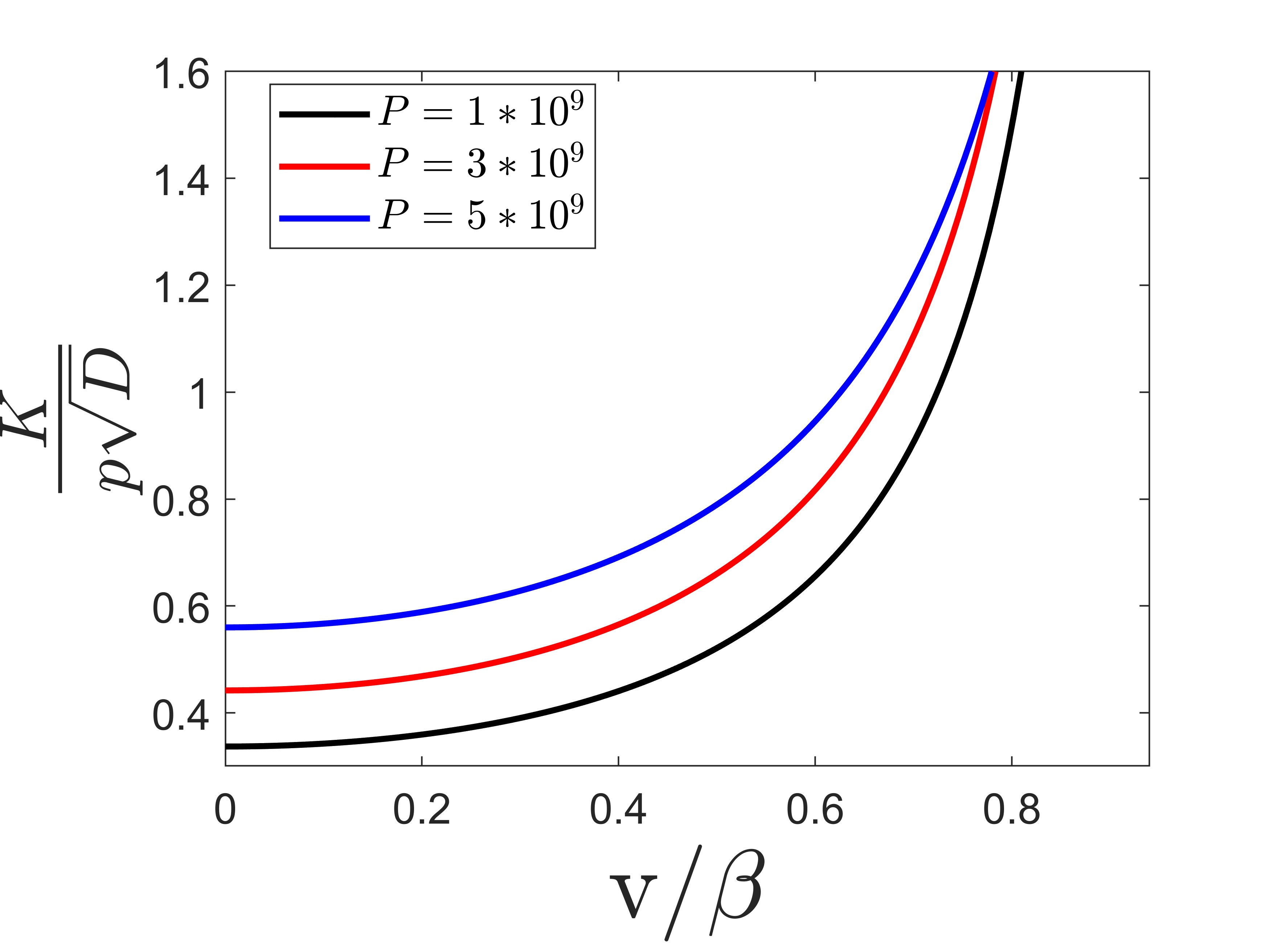}
\caption{}
\label{Fig_9}
\end{subfigure}
~
\begin{subfigure}[b] {0.38\textwidth}
\includegraphics[width=\textwidth ]{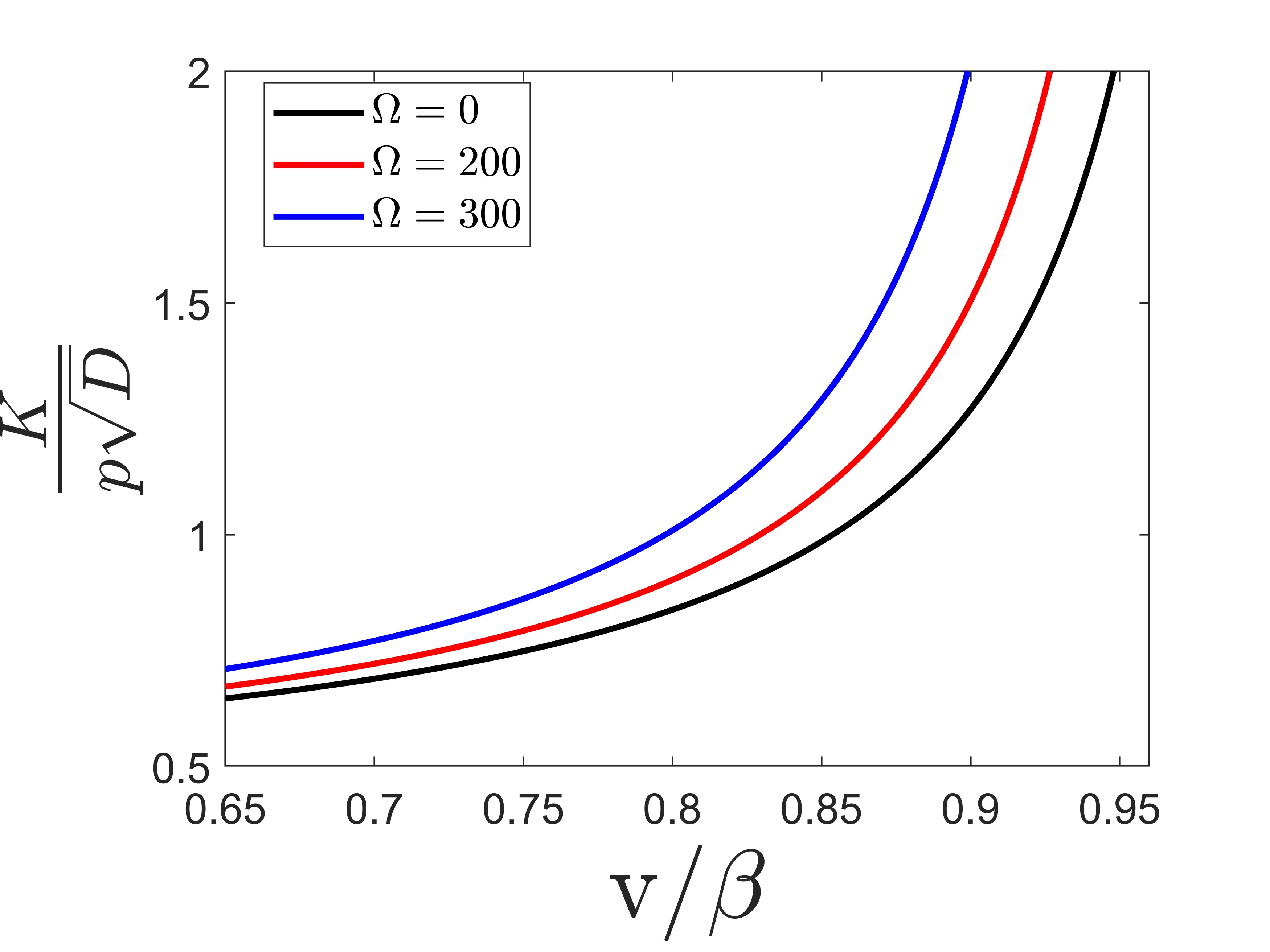}
\caption{}
\label{Fig_10}
\end{subfigure}
~
\begin{subfigure}[b] {0.38\textwidth}
\includegraphics[width=\textwidth ]{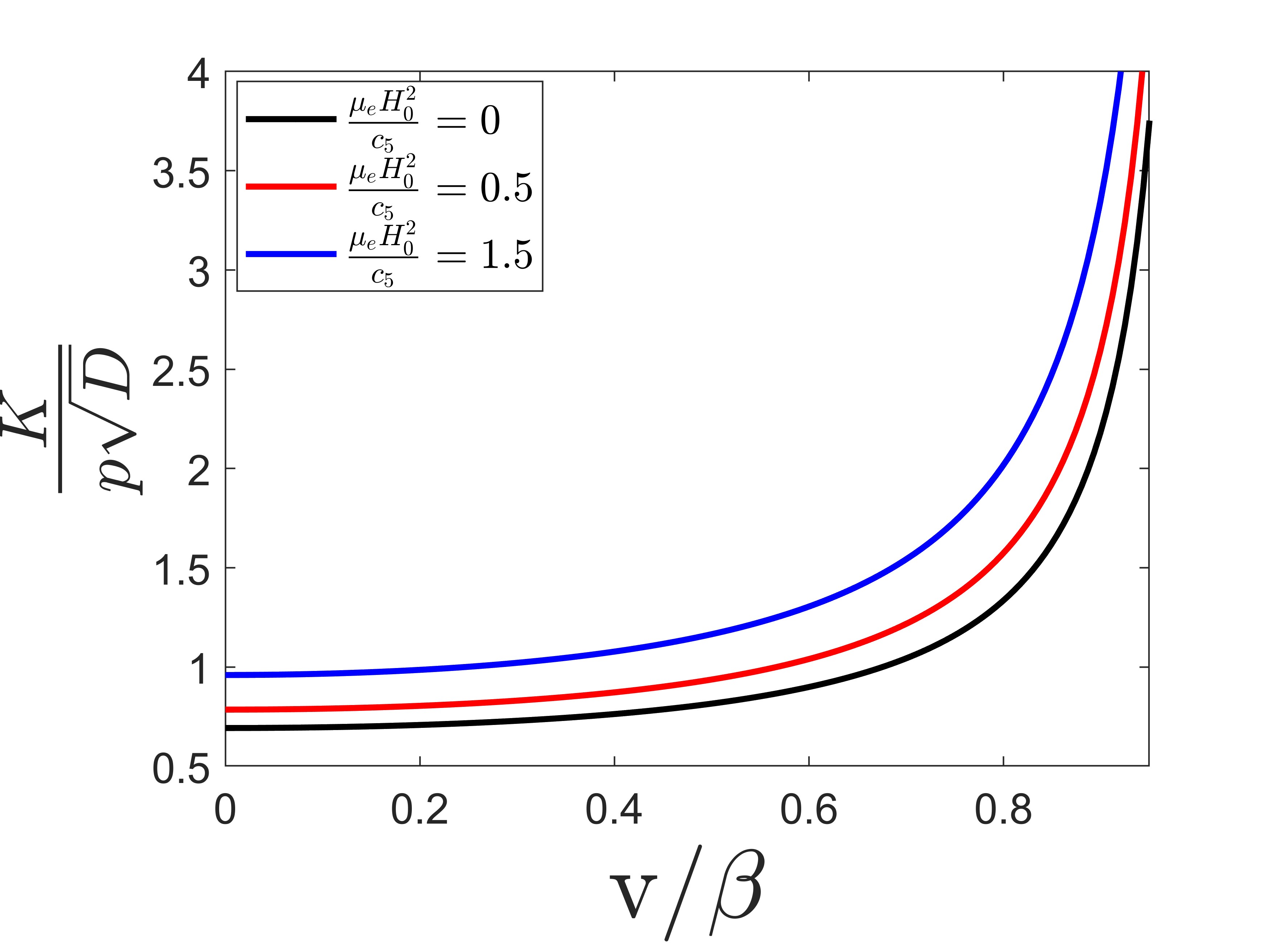}
\caption{}
\label{Fig_11}
\end{subfigure}
~
\begin{subfigure}[b] {0.38\textwidth}
\includegraphics[width=\textwidth ]{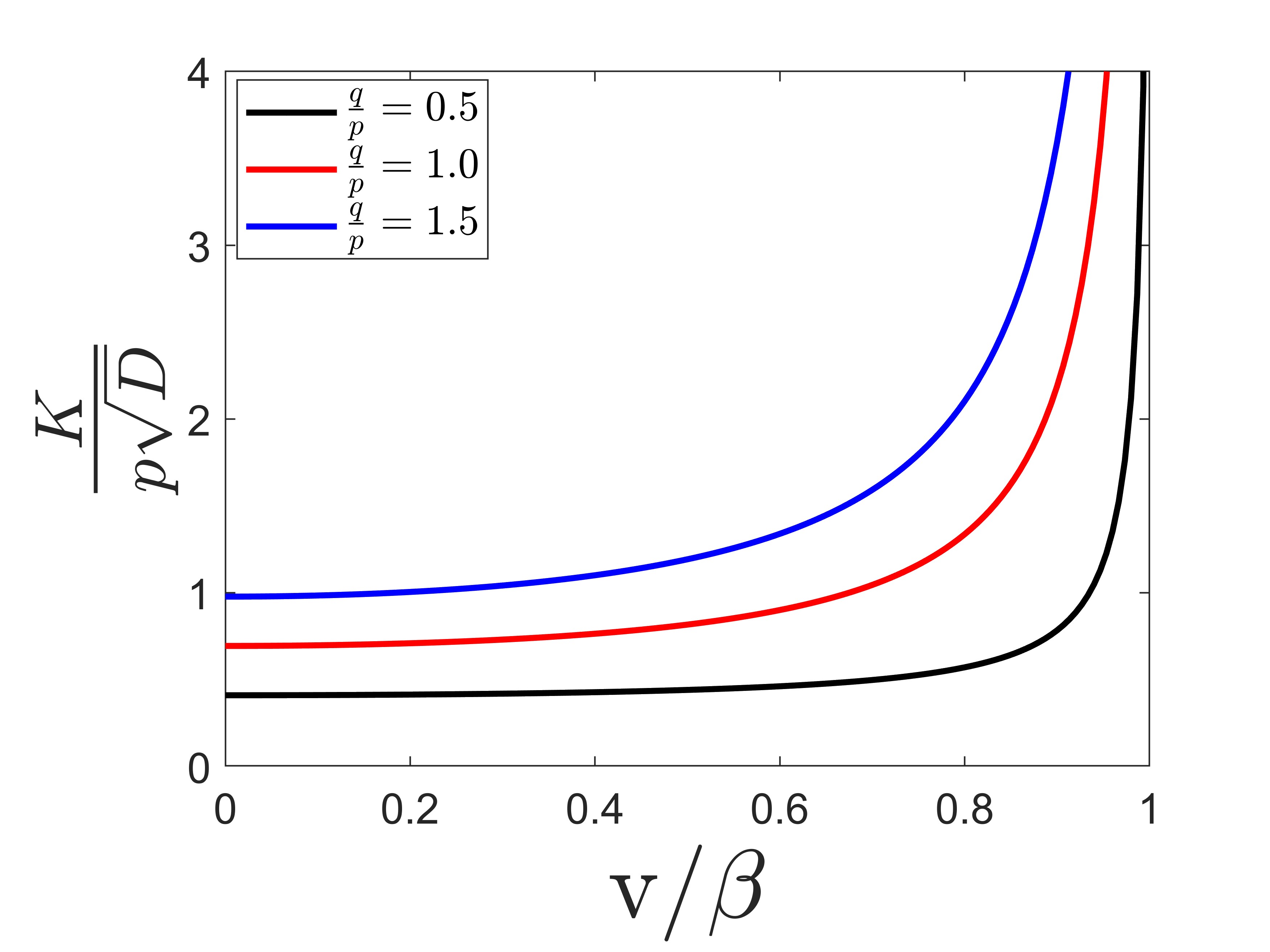}
\caption{}
\label{Fig_12}
\end{subfigure}
~
\begin{subfigure}[b] {0.38\textwidth}
\includegraphics[width=\textwidth ]{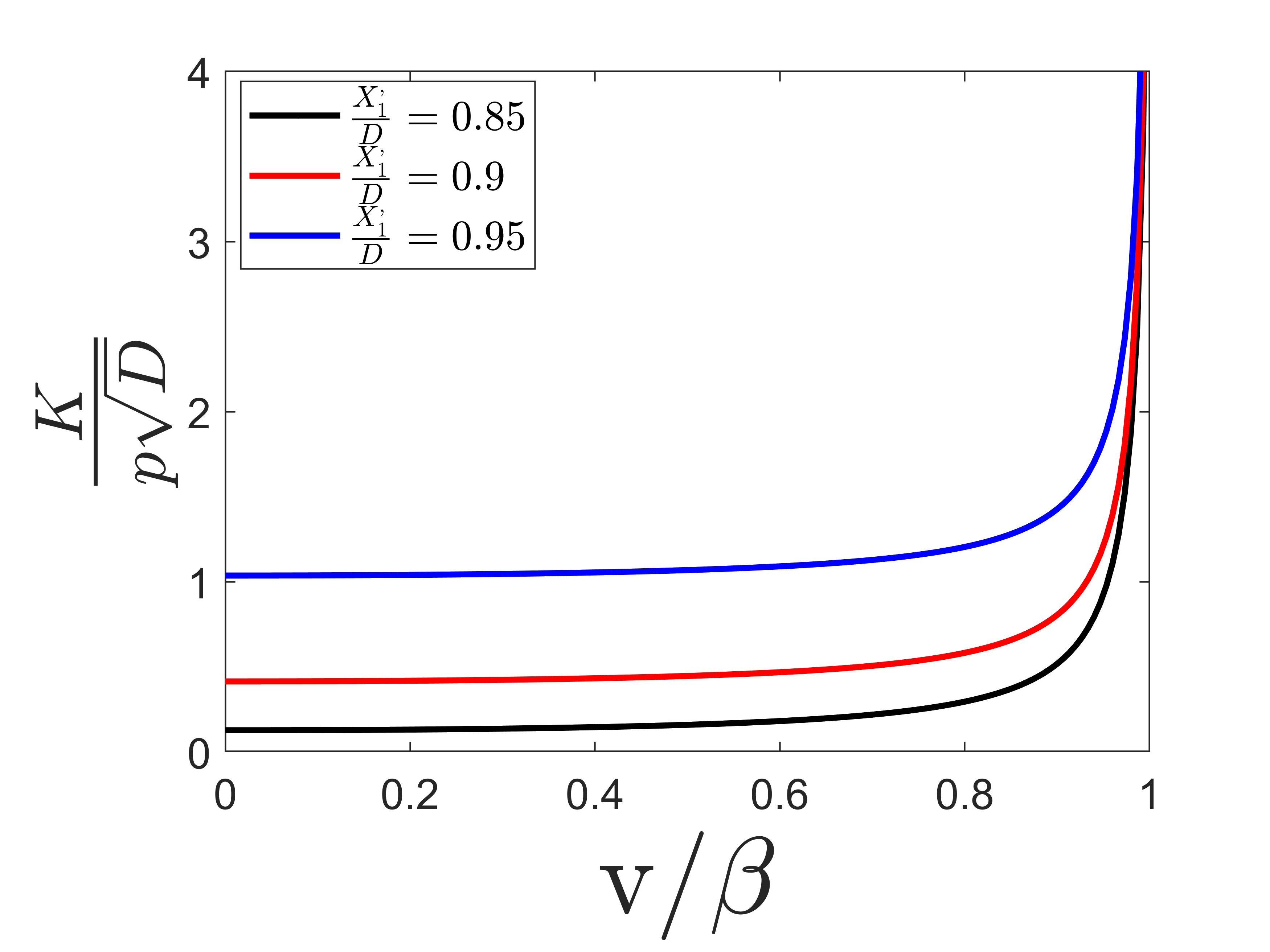}
\caption{}
\label{Fig_13}
\end{subfigure}
\caption{Impact of dimensionless stress intensity factor ($K/p\sqrt{D}$) with dimensionless crack velocity ($v/\beta$) in case of carbon material, illustrating the influence of dimensionless parameters such as (i) crack length $(D/h)$, (ii) initial compressive stress $(P)$, (iii) uniform angular velocity $(\Omega)$, (iv) magnetoelastic coupling parameter (${\mu_e H^2_0}/{c_5}$), (v) punch pressure $(q/p)$, and (vi) point load position ($X^{'}_1/D$). } 
\label{Figure 2}
\efg
%%%=============isotropic Graphs=========================
\bfg[htbp]
\centering
\begin{subfigure}[b] {0.38\textwidth}
\includegraphics[width=\textwidth ]{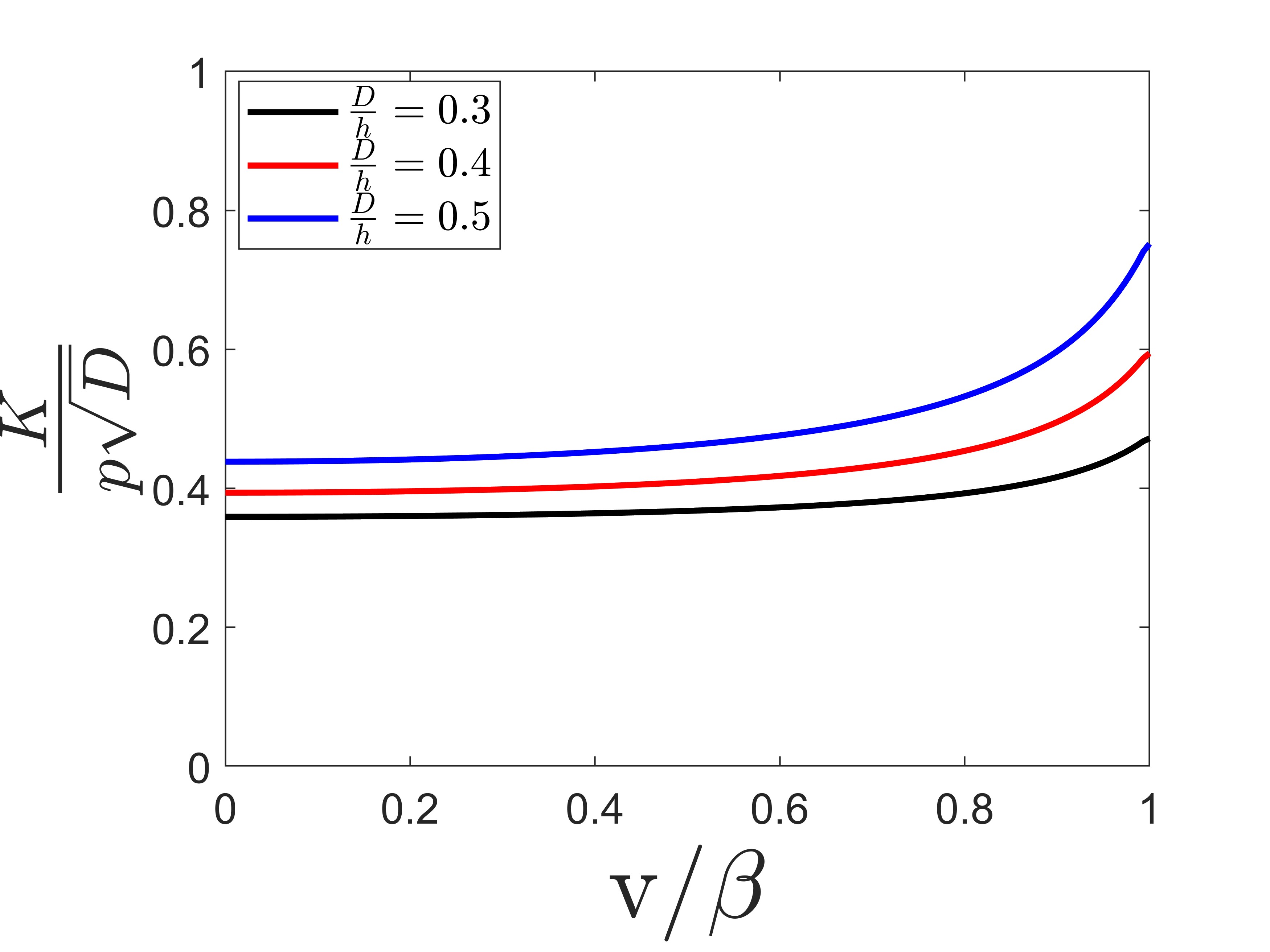}
\caption{}
\label{Fig_14}
\end{subfigure}
~
\begin{subfigure}[b] {0.38\textwidth}
\includegraphics[width=\textwidth ]{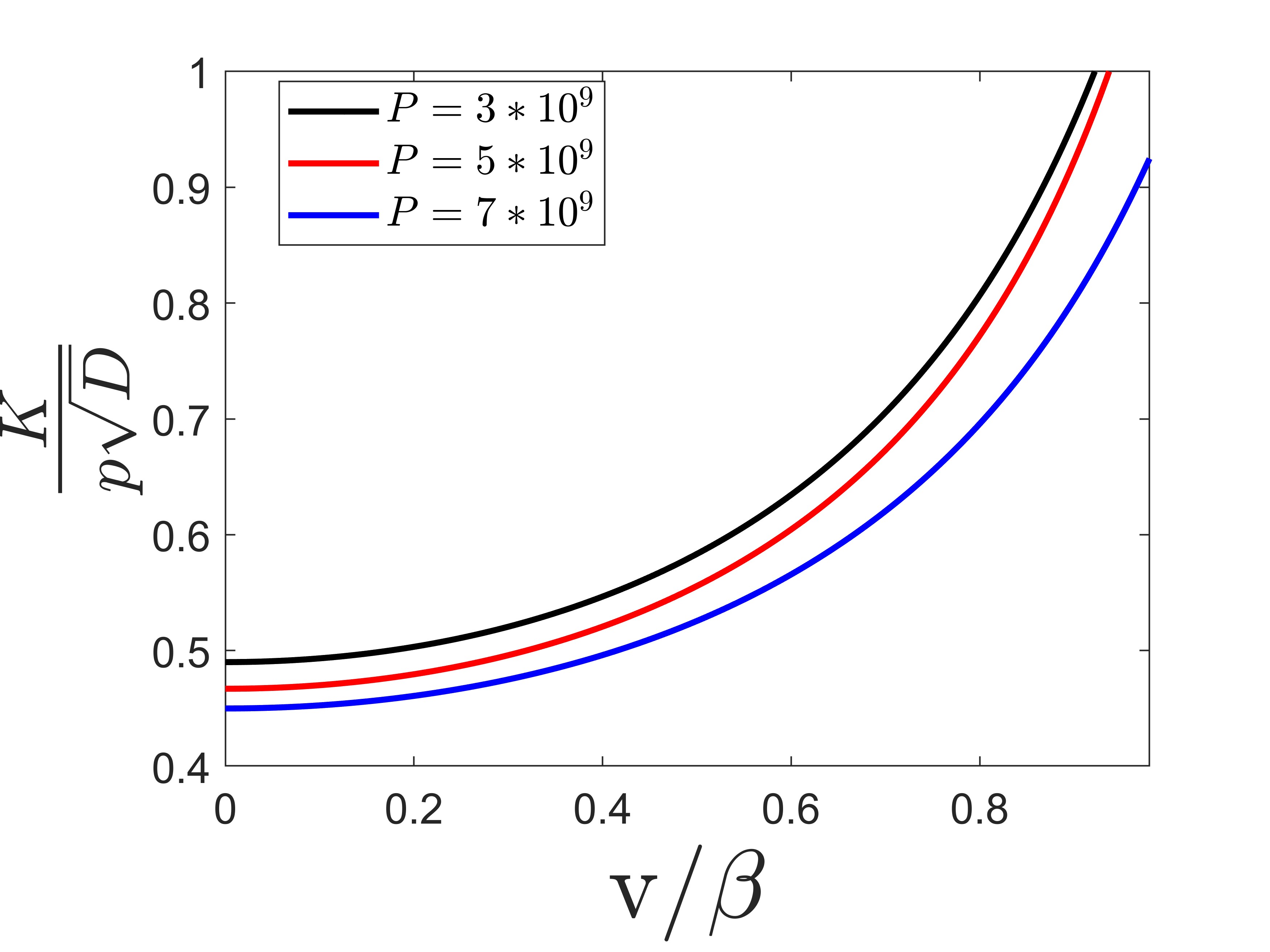}
\caption{}
\label{Fig_15}
\end{subfigure}
~
\begin{subfigure}[b] {0.38\textwidth}
\includegraphics[width=\textwidth ]{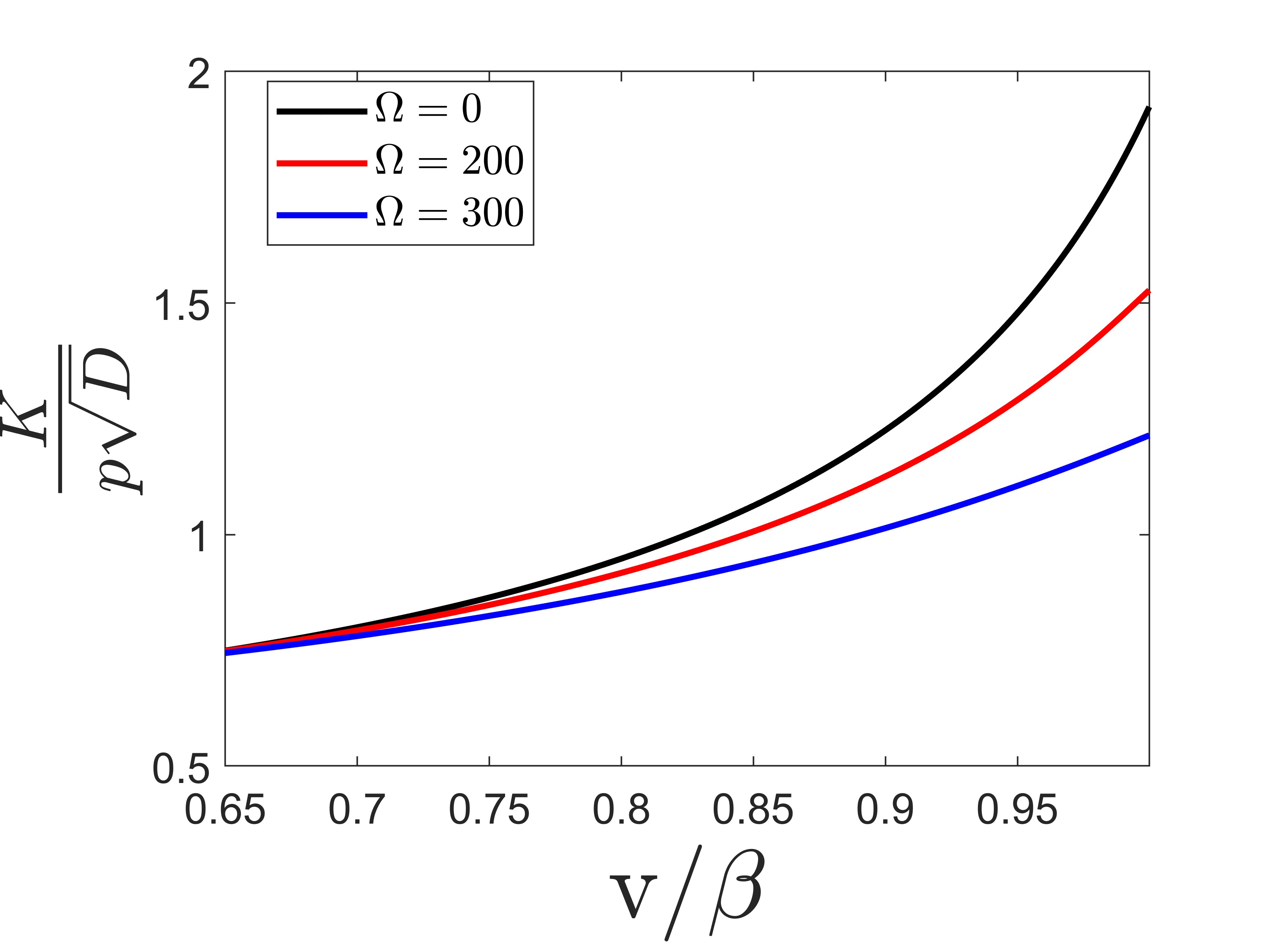}
\caption{}
\label{Fig_16}
\end{subfigure}
~
\begin{subfigure}[b] {0.38\textwidth}
\includegraphics[width=\textwidth ]{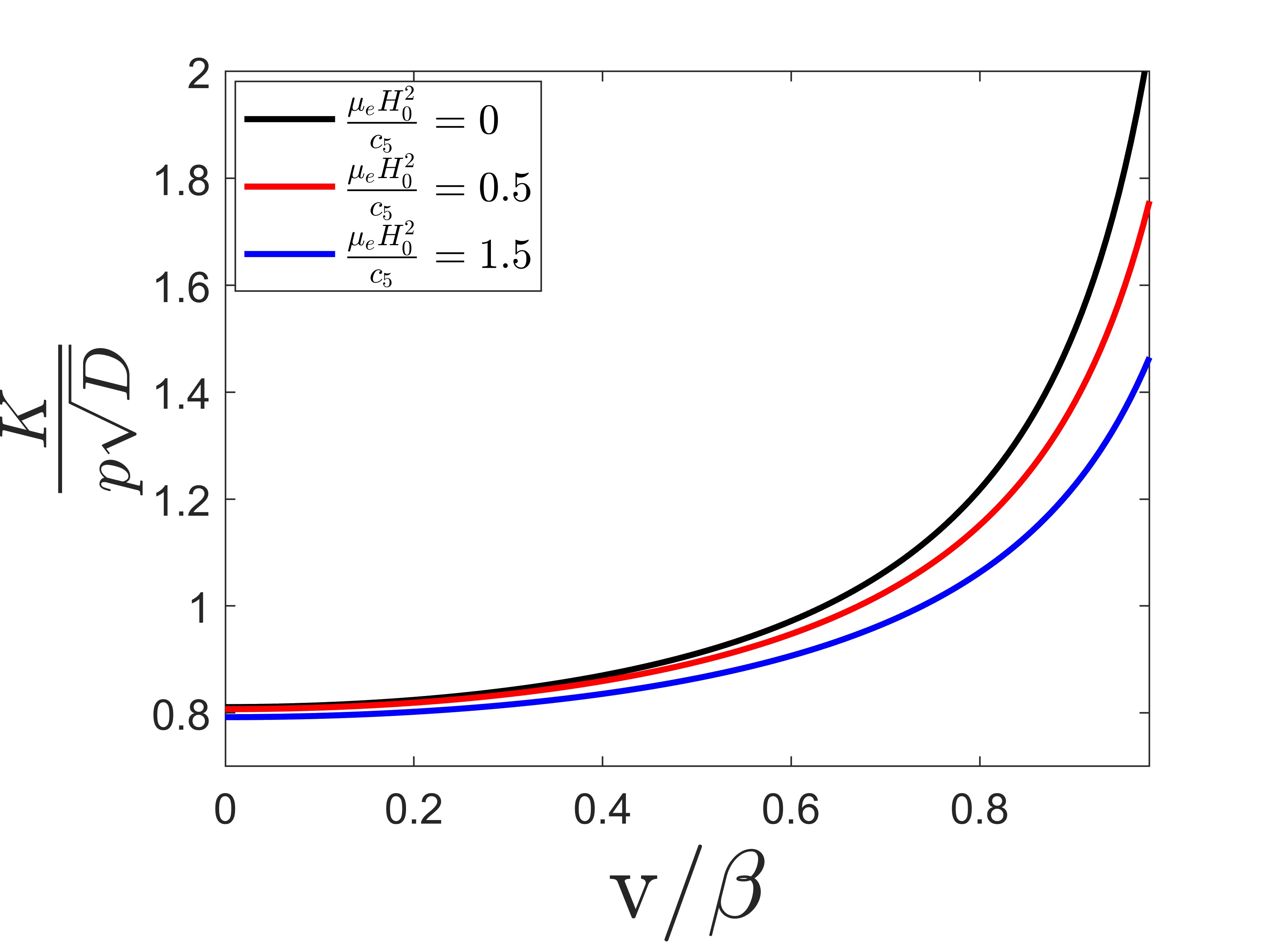}
\caption{}
\label{Fig_17}
\end{subfigure}
~
\begin{subfigure}[b] {0.38\textwidth}
\includegraphics[width=\textwidth ]{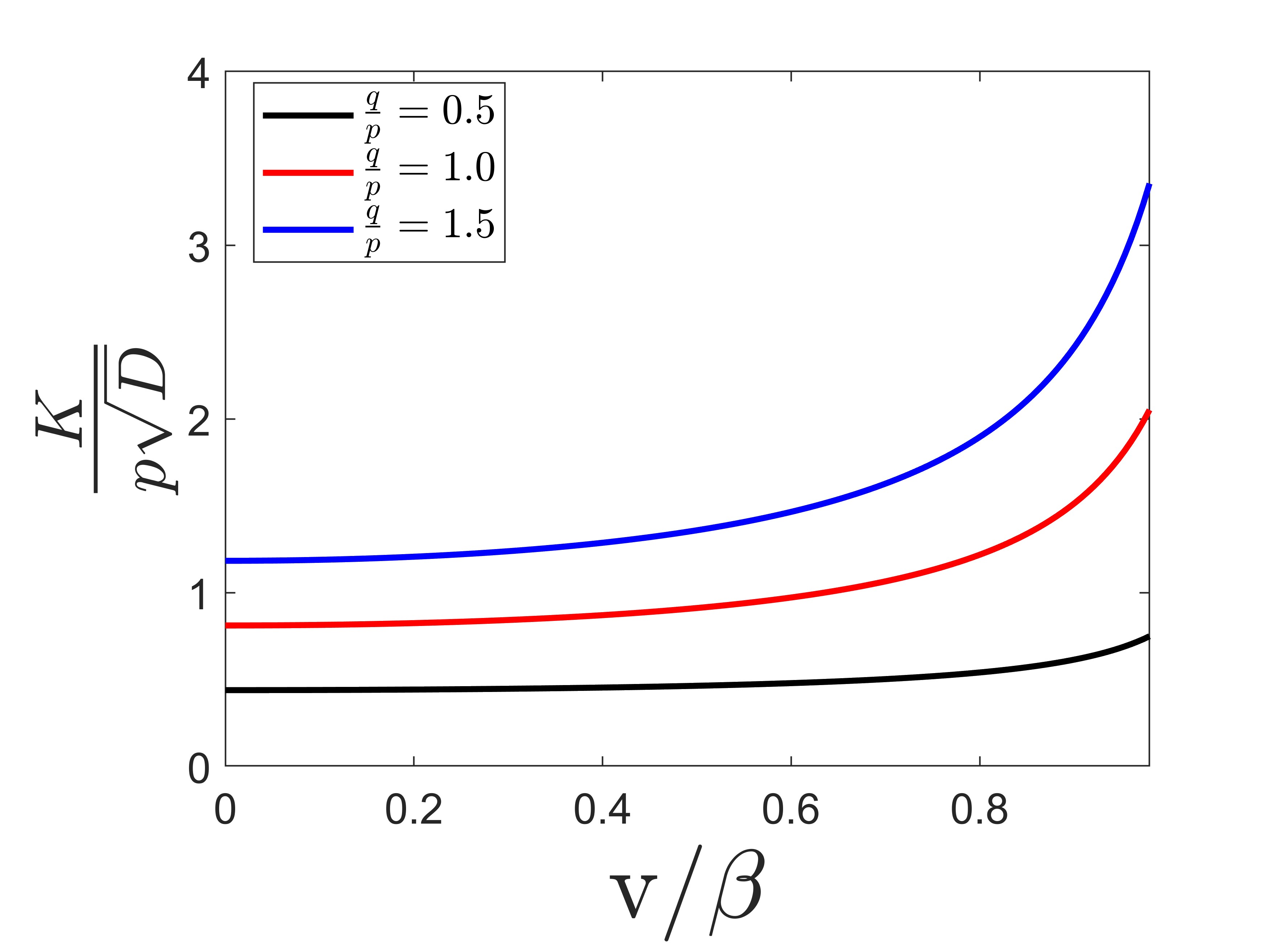}
\caption{}
\label{Fig_18}
\end{subfigure}
~
\begin{subfigure}[b] {0.38\textwidth}
\includegraphics[width=\textwidth ]{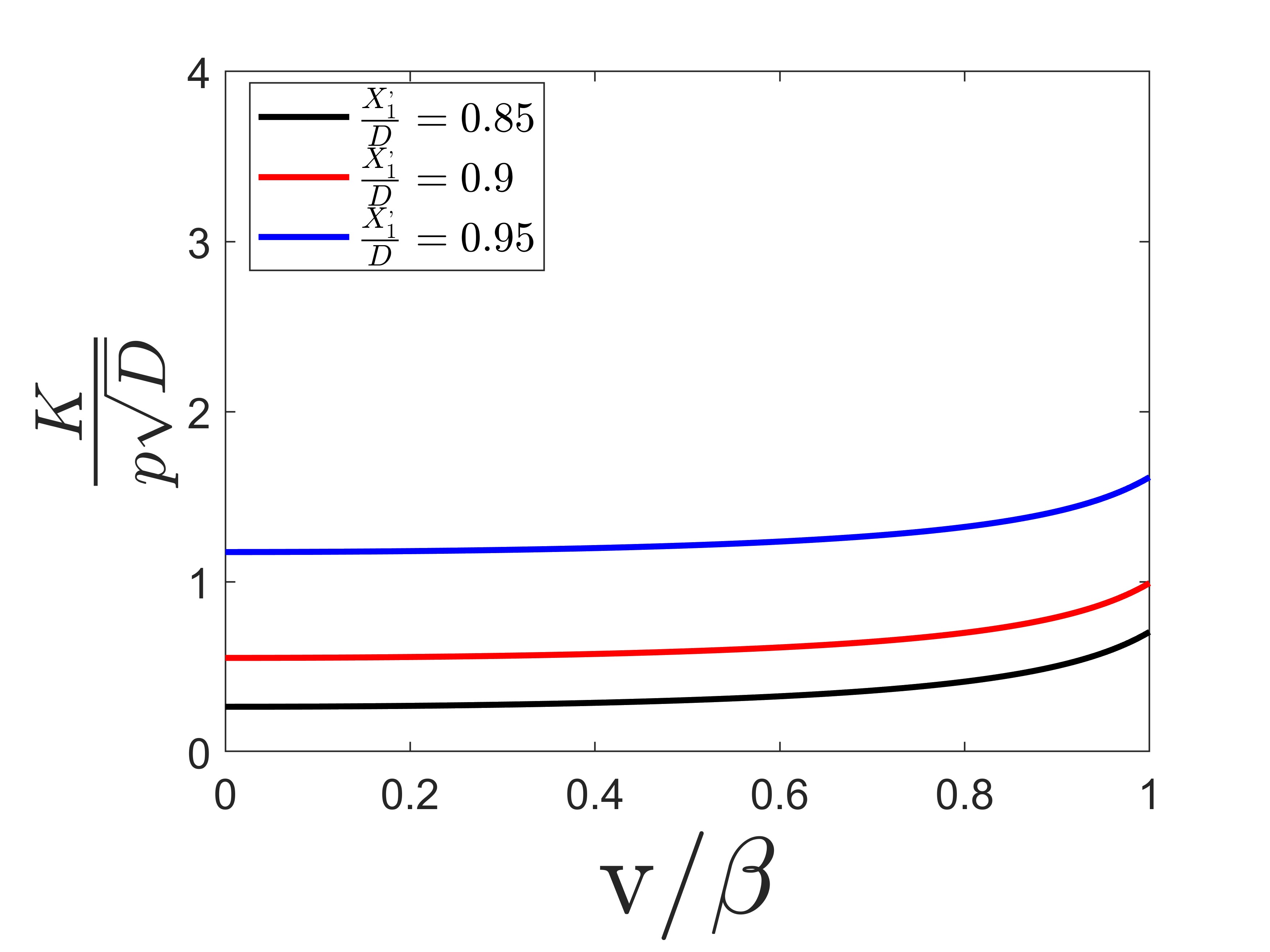}
\caption{}
\label{Fig_19}
\end{subfigure}
\caption{Impact of dimensionless stress intensity factor ($K/p\sqrt{D}$) with dimensionless crack velocity ($v/\beta$) in case of isotropic material, illustrating the influence of dimensionless parameters such as (i) crack length $(D/h)$, (ii) initial compressive stress $(P)$, (iii) uniform angular velocity $(\Omega)$, (iv) magnetoelastic coupling parameter (${\mu_e H^2_0}/{c_5}$), (v) punch pressure $(q/p)$, and (vi) point load position ($X^{'}_1/D$). } 
\label{Figure 3}
\efg

%%%=============carbon Graphs=========================
\bfg[htbp]
\centering
\begin{subfigure}[b] {0.38\textwidth}
\includegraphics[width=\textwidth ]{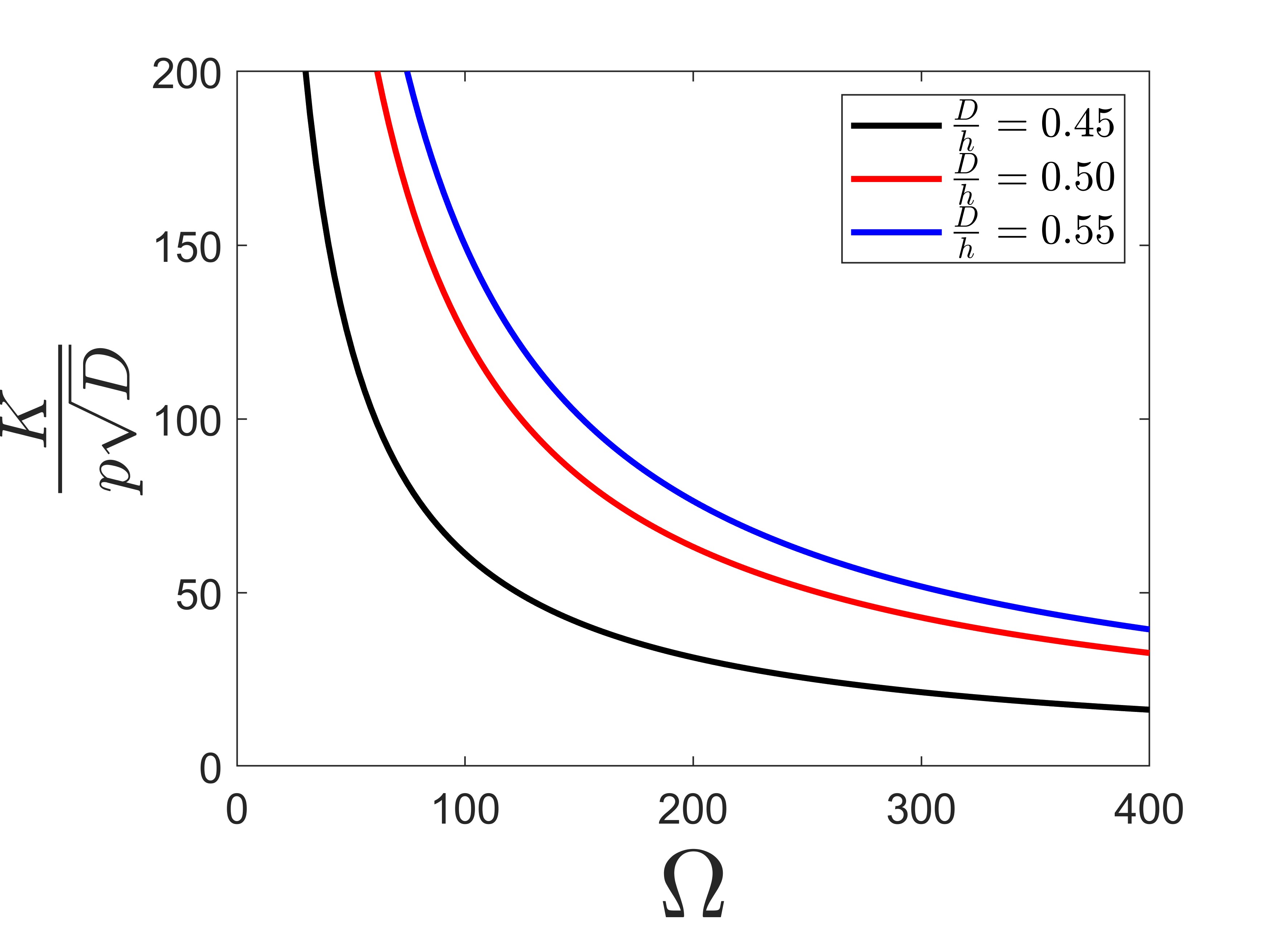}
\caption{}
\label{Fig_20}
\end{subfigure}
~
\begin{subfigure}[b] {0.38\textwidth}
\includegraphics[width=\textwidth ]{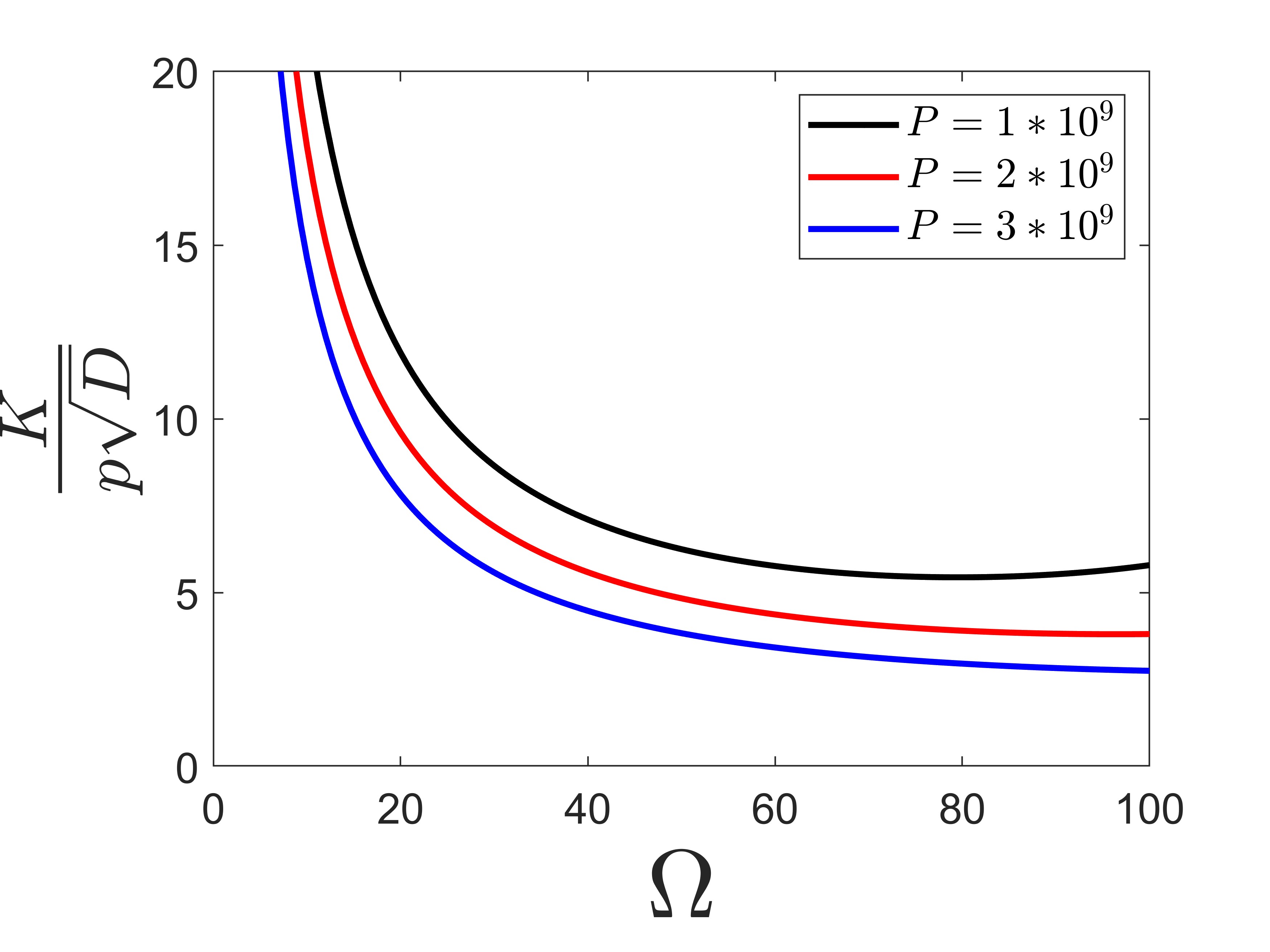}
\caption{}
\label{Fig_21}
\end{subfigure}
~
\begin{subfigure}[b] {0.38\textwidth}
\includegraphics[width=\textwidth ]{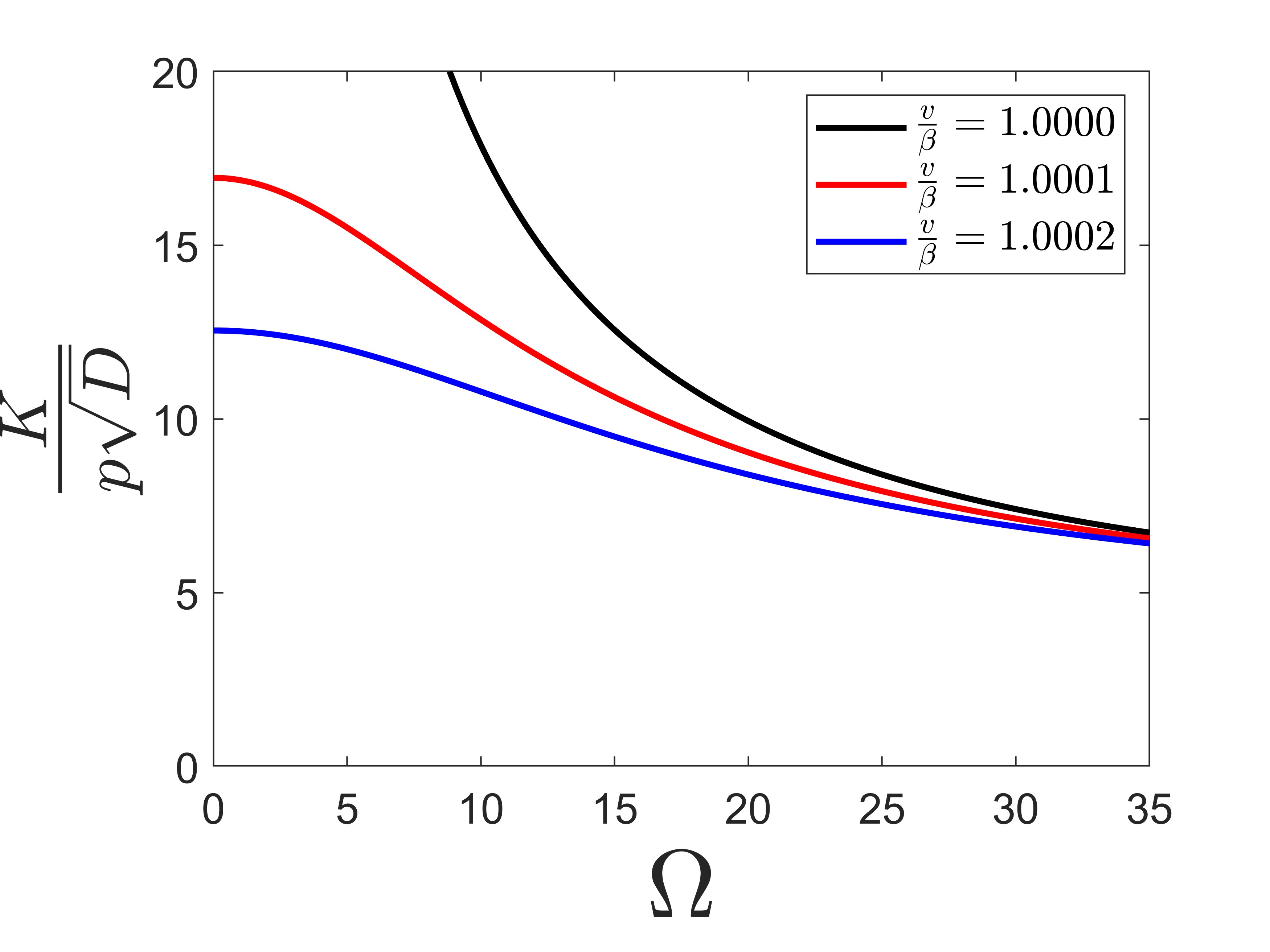}
\caption{}
\label{Fig_22}
\end{subfigure}
~
\begin{subfigure}[b] {0.38\textwidth}
\includegraphics[width=\textwidth ]{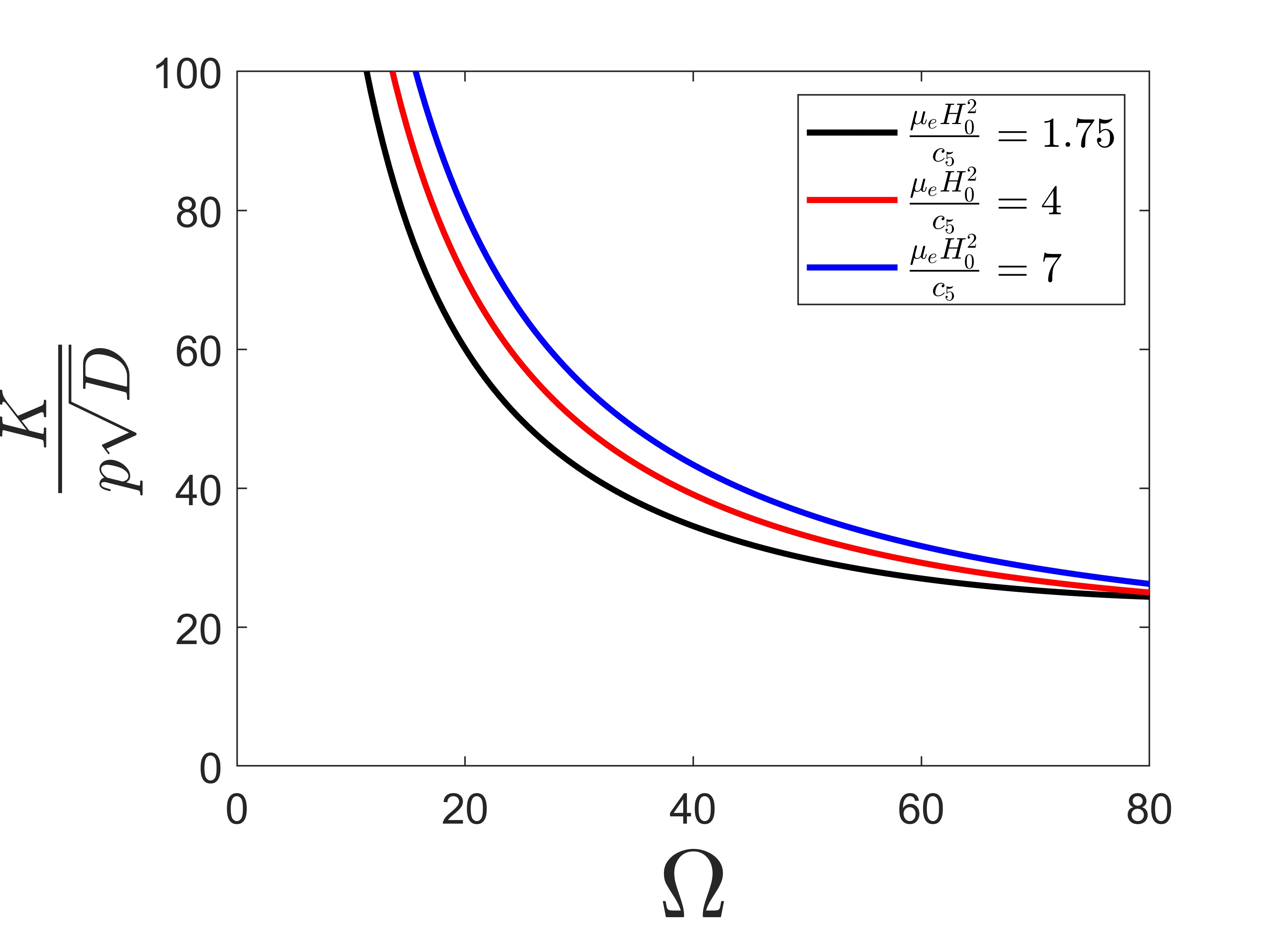}
\caption{}
\label{Fig_23}
\end{subfigure}
~
\begin{subfigure}[b] {0.38\textwidth}
\includegraphics[width=\textwidth ]{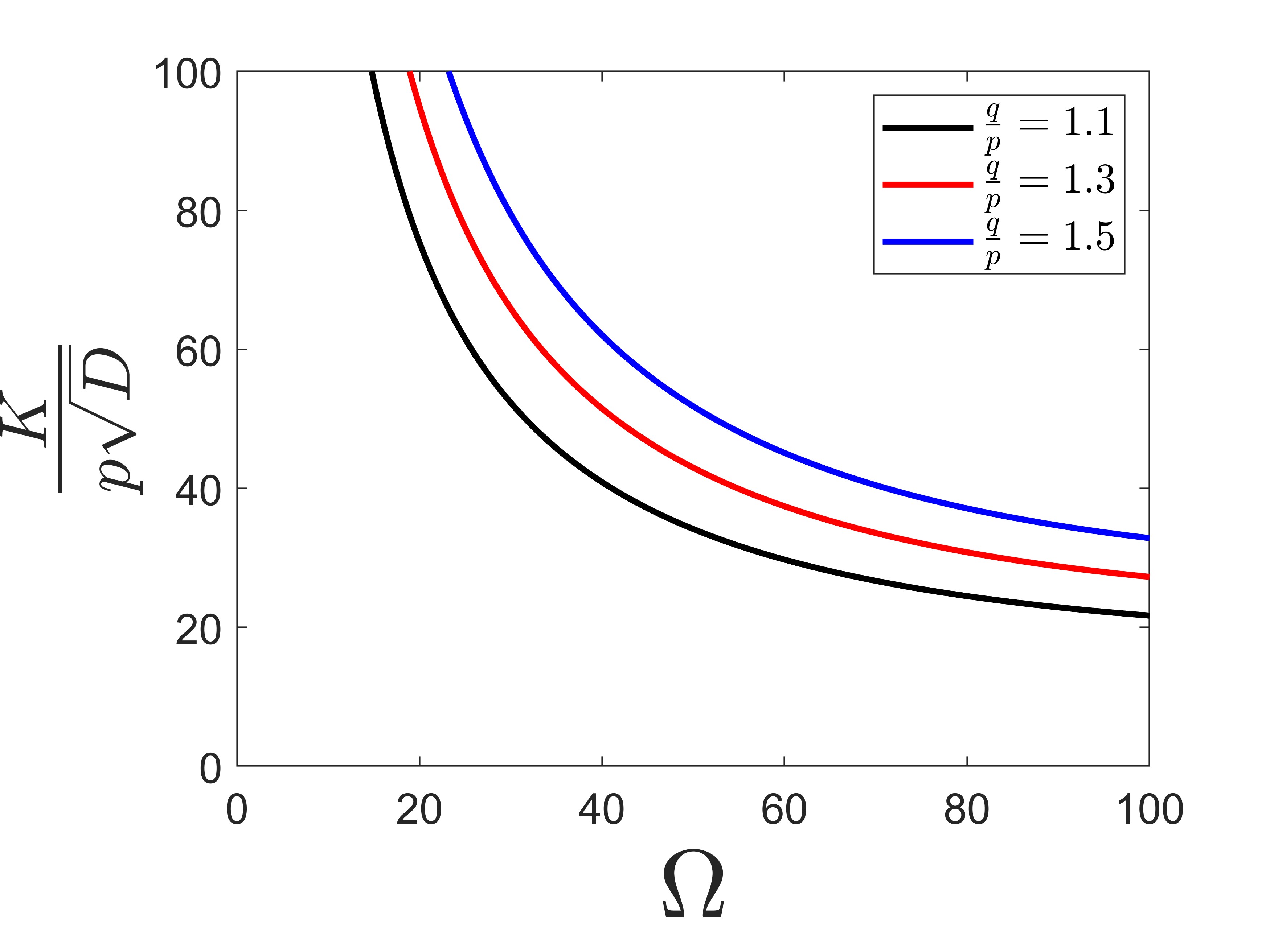}
\caption{}
\label{Fig_24}
\end{subfigure}
~
\begin{subfigure}[b] {0.38\textwidth}
\includegraphics[width=\textwidth ]{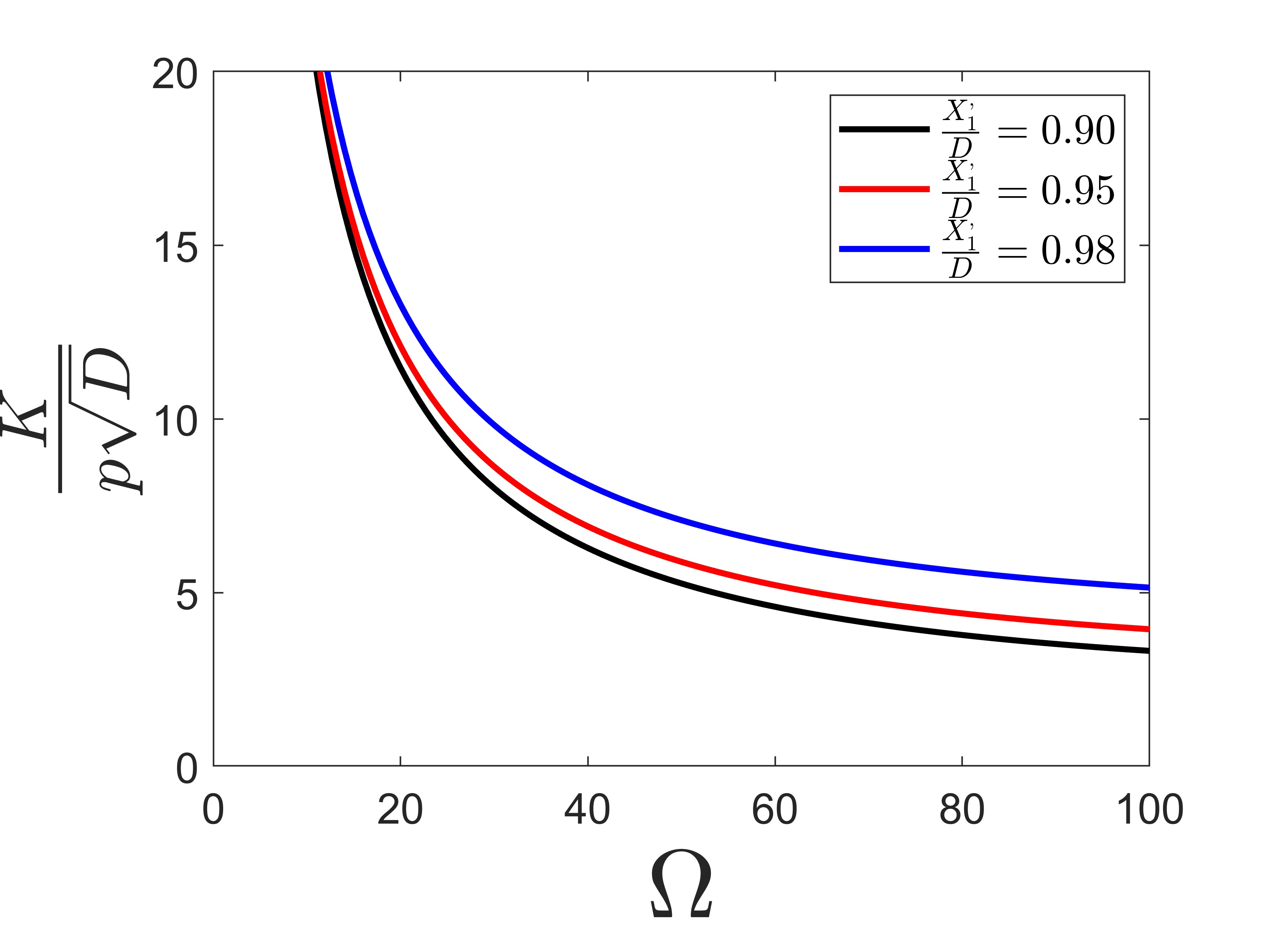}
\caption{}
\label{Fig_25}
\end{subfigure}
\caption{Impact of dimensionless stress intensity factor ($K/p\sqrt{D}$) with uniform angular velocity ($\Omega$) in case of carbon material, illustrating the influence of dimensionless parameters such as (i) crack length $(D/h)$, (ii) initial compressive stress $(P)$, (iii) dimensionless velocity of crack ($v/\beta$), (iv) magnetoelastic coupling parameter (${\mu_e H^2_0}/{c_5}$), (v) punch pressure $(q/p)$, and (vi) point load position ($X^{'}_1/D$).} 
\label{Figure 4}
\efg

%%%=============steel Graphs=========================
\bfg[htbp]
\centering
\begin{subfigure}[b] {0.38\textwidth}
\includegraphics[width=\textwidth ]{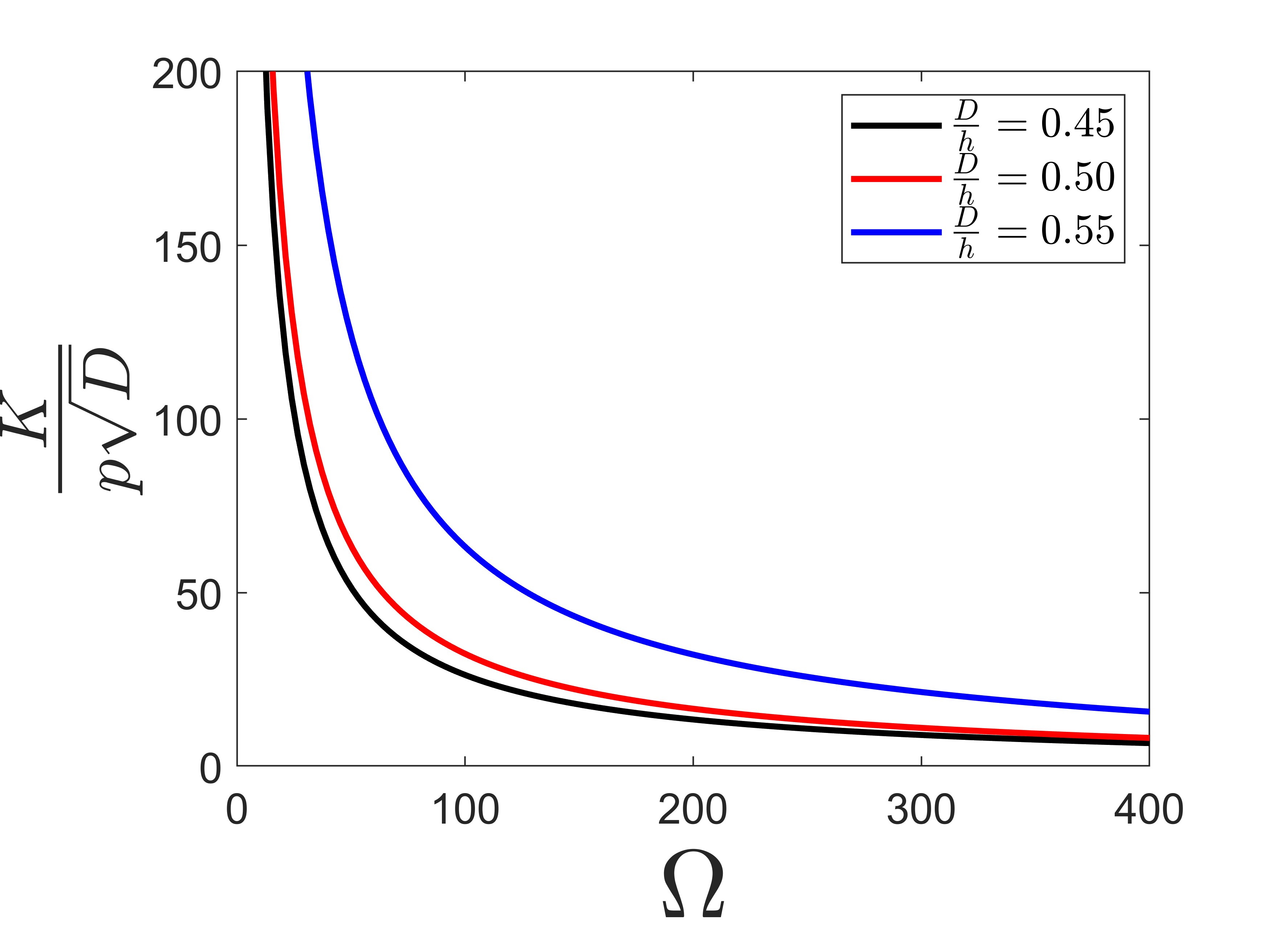}
\caption{}
\label{Fig_26}
\end{subfigure}
~
\begin{subfigure}[b] {0.38\textwidth}
\includegraphics[width=\textwidth ]{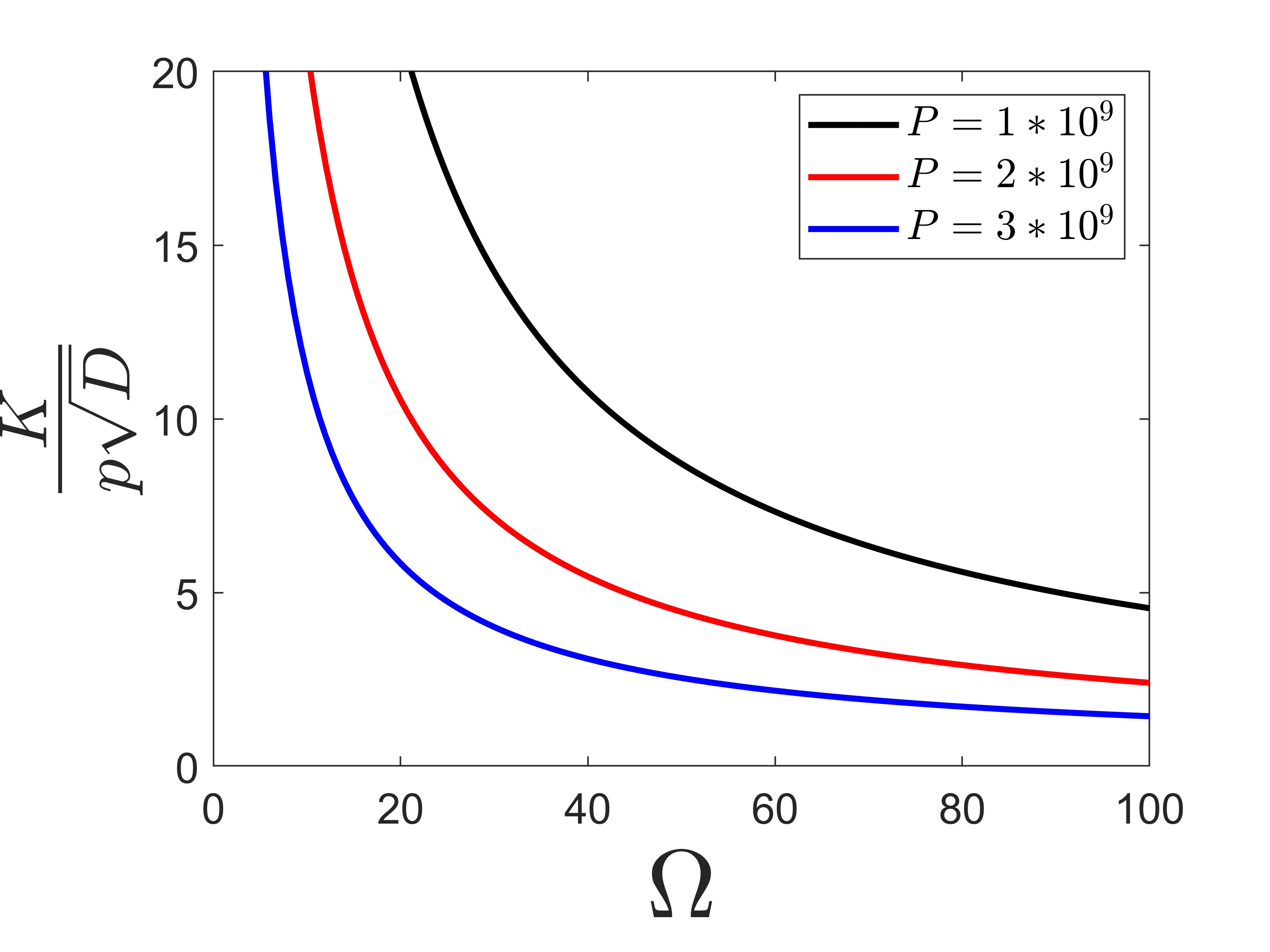}
\caption{}
\label{Fig_27}
\end{subfigure}
~
\begin{subfigure}[b] {0.38\textwidth}
\includegraphics[width=\textwidth ]{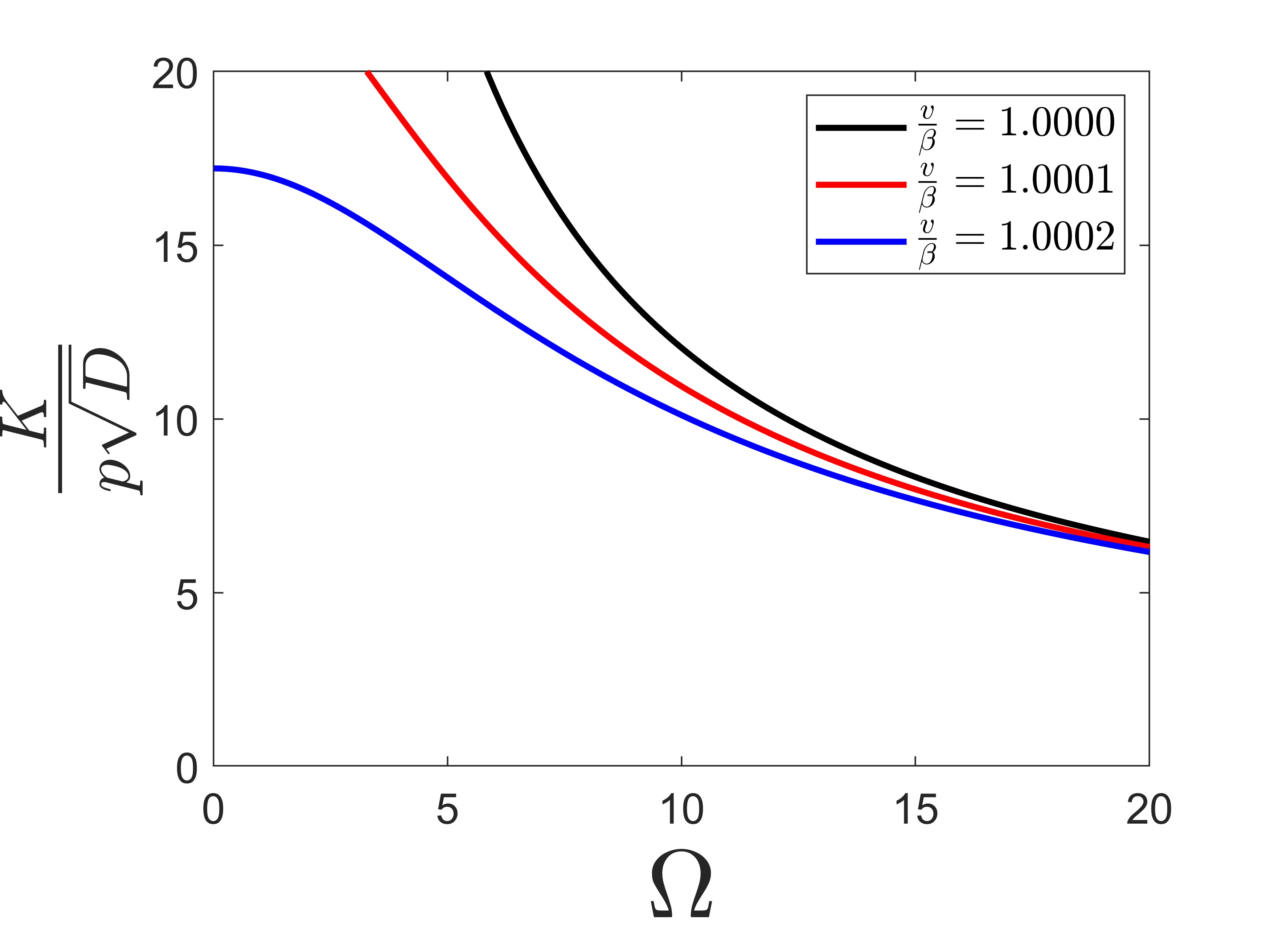}
\caption{}
\label{Fig_28}
\end{subfigure}
~
\begin{subfigure}[b] {0.38\textwidth}
\includegraphics[width=\textwidth ]{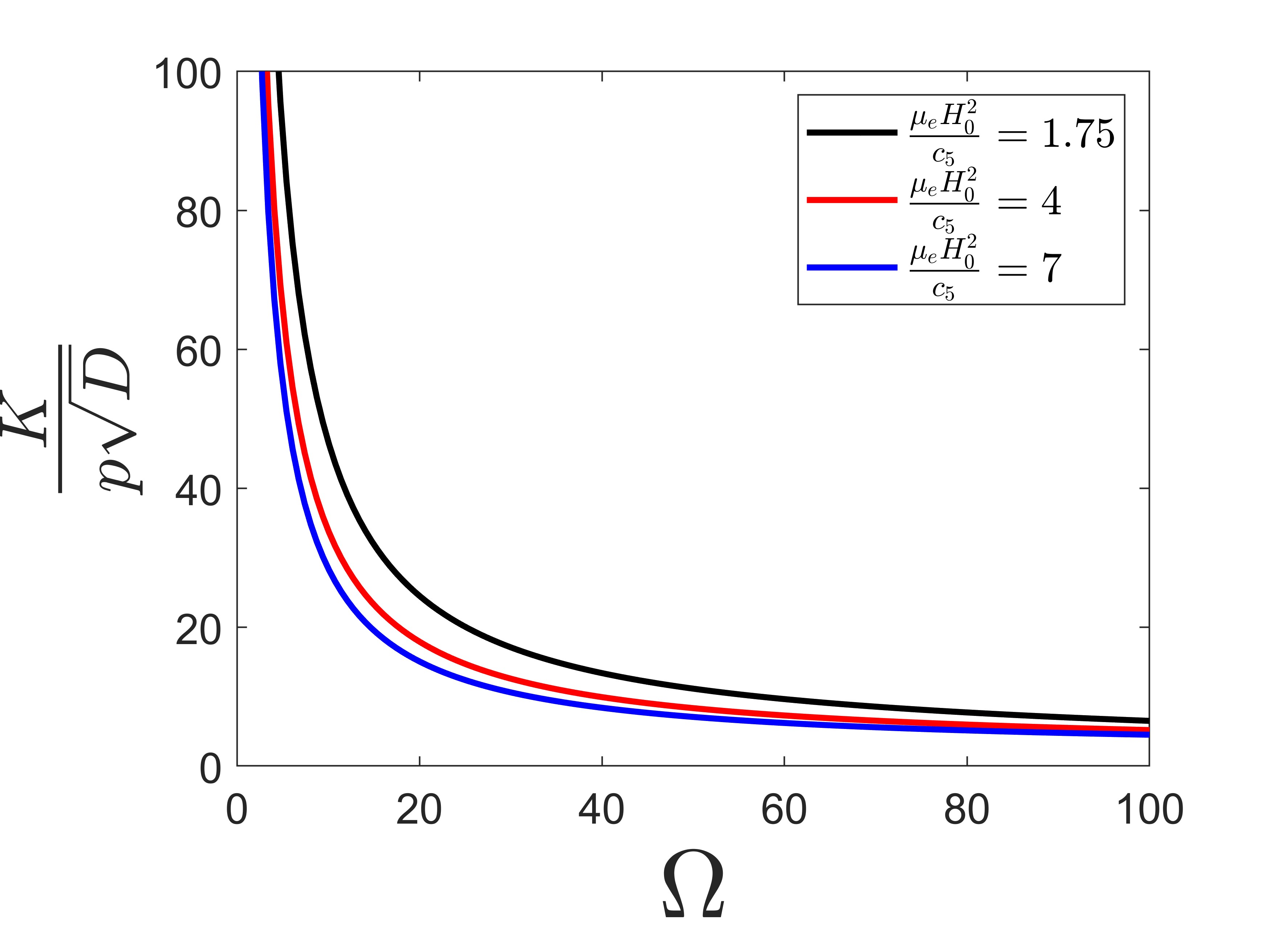}
\caption{}
\label{Fig_29}
\end{subfigure}
~
\begin{subfigure}[b] {0.38\textwidth}
\includegraphics[width=\textwidth ]{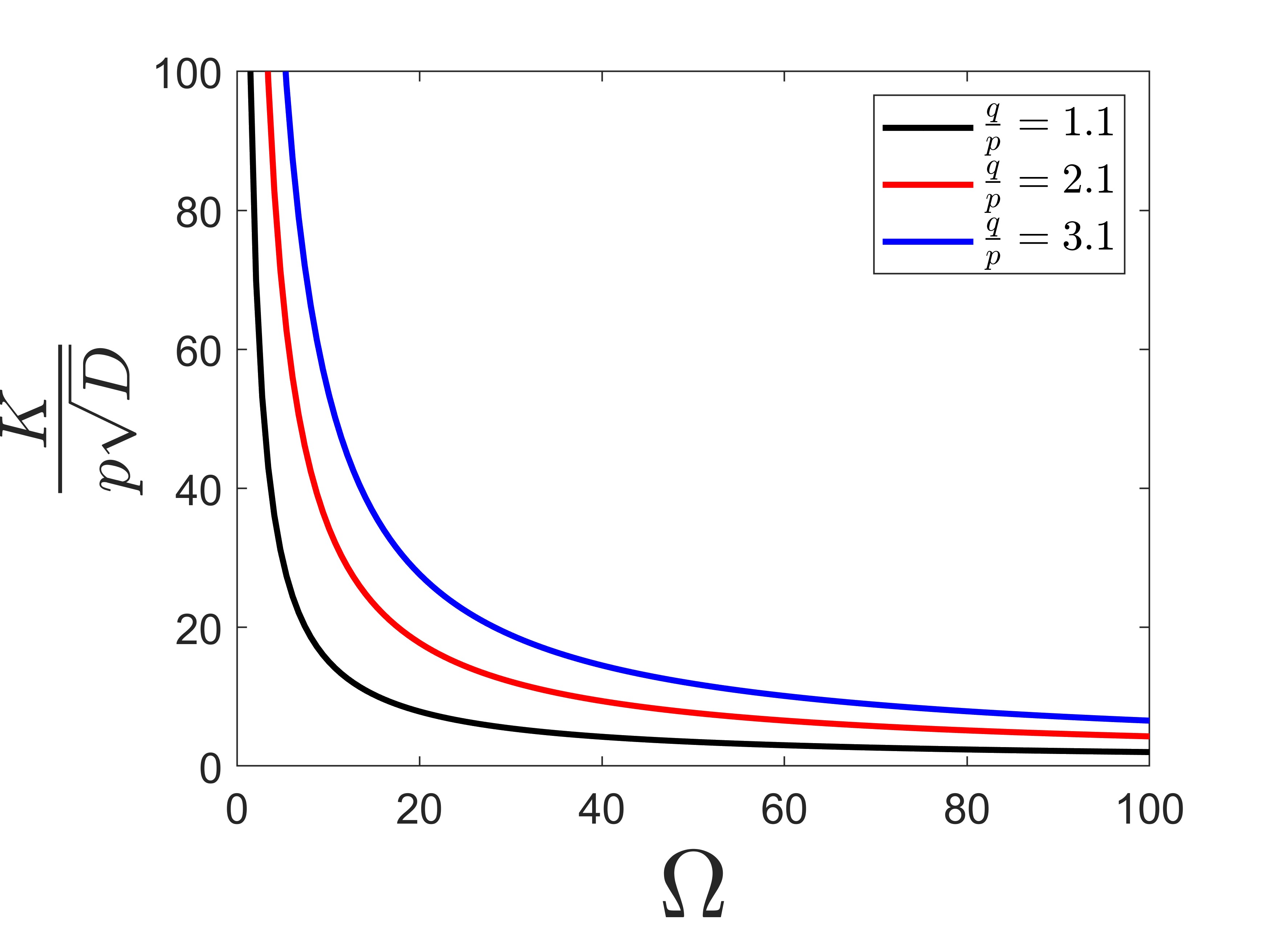}
\caption{}
\label{Fig_30}
\end{subfigure}
~
\begin{subfigure}[b] {0.38\textwidth}
\includegraphics[width=\textwidth ]{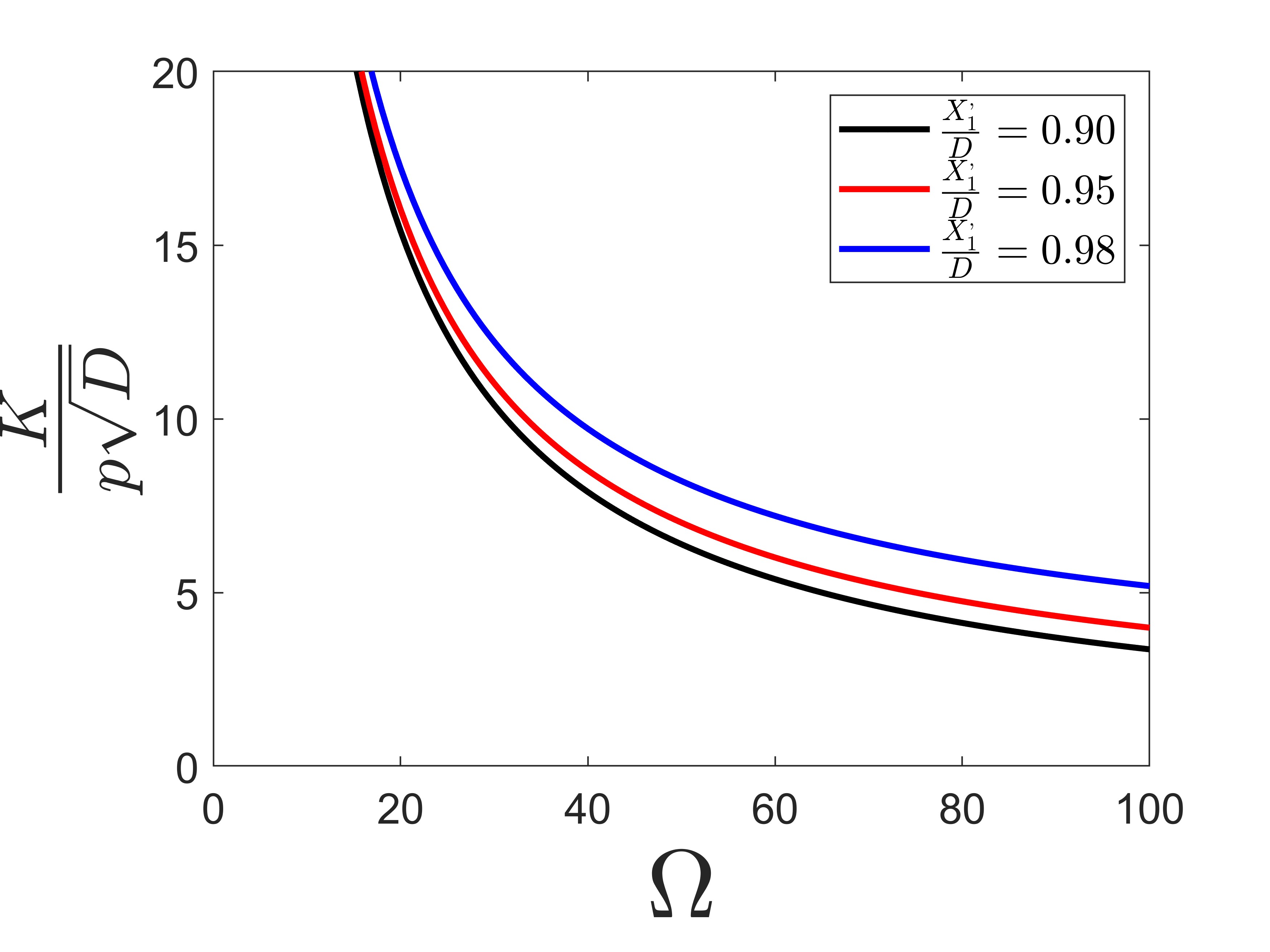}
\caption{}
\label{Fig_31}
\end{subfigure}
\caption{Impact of dimensionless stress intensity factor ($K/p\sqrt{D}$) with uniform angular velocity ($\Omega$) in case of steel, illustrating the influence of dimensionless parameters such as (i) crack length $(D/h)$, (ii) initial compressive stress $(P)$, (iii) dimensionless velocity of crack ($v/\beta$), (iv) magnetoelastic coupling parameter (${\mu_e H^2_0}/{c_5}$), (v) punch pressure $(q/p)$, and (vi) point load position ($X^{'}_1/D$).}
\label{Figure 5}
\efg

%%%=============isotropic Graphs=========================
\bfg[htbp]
\centering
\begin{subfigure}[b] {0.38\textwidth}
\includegraphics[width=\textwidth ]{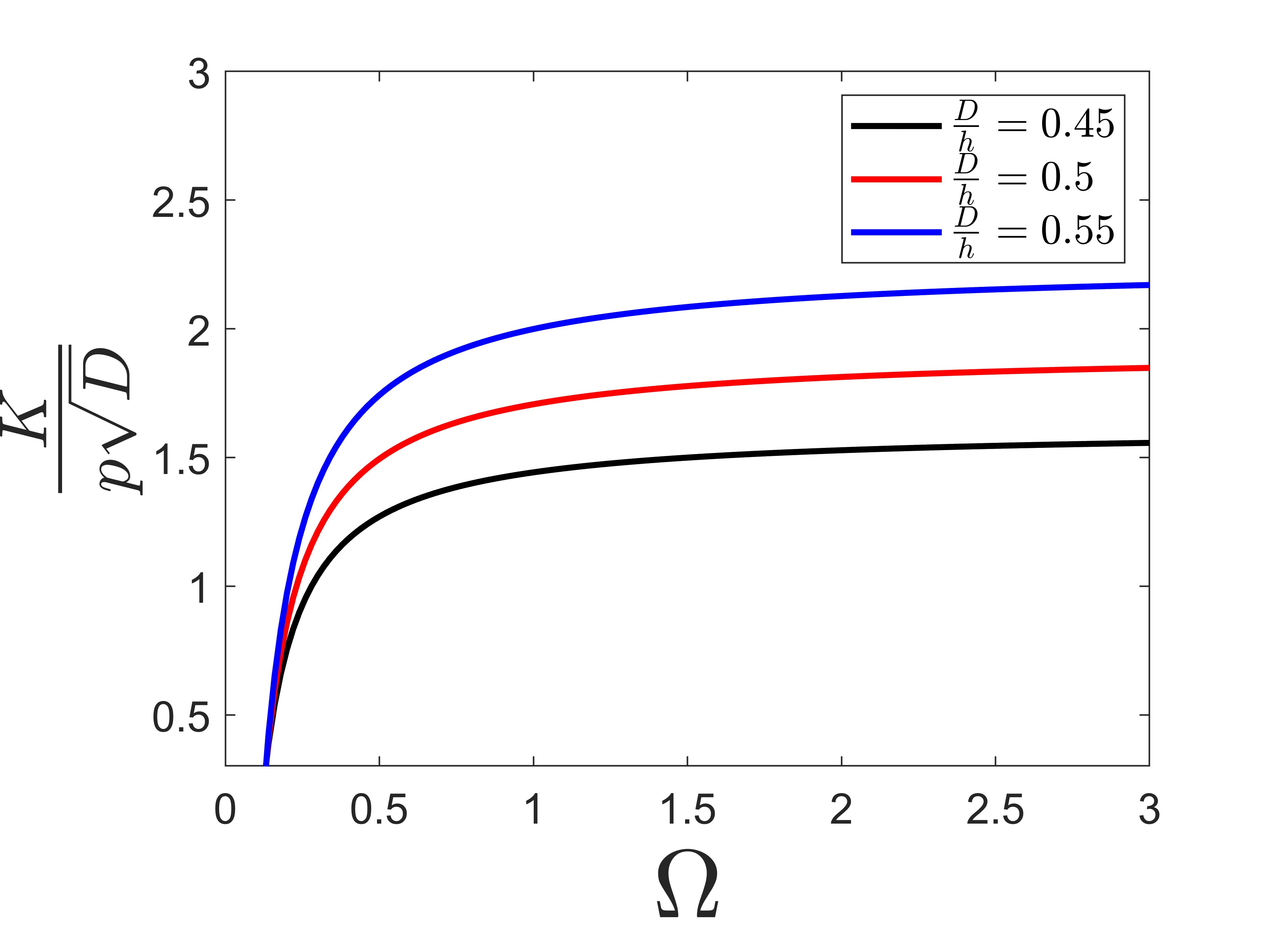}
\caption{}
\label{Fig_32}
\end{subfigure}
~
\begin{subfigure}[b] {0.38\textwidth}
\includegraphics[width=\textwidth ]{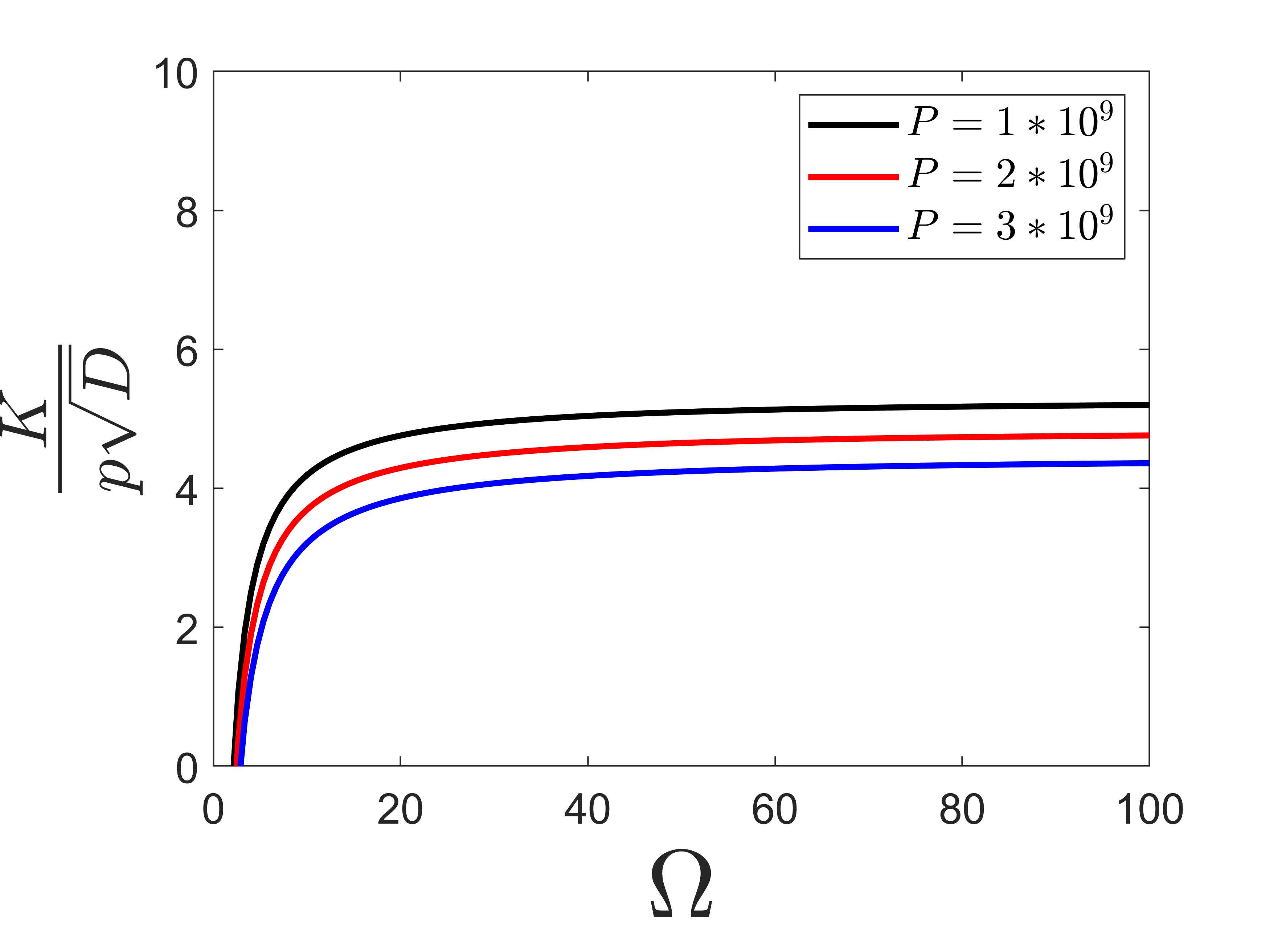}
\caption{}
\label{Fig_33}
\end{subfigure}
~
\begin{subfigure}[b] {0.38\textwidth}
\includegraphics[width=\textwidth ]{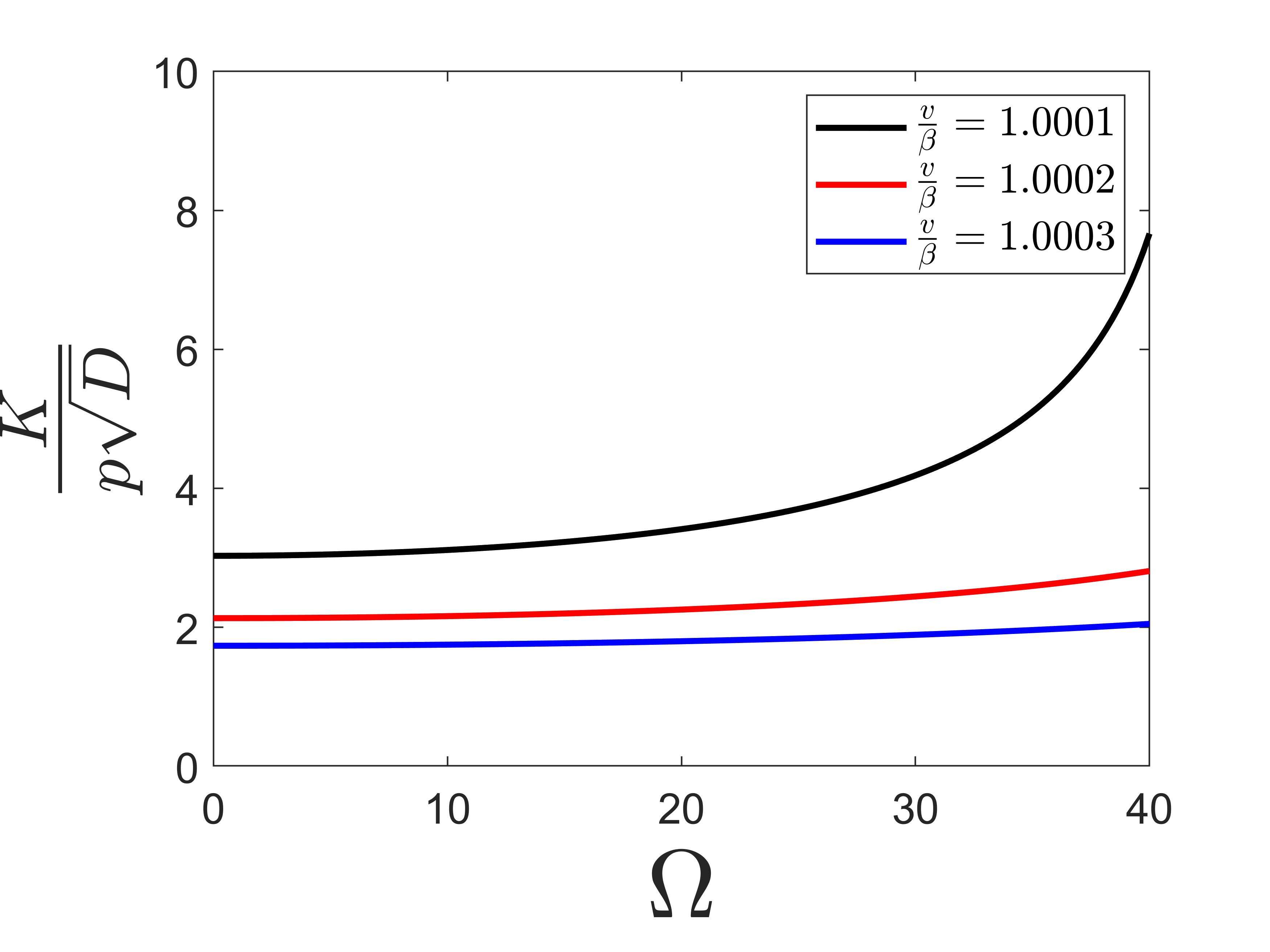}
\caption{}
\label{Fig_34}
\end{subfigure}
~
\begin{subfigure}[b] {0.38\textwidth}
\includegraphics[width=\textwidth ]{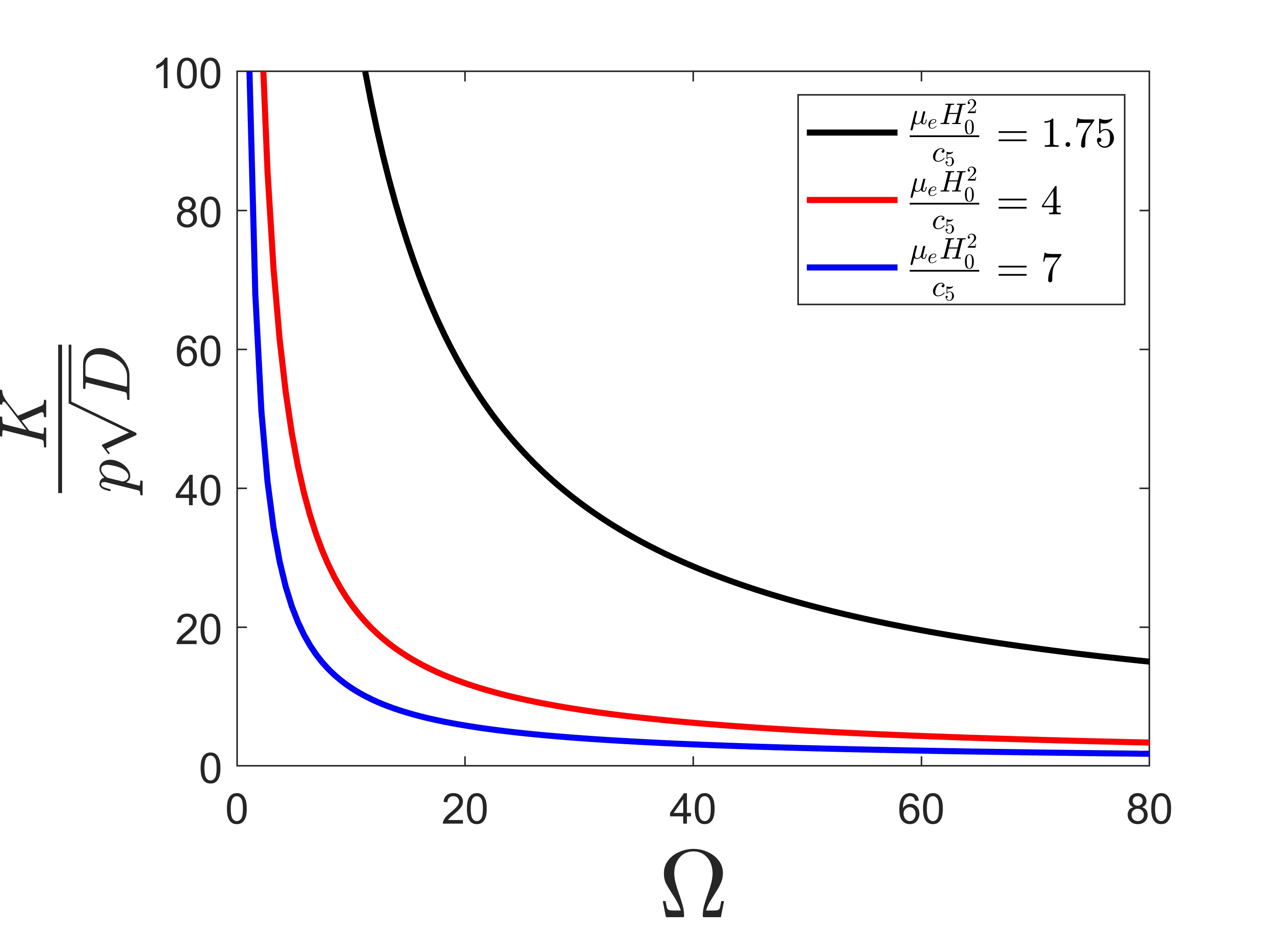}
\caption{}
\label{Fig_35}
\end{subfigure}
~
\begin{subfigure}[b] {0.38\textwidth}
\includegraphics[width=\textwidth ]{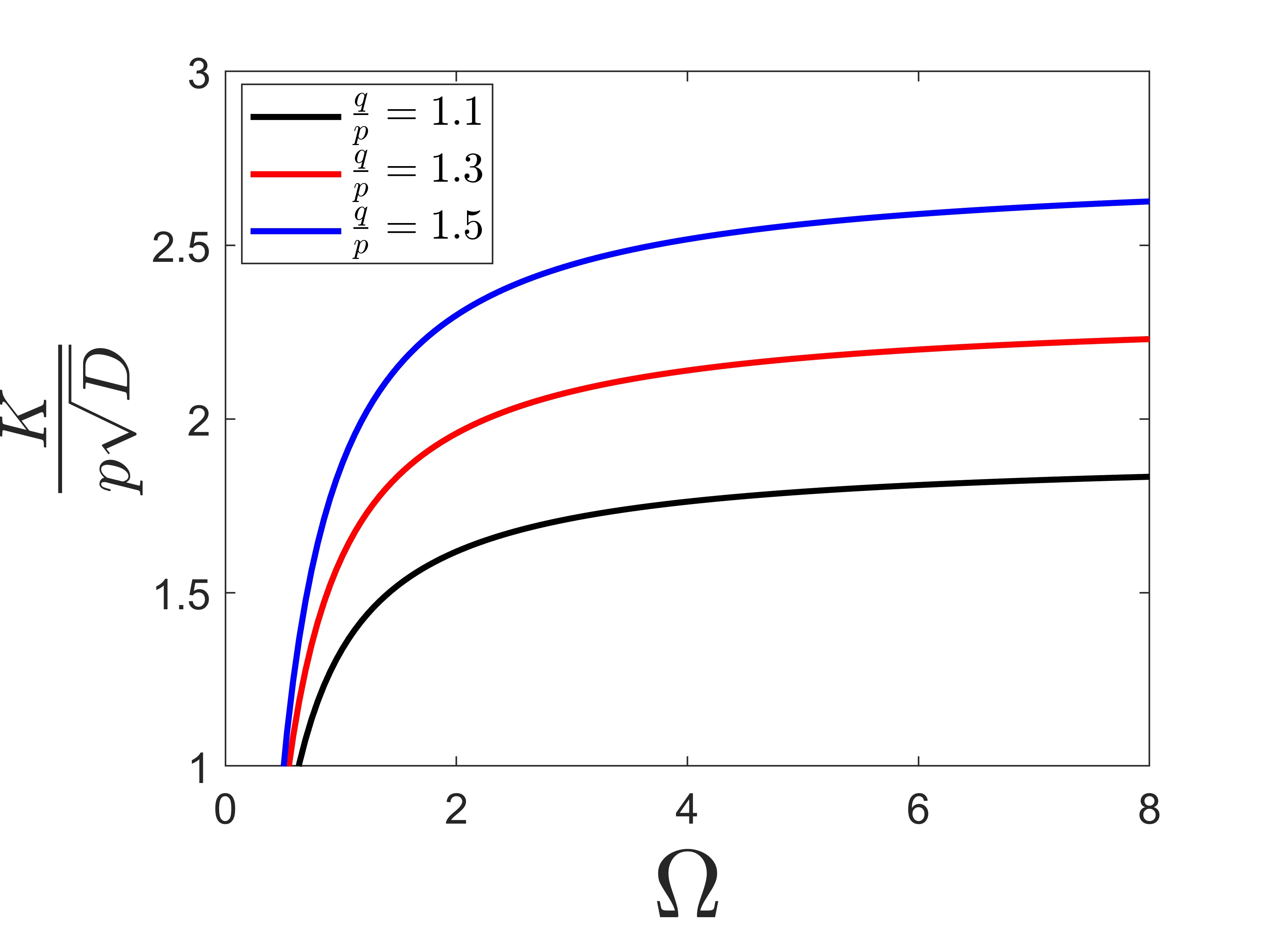}
\caption{}
\label{Fig_36}
\end{subfigure}
~
\begin{subfigure}[b] {0.38\textwidth}
\includegraphics[width=\textwidth ]{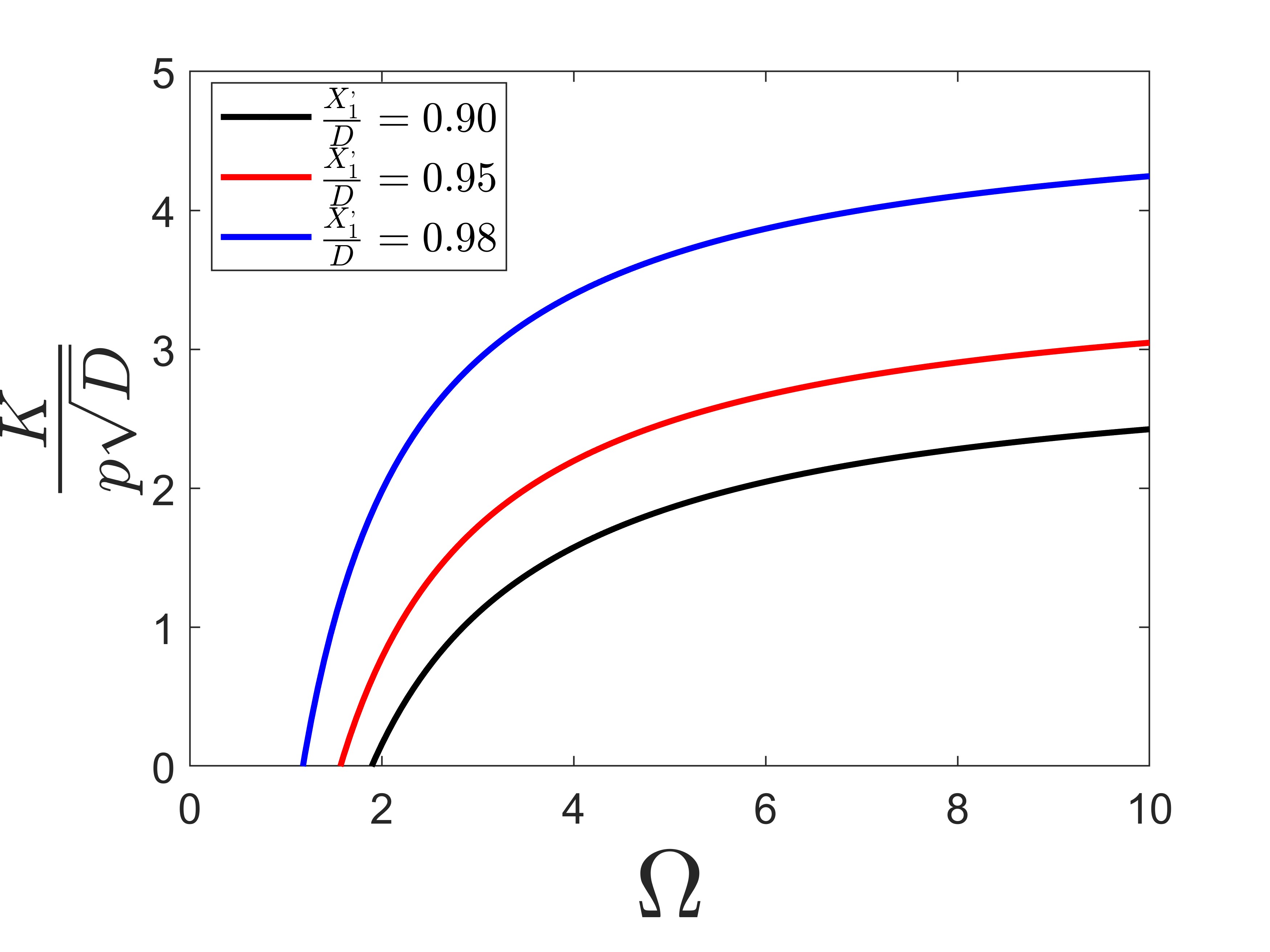}
\caption{}
\label{Fig_37}
\end{subfigure}
\caption{Impact of dimensionless stress intensity factor ($K/p\sqrt{D}$) with uniform angular velocity ($\Omega$) in case of isotropic material, illustrating the influence of dimensionless parameters such as (i) crack length $(D/h)$, (ii) initial compressive stress $(P)$, (iii) dimensionless velocity of crack ($v/\beta$), (iv) magnetoelastic coupling parameter (${\mu_e H^2_0}/{c_5}$), (v) punch pressure $(q/p)$, and (vi) point load position ($X^{'}_1/D$).}
\label{Figure 6}
\efg

\clearpage
%%%%=======================================================%%%%
\section{Special cases}
Figure \ref{Figure 7} has been plotted as a special case where 
Figures \ref{Fig_38}, \ref{Fig_39}, and \ref{Fig_40} depict various aspects of stress intensity factor behavior in different scenarios.

Figure \ref{Figure 7} shows the stress intensity factor plotted against dimensionless crack velocity in two scenarios: with and without initial stress and rotation parameters.

Figures \ref{Fig_38}, \ref{Fig_39}, and \ref{Fig_40} represent graphs for carbon fiber, steel, and isotropic material, respectively.

Observations from these figures reveal that:
\begin{itemize}
    \item In Figure \ref{Fig_38}, the stress intensity factor is notably higher when initial stress and rotation parameters are present, particularly for carbon fiber.
    \item However, in Figures \ref{Fig_39} and \ref{Fig_40}, the stress intensity factor decreases when initial stress and rotation parameters are applied, specifically for steel and isotropic material configurations.
\end{itemize}

%%%=======================================================%%%%
\bfg[htbp]
\centering
\begin{subfigure}[b] {0.38\textwidth}
\includegraphics[width=\textwidth ]{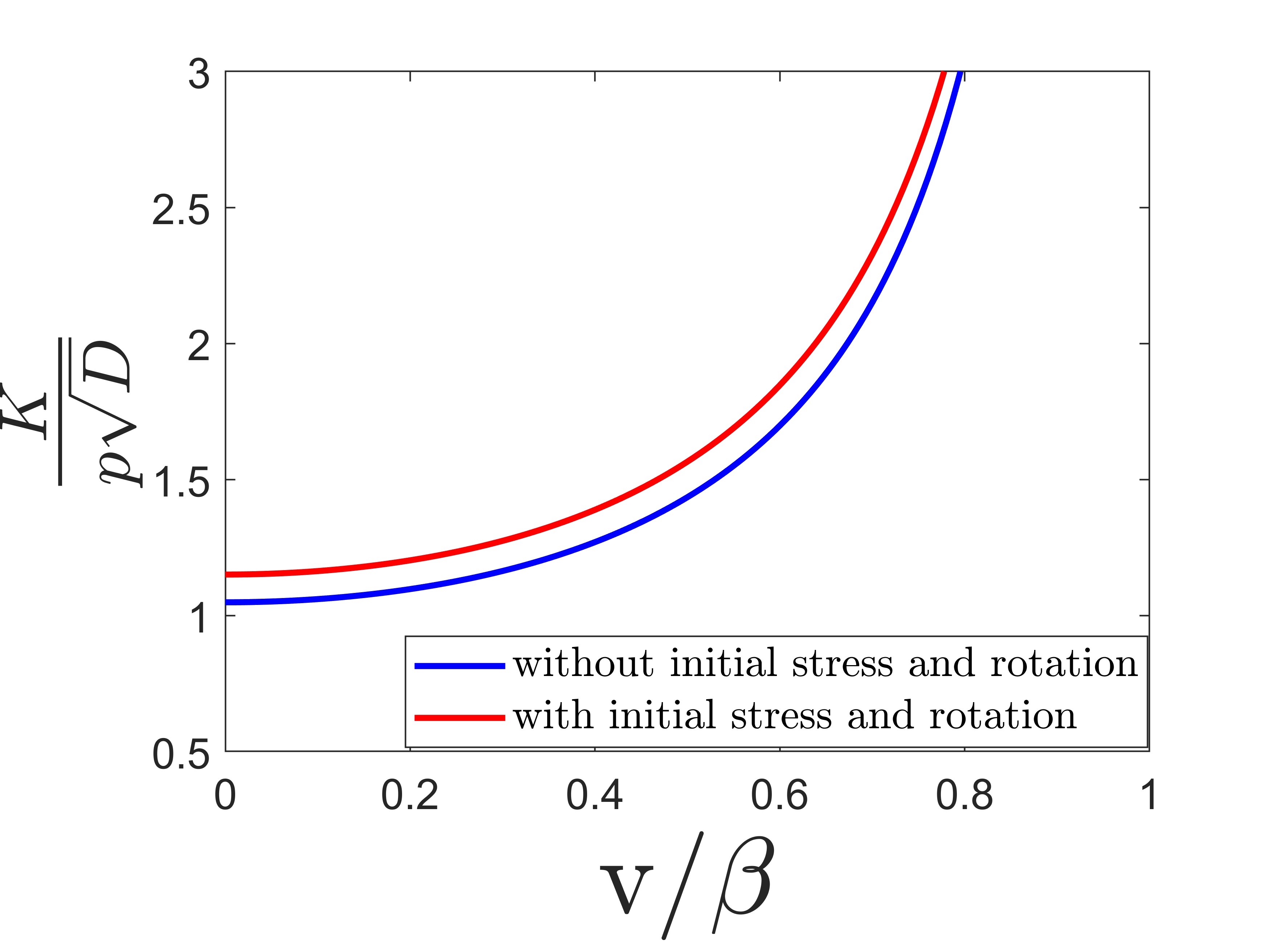}
\caption{}
\label{Fig_38}
\end{subfigure}
~
\begin{subfigure}[b] {0.38\textwidth}
\includegraphics[width=\textwidth ]{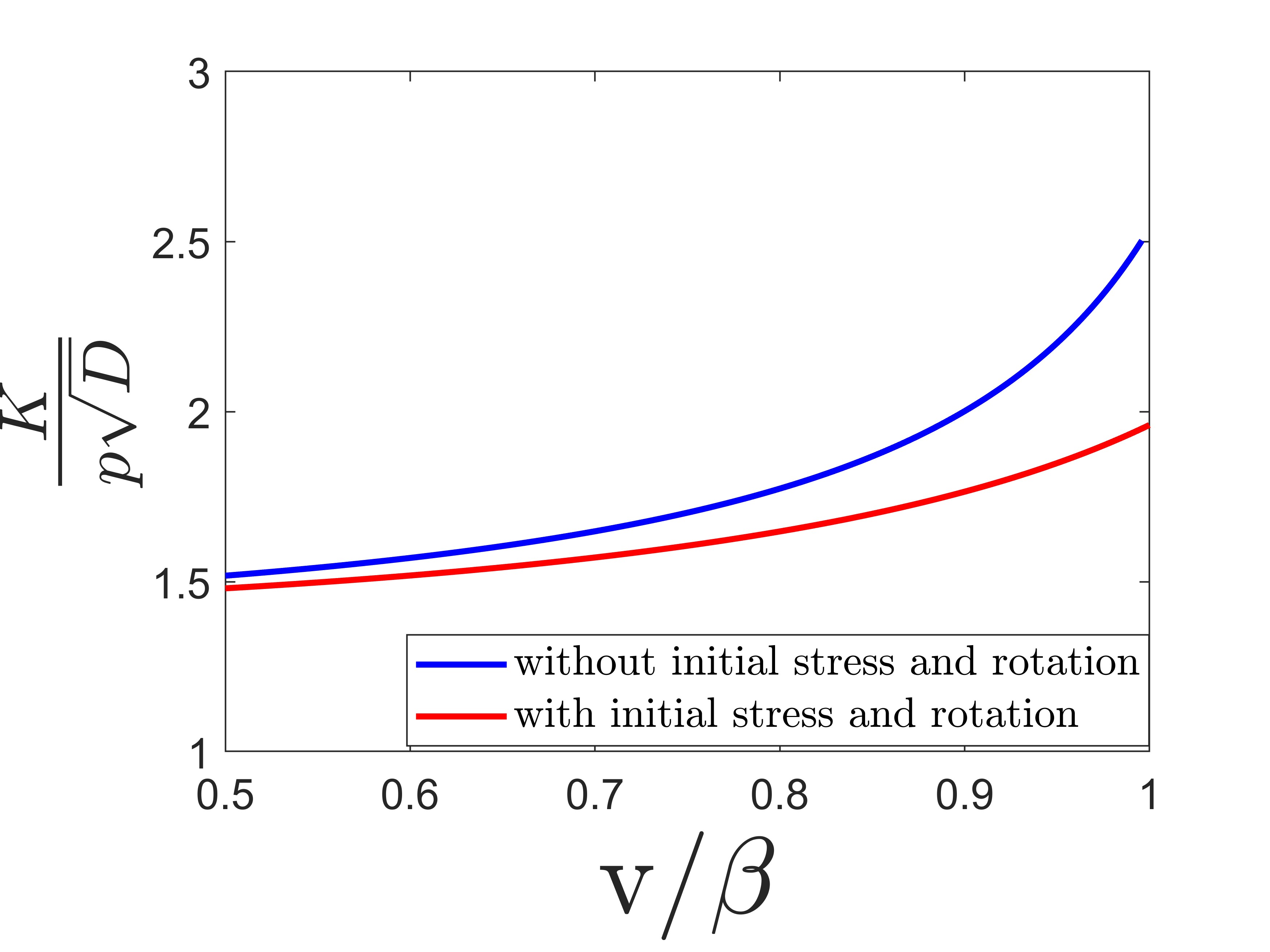}
\caption{}
\label{Fig_39}
\end{subfigure}
~
\begin{subfigure}[b] {0.38\textwidth}
\includegraphics[width=\textwidth ]{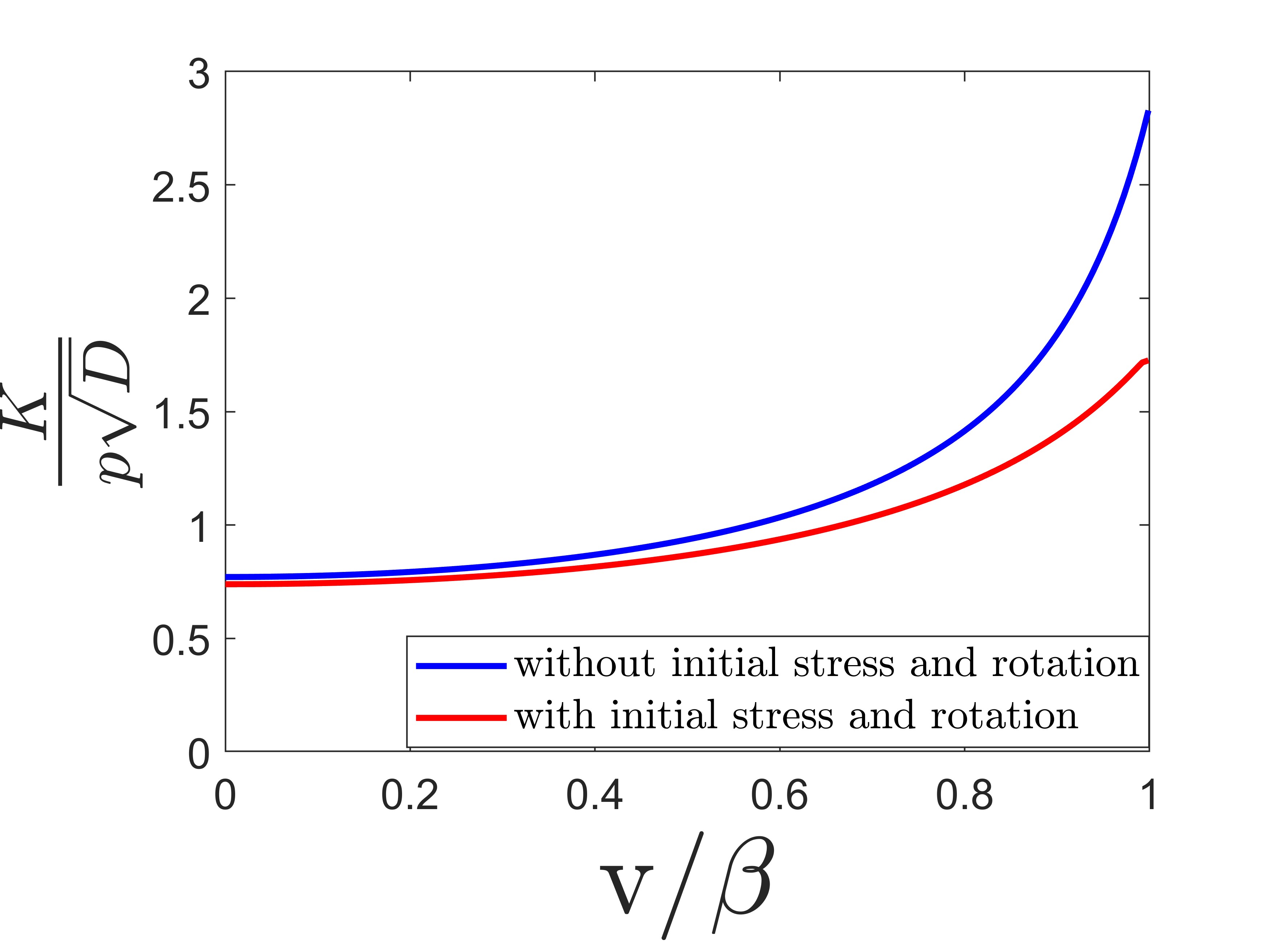}
\caption{}
\label{Fig_40}
\end{subfigure}
\caption{Impact of initial pressure $(P)$ and angular velocity $(\Omega)$ on dimensionless stress intensity factor $(K/p\sqrt{D})$ across various materials (i) carbon fiber (ii) steel (iii) isotropic material}
\label{Figure 7}
\efg
%%%====== END graphs here ============================

\section{Conclusions}
The proposed study studied the crack propagation analysis in an initially stressed self-reinforced rotating material medium. The closed expression has been obtained for the present study. 
A distinct mathematical technique is utilized to streamline the resolution of a pair of singular integral equations featuring First-order singularities. These obtained equations help us understand how the fracture behaves. Point loading at the crack edge is addressed using the Dirac-delta function. The study derives the SIF expression at the crack tip under constant loading conditions using the properties of the Hilbert transformation.
Through numerical simulations, the study visually demonstrates the significant impacts of various parameters such as magnetoelastic interaction, rotation, intensity force without punch pressure, crack length without dimension, initial stress, load positions, and dimensionless crack speed on the stress intensity factor for both initially stressed self-reinforced rotating materials and isotropic materials.

Additionally, the results of the current investigation can be outlined as follows
\begin{itemize}
    \item  The SIF at the front of a crack's progression is significantly affected by the rotation parameter as the considered plane moves within the analyzed slab.
    \item The dimensionless crack length directly influences the SIF at the crack's front, with SIF increasing as the crack length grows. 
    \item The SIF for a crack placed at the center increases as the force of the pressure on the slab's edges rises.
    \item As the magnitude of various point loads acting on the strip boundaries increases the stress intensity factor for a centrally positioned crack under consistent point loading also rises.
    \item The comparison shows that the stress intensity factor is higher in a steel strip than in the isotropic material strip and the carbon fiber/epoxy resin strip.
\end{itemize}
This study will be useful in fracture mechanics by examining crack growth, stress-corrosion crack growth behaviors, and structures' durability, as well as in analyzing and designing fracture-tolerant materials in various reinforced structures like bridges, slopes, tunnels, and buildings.\\

% \noindent {\bf Acknowledgements}\\
% The authors thank the National Institute of Technology Hamirpur (Department of Mathematics and Scientific Computing), Hamirpur, India, for providing a research fellowship to Ms. Diksha.\\

\noindent {\bf Conflicts of interest}\\
The authors state that they have no conflicts of interest.\\

\noindent {\bf Code availability }\\
The present work exclusively employs code developed by the authors in MATLAB, with no utilization of third-party code. 

\clearpage
\bibliography{sources}
\end{document}